\documentclass[11pt,pmlr,cleveref]{jmlr}
\jmlrproceedings{}{}
\jmlrpages{}

\usepackage{algorithm}
\usepackage{enumitem}
\usepackage{caption}
\usepackage{booktabs}
\usepackage[normalem]{ulem}
\usepackage{tablefootnote}
\usepackage{titletoc}
\usepackage{float}

\numberwithin{algorithm}{section}

\newenvironment{proofof}[1]{\begin{proof}\textbf{of {#1}}}{\end{proof}}

\newcommand{\mc}[1]{\ensuremath{\mathcal{#1}}}

\newcommand{\mbb}[1]{\ensuremath{\mathbb{#1}}}
\newcommand{\norm}[1]{\ensuremath{\left\lVert #1\right\rVert}}

\DeclareMathOperator{\TV}{TV}

\newcommand{\num}{N}
\newcommand{\len}{T}
\newcommand{\tranmat}{P}
\newcommand{\eps}{\varepsilon}
\newcommand{\del}{\delta}

\newcommand{\event}{\mc{E}}

\newcommand{\esterr}{\alpha}

\newcommand{\EE}{\mathbb{E}}
\newcommand{\PP}{\mathbb{P}}

\newcommand{\cA}{\mathcal{A}}
\newcommand{\cB}{\mathcal{B}}
\newcommand{\cM}{\mathcal{M}}
\newcommand{\cP}{\mathcal{P}}
\newcommand{\Del}{\Delta}
\newcommand{\Unif}{\mathrm{Unif}}
\newcommand{\Pois}{\mathrm{Pois}}
\newcommand{\iid}{\mathrm{i.i.d.}}

\title[Differentially Private Learning of Markov Chains]{Differentially Private Learning of Markov Chains: Near-Optimal Sample Complexity for Item and User Level Privacy}
\usepackage{times}

\author{
 \Name{Devansh Gupta} \Email{guptadev@usc.edu}\\
 \addr University of Southern California
 \AND
 \Name{Crist\'obal Guzm\'an} \Email{crguzmanp@uc.cl}\\
 \addr Pontificia Universidad Cat\'olica de Chile
 \AND
 \Name{Enayat Ullah} \Email{enayat@meta.com}\\
 \addr Meta
}

\begin{document}

\maketitle

\begin{abstract}
Sequential data such as documents, conversations, and trajectories are naturally
partitioned across users, motivating privacy guarantees that protect each user's
entire trajectory. We initiate the study of privately estimating the
transition matrix of a discrete ergodic Markov chain from $N$ independent trajectories of
length $T$, with arbitrary initial distributions, under maximum row-wise $\ell_1$
error. Besides $N$, $T$, and the number of states, our bounds depend on the minimum stationary probability $\pi_{\min}$ and
the pseudo-spectral gap $\gamma_{ps}$, which govern state coverage and mixing.
Under item-level privacy, we give upper and lower bounds that separate the costs
of sampling, state occupancy and mixing, and privacy.
Under user-level privacy,
we reduce the problem for sufficiently long trajectories to private
product-distribution estimation.
Along the way, we establish optimal
user-level private bounds for estimating products of discrete distributions,
which may be of independent interest.
In the user-level setting, our bounds are nearly tight in the large $T$ regime, whereas in the item-level setting, our bounds as tight upto a non-private term.
\end{abstract}


\section{Introduction}
Sequential data arise in a wide range of applications, including language
modeling, mobility analysis, healthcare, recommendation systems, and web
navigation. In many such settings, the data are naturally partitioned across
individuals: each user contributes a document, a conversation, or a trajectory
of interactions. Since the observations within a user's trajectory may
collectively reveal sensitive information, it motivates protection of privacy of the user's
entire trajectory rather than each observation in isolation. This issue is
particularly relevant to modern language models, which are often trained or
adapted using collections of user-generated documents and conversations.

Differential privacy (DP) \citep{dwork2006calibrating, dwork2006differential} formalizes this guarantee of protection by requiring the
distribution of an algorithm's output to change little when one protected unit
is replaced. The choice of unit is crucial: \emph{item-level} DP protects a
single interaction, whereas \emph{user-level} DP protects all observations
contributed by one individual jointly. For sequential data, protecting one
item may still reveal patterns across a document, conversation, or trajectory,
making user-level privacy the natural goal.

A substantial literature on user-level private statistics studies
\emph{grouped i.i.d.}\ data, where each user contributes repeated independent
samples and within-user concentration can reduce the cost of privacy
\citep{liu2020discrete,levy2021user,esfandiari2022tight,ghazi2023user}.
Markov trajectories differ in two important respects: observations within a
trajectory are dependent, and the number of transitions observed from each
state is random. Consequently, private estimation must account for temporal
dependence, mixing, and state coverage in addition to within-user
concentration \citep{pmlr-v98-wolfer19a,MarkovChainEstimationBernoulli}.

Motivated by this setting, we study the problem of privately learning a
discrete Markov chain. We observe $N$ independent trajectories, each of length $T$,
from the same ergodic Markov chain on $S$ states and seek to estimate its
transition matrix $P$. Accuracy is measured by the maximum row-wise $\ell_1$
error, so every conditional next-state distribution must be learned
simultaneously. We parameterize the difficulty of the chain by its minimum
stationary probability $\pi_{\min}$ and pseudo-spectral gap $\gamma_{ps}$,
which control state coverage and mixing, respectively. We make no stationarity
assumption on the starting states: every trajectory may begin from an
arbitrary initial distribution. For a target error $\alpha$, privacy parameters
$(\varepsilon,\delta)$, and a fixed trajectory length $T$, we ask:
\begin{quote}
\emph{How many users $N$ are necessary and sufficient to estimate $P$
accurately under item-level or user-level differential privacy?}
\end{quote}
Formal definitions and the complete problem formulation are given in Section~\ref{sec:prelims}.

\begin{table}[t]
    \centering
    \begin{tabular}{ccc}
        \toprule
        & \textbf{Upper Bound} & \textbf{Lower Bound} \\
        \midrule
        \textbf{Item-Level DP}
            & $\underset{\text{\Cref{thm:item-level-dp-upperbound-informal,thm:pure-dp-item-level-upper-bound:utility,thm:pure-dp-item-level-upper-bound:privacy}}}{\Tilde{O}\left(\frac{S}{\pi_{\min}T\alpha^2} + \frac{1}{\gamma_{ps} \pi_{\min}T} + \frac{S}{\varepsilon \pi_{\min}T\alpha}\right)}$
            & {$\underset{\text{\Cref{thm:item-level-dplowerbound-informal,thm:item-level-lowerbound-formal}}}{\Tilde{\Omega}\left(\frac{S}{\pi_{\min}T\alpha^2} + \frac{S}{\varepsilon \pi_{\min}T\alpha}\right)}$} \\

        $\underset{T = \tilde{\Omega}\left((\gamma_{ps} \pi_{\min})^{-1}\right)}{\text{\textbf{User-Level $(\varepsilon, \delta)$-DP}}}$
        & $\underset{\text{\Cref{thm:user-level-markov-approx-upper-informal,thm:user-level-markov-approx-upper-formal-privacy,thm:user-level-markov-approximate}}}{\tilde{O}\left(\frac{\log\left(\frac{1}{\delta}\right)}{\varepsilon} + \frac{S}{\pi_{\min}T\alpha^2} + \frac{S^{3/2}\log^{3/2}\left(\frac{1}{\delta}\right)}{\alpha\eps\sqrt{\pi_{\min}T}}\right)}$
        & $\underset{\text{\Cref{thm:user-level-markov-approx-lower-informal,thm:markov-chain-estimation-approx}}}{\tilde{\Omega}\left(\frac{M_\delta}{\varepsilon} + \frac{S}{\pi_{\min}T\alpha^2} + \frac{S^{3/2}}{\alpha\eps\sqrt{\pi_{\min}T}}\right)}$ \\

        $\underset{T = \tilde{\Omega}\left((\gamma_{ps} \pi_{\min})^{-1}\right)}{\text{\textbf{User-Level $\varepsilon$-DP}}}$
        & $\underset{\text{
        \Cref{thm:user-level-markov-pure-upper-formal-privacy,thm:user-level-markov-pure}
        }}{\tilde{O}\left(\frac{S^2}{\alpha\eps\sqrt{\pi_{\min}T}} + \frac{S^2}{\varepsilon} + \frac{S}{\pi_{\min}T\alpha^2}\right)}$
        & $\underset{\text{
        \Cref{thm:user-level-markov-pure-lower-formal}
        }}{\Omega\left(\frac{S^2}{\varepsilon} + \frac{S^{3/2}}{\alpha\eps\sqrt{\pi_{\min}T}} + \frac{S}{\pi_{\min}T\alpha^2}\right)}$ \\
        \bottomrule
    \end{tabular}
    \caption{Comparison of user-complexity upper and lower bounds under different privacy models.
    $M_\delta := \min\left\{S^2,\log\left(\frac{1}{\delta}\right)\right\}$. $\tilde{O}$ and $\tilde{\Omega}$ omit poly-logarithmic dependencies.
    Item-level bounds match in the sampling and privacy terms, while the upper-bound mixing term $1/(\gamma_{ps}\pi_{\min}T)$ has no counterpart in the displayed lower bound \tablefootnote{\label{fn:item-dp} The displayed item-level upper bound is achieved under pure \(\varepsilon\)-DP and therefore also holds under \((\varepsilon,\delta)\)-DP for every \(\delta\). It is reported as a \(\delta\)-uniform upper bound; see Appendix~\ref{ssec:approxDP} for the dedicated \(\delta\)-dependent refinement.}. For $(\varepsilon, \delta)$-DP ,the user-level, bounds are tight except for dependence on $\log\left(1/\delta\right)$. For $\varepsilon$-DP, the sampling and $S^2/\varepsilon$ terms match, while the privacy-accuracy interaction term is loose by a factor of $\sqrt{S}$.
    }
    \label{tab:privacy-bounds}
\end{table}

\paragraph{Our contributions.}
We initiate the study of privately estimating an unknown transition matrix
directly from Markovian trajectories.

\begin{itemize}[leftmargin=1em]
    \item \textbf{Item-level private Markov-chain estimation.}
    We establish upper and lower bounds for transition-matrix estimation under item-level privacy\textsuperscript{\ref{fn:item-dp}}, separating the costs of ordinary sampling, state occupancy and mixing, and privacy, as summarized in Table~\ref{tab:privacy-bounds}. The ordinary-sampling and privacy-accuracy interaction terms match up to logarithmic factors.

    \item \textbf{User-level private Markov-chain estimation.}
    For sufficiently long trajectories, we reduce user-level private Markov-chain estimation to private product-distribution estimation for both upper and lower bounds. The resulting Markov-chain rates under approximate and pure DP are summarized in Table~\ref{tab:privacy-bounds}
    respectively. Under approximate DP, our upper and lower bounds match up to log factors, as observed in Table~\ref{tab:privacy-bounds}.

    \item \textbf{User-level private product-distribution estimation.}
    As an intermediate result of independent interest, we establish upper and lower bounds for estimating \(d\)-fold products of \(k\)-ary distributions from \(m\) samples per user under maximum row-wise \(\ell_1\) loss under pure and approximate DP. At the user level, \citet{ghazi2023user} study learning $d$-fold $k$-ary product distributions in joint total variation under pure DP. For $k=2$, \citet{esfandiari2022tight} give approximate-DP guarantees for global $\ell_2$ mean estimation, which apply to Bernoulli products and yield bounds under maximum row-wise \(\ell_1\) loss. To our knowledge, ours is the first near-tight characterization of $d$-fold $k$-ary product estimation under maximum row-wise $\ell_1$ loss in the approximate user-level DP setting, and the first lower bound capturing the joint dependence on $k$ and $d$. These results may be of independent interest and also underpin our Markov-chain guarantees under user-level privacy.
\end{itemize}

\subsection{Overview of techniques}

Our proofs use the row-wise structure of transition-matrix estimation
together with reductions that preserve the relevant neighboring
relation. At the item level, this leads to privatizing transition
counts directly. At the user level, sufficiently long trajectories can
instead be related to repeated sampling from a product distribution.

\paragraph{Item-level private learning of Markov chains.}
In this case, we aggregate the transition-count matrix across
all trajectories; changing one observation affects only a constant
number of counts, so this matrix has constant global sensitivity.
Adding Laplace or Gaussian noise and then normalizing the rows gives the
private estimator.
The analysis combines row-wise distributional concentration with Markov-chain occupation-time concentration from \citet{leskela2026robust}. Their role is to separate
ordinary sampling error and privacy noise from the mixing and coverage
requirements. For the lower bound, a rare-row embedding places an
arbitrary categorical distribution in the transitions from a state of
stationary mass \(\pi_{\min}\). An adjacency-preserving coupling
transfers private categorical-estimation lower bounds to this row.

Under user-level privacy, longer trajectories should provide an
advantage: a length-\(T\) trajectory contains about \(T\pi_i\)
transitions from state \(i\), so each user's empirical transition
profile becomes more concentrated as \(T\) grows. A mechanism based
only on global sensitivity misses this gain, since replacing one user
can alter all \(T\) transitions. Our approach instead exploits this
within-user concentration once trajectories are sufficiently long.

\paragraph{User-level private learning of Markov chains.}
The upper bound combines occupation-time concentration \citep{MarkovChainEstimationBernoulli} with an
adjacency-preserving reduction to product-distribution estimation.
Beyond the mixing threshold, each trajectory supplies
\(\Theta(\pi_{\min}T)\) outgoing transitions from every state. A
transition-stack coupling converts these transitions into a user block
of repeated samples from the product of the transition rows. The role
of mixing is therefore to guarantee uniform row coverage; the remaining
privacy and estimation problem is handled by our product-distribution estimator.

For the converse, we again reduce to the complexity of product distribution estimation. We construct a two-layer Markov
chain, embedding the unknown product marginals into designated transitions,
while a data-independent trajectory skeleton makes the simulation
user-wise and hence adjacency preserving. These reductions transfer
both upper and lower bounds and show that, in the long-trajectory
regime, user-level Markov-chain estimation reduces to
product-distribution estimation up to logarithmic losses.
This yields matching approximate-DP bounds up to logarithmic factors;
the pure-DP results inherit the remaining product-distribution gap.

\paragraph{User-level private product-distribution estimation.}
For the approximate-DP upper bound, we use concentration-aware private
mean estimation. Within-user averaging makes the empirical distribution matrices concentrate in a Frobenius ball. The Friendly-Core framework
\citep{tsfadia2022friendlycore} uses a stable filter to certify this
small diameter before applying Gaussian averaging. Its role is to
replace the worst-case sensitivity of an arbitrary user block by the
smaller scale revealed by the \(m\) samples within each user.

For the approximate-DP lower bound, we introduce a paired Poisson
estimation problem. In each marginal, categories are grouped into pairs
with independent Poisson counts with means \(m(q+\theta_{j,b})\) and
\(m(q-\theta_{j,b})\). The opposite perturbations
keep every marginal normalized while providing a convenient product
structure across the \(d\) marginals. We prove a lower bound for this
intermediate problem using the Poissonization and fingerprinting
techniques of \citet{esfandiari2022tight} and \citet{pmlr-v99-kamath19a}. The ordinary sampling and small-user
terms follow from standard nonprivate and packing arguments. For pure
DP, row-wise private estimation gives the upper bounds and product packings give the
corresponding lower bounds.

\subsection{Related Work} Estimation of Markov chains is a well-studied area, spanning classical asymptotic inference \citep{anderson1957statistical,billingsley1961statistical} and finite-sample transition-matrix estimation from a single trajectory \citep{hao2018learning,pmlr-v98-wolfer19a,MarkovChainEstimationBernoulli, leskela2026robust}.
In parallel, user-level differential privacy has been studied for grouped i.i.d.\ data, particularly for mean and discrete-distribution estimation \citep{liu2020discrete,levy2021user,esfandiari2022tight,ghazi2021correlated,ghazi2023user}.
Most closely related to our setting are mechanisms for privatizing stochastic matrices \citep{fallin2023differential,benvenuti2026modeling}.
Unlike these works, we estimate an unknown transition matrix directly from Markovian trajectories under both item- and user-level privacy.
A more detailed overview of these works can be found in Appendix~\ref{sec:relatedwork}.

\subsection{Preliminaries}
\label{sec:prelims}
We begin by introducing the notation and preliminary definitions needed to understand the problem setup and the results in this paper. Let $[S] := \{1,\ldots,S\}$ denote the state space, where $S$ is the number of states of the Markov chain. We regard probability distributions on $[S]$ as row vectors. For any matrix $P \in \mathbb{R}^{k \times d}$, the notation $P_{i,j}$ denotes its entry in the $i^{\text{th}}$ row and $j^{\text{th}}$ column, and $P_{(i,\cdot)}$ denotes its $i^{\text{th}}$ row whereas $P_{(\cdot, j)}$ represents the $j^{\text{th}}$ column. The vector $\mathbf{e}_i \in \mathbb{R}^S$ denotes the $i^{\text{th}}$ standard basis vector. $\Delta(k)$ represents the $k$-dimensional standard simplex, and for any $p \in \Delta(k)$, $X \sim p$ corresponds to sampling an element from $[k]$ given that $p$ is the probability distribution over $[k]$. $a\gtrsim b$ means that there exists a universal constant $C>0$ such that $a \geq Cb$ and $a\lesssim b$ means there exists a universal constant $C>0$ such that $a\leq Cb$.

Standard vector norms are denoted by $\norm{\mu}_1 := \sum_i |\mu_i|$, $\norm{\mu}_2 := \left(\sum_i \mu_i^2\right)^{1/2}$, and $\norm{\mu}_\infty := \max_i |\mu_i|$. For two probability distributions $\mu$ and $\nu$ on $[S]$, their total variation distance is $d_{\mathrm{TV}}(\mu,\nu) := \frac{1}{2}\norm{\mu-\nu}_1 = \frac{1}{2}\sum_{i=1}^S |\mu_i-\nu_i|$. For matrices, $\norm{P}_2$ denotes the spectral norm, while $\norm{P}_\infty := \max_i \norm{P_{(i,\cdot)}}_1$ denotes the maximum row-sum norm. These are the operator norms induced by the $\ell_2$ and $\ell_\infty$ norms on $\mathbb{R}^S$, respectively. A matrix $P \in \mathbb{R}^{k \times d}$ is a stochastic matrix if $P_{(i,\cdot)} \in \Delta(k)$ for each $i \in [d]$. For two stochastic matrices $P$ and $Q$,
\[
    \rho(P,Q) = \norm{P-Q}_\infty
    =
    \max_{i\in[S]}
    \norm{P_{(i,\cdot)}-Q_{(i,\cdot)}}_1
    =
    2\max_{i\in[S]}
    d_{\mathrm{TV}}\!\left(
        P_{(i,\cdot)},Q_{(i,\cdot)}
    \right).
\]

In this work, we consider ergodic Markov chains on the state space $[S]$, with transition matrix $P$ and stationary distribution $\pi$. The difficulty of estimating $P$ is parameterized by three basic quantities: the number of states $S$; the minimum stationary mass $\pi_{\min}:=\min_{i\in[S]}\pi_i$, which controls how often the least frequently visited state is observed at equilibrium; and the pseudo-spectral gap $\gamma_{\mathrm{ps}}(P)$, which quantifies the rate at which the chain mixes. These Markov chain notions are standard \citep{Levin:2017}, with the pseudo-spectral gap following \citet{paulinconcenteration}. For completeness, formal definitions and further discussion are provided in Appendix~\ref{app:markov-preliminaries}.

\paragraph{Problem formulation.}
Let $\mathcal{C}(S,\pi_{\min},\gamma_{ps}) :=\left\{ P:\ P \text{ is ergodic}, \pi_{\min}(P)\geq\pi_{\min}, \gamma_{ps}(P)\geq\gamma_{ps} \right\}$.
For each \(u\in[\num]\), let
\(X^{(u)}_{1:\len}=(X^{(u)}_1,\ldots,X^{(u)}_{\len})\) be an independent
length-\(\len\) trajectory from an ergodic Markov chain on \([S]\) with
transition matrix \(\tranmat \in \mathcal{C}(S,\pi_{\min},\gamma_{ps})\), started from an arbitrary initial
distribution, and write $D=\bigl(X^{(1)}_{1:\len},\ldots,X^{(\num)}_{1:\len}\bigr)$. An estimator $\mathcal M$ outputs a row-stochastic matrix
\(\widehat{\tranmat}=\mathcal M(D)\). Following \citet{MarkovChainEstimationBernoulli}, we measure its error using the maximum
row-wise \(\ell_1\) distance $\rho(\widehat{\tranmat},\tranmat)$. This loss ensures that every state’s next-state distribution, including those of rare states, is accurate in total variation, which is half the corresponding row-wise \(\ell_1\) distance.
For accuracy \(\esterr\in(0,1)\) and failure probability \(f\in(0,1)\),
we say that $\mathcal M$ is $(\esterr,f)$-accurate if
\[
    \inf_{\tranmat \in \mathcal{C}(S, \pi_{\min}, \gamma_{ps})}
    \mathbb{P}\!\left(
        \rho(\tranmat,\widehat{\tranmat})\leq\esterr
    \right)
    \geq 1-f.
\]

This guarantee is required uniformly over arbitrary initial distributions. The probability is over both the observed trajectories and the internal randomness of the estimator.

\begin{definition}[Differential privacy \citep{dwork2006calibrating}]
Let $\varepsilon \ge 0$ and $\delta \in [0,1)$. A randomized
algorithm $\mathcal{A} \colon \mathcal{S} \to \mathcal{Y}$
is
$(\varepsilon,\delta)$-differentially private (DP) if, for every pair
of neighboring datasets $S,S' \in \mathcal{S}$ and every
measurable set $V \subseteq \mathcal{Y}$,
\[
    \Pr\!\left[\mathcal{A}(S) \in V\right]
    \le
    e^{\varepsilon}
    \Pr\!\left[\mathcal{A}(S') \in V\right]
    + \delta,
\]
where the probabilities are taken over the internal randomness
of $\mathcal{A}$. When $\delta=0$, we say that $\mathcal{A}$
satisfies pure $\varepsilon$-DP.
\end{definition}

Our goal is to estimate the transition-matrix using an estimator $\mathcal{M}$ that is $(\varepsilon, \delta)$-DP (or $\varepsilon$-DP). In the preceding definition, the input space of the algorithm is $([S]^{\len})^{\num}$, and its output space is the set of row-stochastic $S\times S$ matrices.  We consider two neighboring relations on this common input space, corresponding to two different units of privacy:
\begin{enumerate}
    \item In the \emph{item-level setting}, two datasets $D$ and $D'$ are neighbors if there exists at most one $u \in [N]$ and $t \in [T]$ such that $X_t^{(u)} \neq {X_t^{(u)}}'$.
    \item In the \emph{user-level setting}, two datasets $D$ and $D'$ are neighbors if there exists at most one $u \in [N]$ such that $X^{(u)}$ and ${X^{(u)}}'$ differ arbitrarily.
\end{enumerate}

In both settings, differential privacy must hold pointwise for every pair of deterministic neighboring inputs. The Markov chain model and the randomness of the observed trajectories enter only in the accuracy guarantee. For a fixed trajectory length $\len$, we seek to characterize the number of trajectories $\num$ required for an $(\eps,\del)$-DP (or pure $\varepsilon$-DP) estimator to satisfy the uniform row-wise estimation guarantee above.

\section{Learning Markov Chains under Item-Level Privacy}
\label{sec:ItemLevelEstimation}

We begin by studying transition matrix estimation under item-level privacy. Recall that datasets \(D\) and \(D'\) are item-level neighbors if they agree on every trajectory except
possibly one and the exceptional trajectory may only differ by one entry. This setting isolates the main statistical and privacy costs before we turn to the user-level privacy model.

The natural non-private estimator is obtained from the empirical transition counts. For \(i,j\in[S]\), let $K_{i,j}$ be the total count of $i \rightarrow j$ transitions occurring across all chains and let $K_i$ be the number of observed departures from state $i$.
Whenever \(K_i>0\), the empirical estimator is
\[
    \widehat P^{\mathrm{np}}_{i,j}:=\frac{K_{i,j}}{K_i}.
\]
If \(K_i=0\), its \(i\)-th row can be set to any fixed probability
vector, such as the uniform distribution. This is the usual estimator
studied by \citet{leskela2026robust}. For a sufficiently
long trajectory, concentration of the occupation counts gives
\(K_i\asymp NT\pi_i\), and the normalized counts in row \(i\) concentrate
around \(P_{(i,\cdot)}\). The non-private analysis
therefore has two parts: controlling the empirical conditional
distribution of the outgoing transition from each state and ensuring
that every state is visited often enough along the dependent
trajectory. These yield, up to logarithmic factors,
$
    N = \tilde{\mathcal{O}}\left(\frac{S}{\pi_{\min}T\alpha^2} + \frac{1}{\gamma_{ps}\pi_{\min}T}\right).
$

\paragraph{The private estimator.} We now turn to the item-level private setting. Our proposed estimator (as shown in Algorithm~\ref{alg:private-transition-estimator-multiple-chains}) retains the same count-and-normalize structure, but first privatizes the empirical transition counts. Directly perturbing \(\widehat P^{\mathrm{np}}\) is inconvenient because its sensitivity depends on the potentially small row counts \(K_i\).
Instead, our estimator perturbs the
raw transition counts, which have constant global sensitivity. Changing one state affects at most its two incident transition records; viewed separately, their joint change has Frobenius norm at most \(2\), while
the aggregated count matrix has \(\ell_1\)-sensitivity at most \(4\). Therefore, adding independent \(\operatorname{Lap}(0,4/\eps)\) noise to the counts gives \(\eps\)-differential privacy. Taking absolute values and normalizing the rows are post-processing operations that produce a valid transition matrix.

\begin{theorem}[Complete version in \Cref{thm:pure-dp-item-level-upper-bound:utility,thm:pure-dp-item-level-upper-bound:privacy}]
\label{thm:item-level-dp-upperbound-informal}
Fix \(0<\alpha,f<1\) and \(0<\eps\leq 1\). Uniformly over
\(P\in\mathcal{C}(S,\pi_{\min},\gamma_{ps})\) and arbitrary initial distributions, there exists an algorithm which
is \(\eps\)-differentially private and achieves $\rho(\widehat P,P) \leq \alpha$
with probability $1 - f$ for
$N
    =
    \widetilde{\mathcal{O}}\!\left(
        \frac{S}{\pi_{\min}T\alpha^2}
        +\max\left\{\frac{1}{\gamma_{ps}}, 1\right\}\frac{1}{\pi_{\min}T}
        +\frac{S}{\pi_{\min}T\alpha\eps}
    \right).
$
\end{theorem}

\begin{proof}{\textbf{Sketch.}}
    Privacy follows from the sensitivity calculation above and the privacy guarantee of the Laplace mechanism. For utility, it is more convenient to analyze the perturbed counts before row normalization. This separates the ordinary statistical error from the additional error due to privacy. In particular, for every row \(i\),
\[
    \norm{\widehat P_{(i,\cdot)}-P_{(i,\cdot)}}_1
    \leq
    \norm{\widehat P^{\mathrm{np}}_{(i,\cdot)}
          -P_{(i,\cdot)}}_1
    +
    \norm{\widehat P_{(i,\cdot)}
          -\widehat P^{\mathrm{np}}_{(i,\cdot)}}_1.
\]
The first term is controlled by the aforementioned non-private concentration
argument. Recall that $K_i$ denotes the number of departures from state $i$. On the event that every \(K_i\) is of order \(NT\pi_i\), we show that the added noise contributes at most
$\widetilde{\mathcal{O}}\!\left(\frac{S}{NT\pi_{i}\eps}\right)$ to the row-wise $\ell_1$ error in row $i$ yielding
an overall additive term of $\widetilde{\mathcal{O}}\!\left(\frac{S}{NT\pi_{\min}\eps}\right)$. Combining it with non-private estimation gives the three terms in the stated sample complexity.
\end{proof}

The complete statement and full proof of this claim can be found in
Appendix~\ref{app:upperbound-item-formal}. Every \(\eps\)-DP
algorithm is also \((\eps,\del)\)-DP for every \(\del\geq 0\), so the estimator above already provides approximate-DP guarantees. In Appendix~\ref{ssec:approxDP}, we also give a Gaussian-mechanism based variant with an explicit \((\eps,\del)\)-DP guarantee that achieves the same sample complexity up to logarithmic factors. This naturally leads to the question of whether the rate can be improved under approximate DP; the following lower bound shows that it is essentially tight.

\begin{theorem}[Complete version in \Cref{thm:item-level-lowerbound-formal}]
\label{thm:item-level-dplowerbound-informal}
For $\varepsilon < 1$ and
$\del = o\!\left(\frac{\varepsilon^{3/2}\alpha}{S^2\log^2\!\left(\frac{S}{\alpha\varepsilon}\right)}\right)$, every item-level \((\eps,\del)\)-differentially private estimator that achieves error at most \(\alpha\) with probability at least \(3/4\), uniformly over
\(\mathcal{C}(S,\pi_{\min},\gamma_{ps})\), requires
$
    N
    =\widetilde{\Omega}\!\left(\frac{S}{\pi_{\min}T\alpha^2} +\frac{S}{\pi_{\min}T\alpha\eps}\right).
$
\end{theorem}

\begin{proof}{\textbf{Sketch.}} The lower bound follows from a reduction to the i.i.d. discrete distribution estimation problem. The idea is that, given $M$ i.i.d. samples from a discrete distribution $p \in \Delta(S-1)$, we can carefully design corresponding Markov chains of total length $\frac{M}{\pi_{\min}}$ with high probability. We do this by embedding $p$ in the least frequently accessed row of the transition matrix $M^{(p)}$. \citet{MarkovChainEstimationBernoulli} use this matrix for a non-private lower bound, but we exploit its structure for a privacy-preserving reduction.
\begin{align*}
    M^{(p)} = \begin{pmatrix}
    \frac{1 - \pi_{\min}}{k} \mathbf{1}_{S-1}\mathbf{1}_{S-1}^\top &\pi_{\min} \mathbf{1}_{S-1} \\
    (1 - \pi_{\min})p & \pi_{\min}
    \end{pmatrix}.
\end{align*}
The key property that we use is that the above Markov chain can go from one row to another row without requiring samples from $p$ unless the last row is explicitly accessed. Thus, if we initialize from a carefully chosen distribution, then we observe that the probability of reaching that row is $\pi_{\min}$ independently. Moreover, this reduction is privacy preserving: changing one i.i.d. sample changes only the state generated after the corresponding visit to the last row. Since all the other rows are identical, the remainder of the two sets of trajectories can be coupled to agree, so item level neighboring i.i.d. datasets produce item-level neighboring trajectories. Hence, if we need to access the row $M$ times, then with high probability, we need a total chain of length $\frac{M}{\pi_{\min}}$. The $\norm{\cdot}_\infty$ guarantee of the Markov chain estimator gives a $d_{\TV}$ error guarantee for estimating $p$ upto constant factors. With this established reduction, a corollary of \citet{esfandiari2022tight} (proven in Appendix~\ref{app:lowerbound-item-formal}) shows that $M = \tilde{\Omega}\left(\frac{S}{\alpha \varepsilon} + \frac{S}{\alpha^2}\right)$ samples are necessary for estimation. Thus, with the reduction, the chain length of $\widetilde{\Omega}\!\left(\frac{S}{\pi_{\min}T\alpha^2} + \frac{S}{\pi_{\min}T\alpha\eps}\right)$ is necessary to achieve error $\alpha$.
\end{proof}

The complete statement and full proof of this claim can be found in
Appendix~\ref{app:lowerbound-item-formal}. Since every \(\eps\)-DP
algorithm is also \((\eps,\del)\)-DP, the same lower bound holds for pure $\varepsilon$-DP algorithms.

\section{Learning Markov Chains under User-Level Privacy}
\label{sec:userlevelest}
In this section, we study the user complexity of learning Markov chains under user-level privacy, where
each user contributes an entire Markov chain trajectory. Recall that datasets \(D\) and \(D'\) are user-level neighbors if they agree on every trajectory except possibly one and the exceptional trajectories may differ arbitrarily.

\paragraph{A direct extension of the item-level estimator.}
Algorithm in Appendix~\ref{app:upperbound-item-formal}
extends directly by
aggregating transition counts across users and recalibrating the noise. A length-\(\len\) trajectory contributes \(\len-1\) transitions, so
replacing one user changes the aggregate count matrix by at most
\(2(\len-1)\) in \(\ell_1\). Thus, adding Laplace
noise calibrated to sensitivity $T$, to the counts before the
same row-normalization step gives user-level pure \(\eps\)-DP, and hence \((\eps,\del)\)-DP, for every  \(\len\). This sensitivity calculation is deterministic; the Markov assumptions only enter in the utility analysis. On the event that the count in row \(i\)
is of order \(\num\len\pi_i\), the factor \(\len\) in the noise cancels after normalization, giving privacy error \(\widetilde{\mathcal O}(S/(\num\pi_{\min}\eps))\).
A complete statement and detailed analysis of this direct item-to-user-level extension are provided in Appendix~\ref{sec:itemtousertrivial}.

\paragraph{The Case of Large $T$.}
The direct extension above degrades as \(\len\) increases,
because the growth in the number of observed transitions is offset by the
corresponding growth in user-level sensitivity. For sufficiently long
trajectories, however, every state is visited often enough that each user
supplies many outgoing transitions from every row of \(\tranmat\). More
precisely, applying the occupation-count concentration inequalities of
\citet{paulinconcenteration} simultaneously over all users and states
shows that, with high probability, every state has at least $m_0:=\Theta\left(\pi_{\min}\len\right)$ departures in every trajectory when \(\len\) exceeds the required mixing
threshold. For
each such departure, the Markov property gives
\(X_{t+1}\mid X_t=j\sim\tranmat_{(j,\cdot)}\), so the outgoing transitions
from state \(j\) reveal samples associated with the \(j\)-th row of the transition matrix.

The reduction, described formally in
Appendix~\ref{ssec:markov-upper}, converts each state's collection to exactly \(m_0\)
entries by truncating excess transitions and padding a deficient
collection with independent uniform draws. It then aligns one entry from
each state into each of \(m_0\) observations in \([S]^S\). This gives a
fixed-size instance of the product-distribution problem whose target marginals are
\(\tranmat_{(1,\cdot)},\ldots,\tranmat_{(S,\cdot)}\). After coupling the reduced data to this ideal product sample, we may invoke the i.i.d. product-distribution estimators under the loss $\rho(\widehat{\tranmat}, \tranmat)$ as a black box.
Finally, the preprocessing is performed separately for each trajectory, and hence replacing one user changes only one output block, preserving user-level adjacency.
Combining this reduction with the product-distribution guarantees (Section~\ref{sec:kd_dist_bounds}) yields the following upper bound.

\begin{theorem}[Complete version in \Cref{thm:user-level-markov-approx-upper-formal-privacy,thm:user-level-markov-approximate}]
\label{thm:user-level-markov-approx-upper-informal}
Let \(0<\alpha,f<1\) and \(0<\eps\leq 1\). For trajectory lengths $\len = \tilde{\Omega}\left(\frac{1}{\gamma_{ps}\pi_{\min}}\right)$,
there exists a user-level \((\eps,\del)\)-differentially private
estimator for which
$
    \num
    =
    \widetilde{\mathcal O}\!\left(
        \frac{S}{\pi_{\min}\len\esterr^2}
        +\frac{S^{3/2}}{\eps\esterr\sqrt{\pi_{\min}\len}}
        +\frac{1}{\eps}
    \right)
$
users suffice to achieve \(\rho(\widehat P,P)\leq\alpha\) with
high probability uniformly over \(\tranmat\in\mathcal C(S,\pi_{\min},\gamma_{ps})\) and
arbitrary initial distributions.
\end{theorem}

The corresponding pure DP result is given in Appendix~\ref{ssec:pure-dp-upper-bounds}. We now discuss the approach to proving a lower bound for this setting.

\paragraph{Reduction and proof sketch.}
The lower bound is transferred from user-level estimation of product
distributions. Let \(Q=(q_1,\ldots,q_S)^\top\) be an \(S\times S\)
row-stochastic matrix and let
\(\mathsf D_Q=\bigotimes_{i=1}^S q_i\). In the source problem, each
user contributes \(m\) independent samples from \(\mathsf D_Q\), and
the goal is to estimate all the marginals \(q_i\) simultaneously. To
turn this into a Markov chain instance, partition the state space into two layers, \(\mathcal A=\{a_1,\ldots,a_S\}\)
\(\mathcal B=\{b_1,\ldots,b_S\}\), and define the corresponding block construction, $P_{\pi_{\min},Q}$, corresponding to the ${\cal A},{\cal B}$ partition.
\[
    P_{\pi_{\min},Q}
    =
    \begin{pmatrix}
        0 & (1-\eta)Q+\eta U\\[1mm]
        \eta U & (1-\eta)I_S
    \end{pmatrix},
    \qquad
    \eta:=\frac{S\pi_{\min}}{1-S\pi_{\min}},
    \qquad
    U:=\frac{1}{S}\mathbf 1\mathbf 1^\top.
\]
Thus, from \(a_i\), the chain moves to the \(\mathcal B\)-layer
according to \((1-\eta)q_i+\eta\operatorname{Unif}([S])\); from a
state \(b_j\), it remains at \(b_j\) with probability \(1-\eta\) and
moves to a uniformly chosen state in \(\mathcal A\) with total
probability \(\eta\). This chain is ergodic, and the parameters are
chosen so that every \(a_i\) has stationary mass exactly
\(\pi_{\min}\). Consequently, a length-\(T\) trajectory visits each
\(a_i\) on the order of \(\pi_{\min}T\) times, and its outgoing
transitions encode samples from \(q_i\).

The simulation is performed separately for each user. First generate
the sequence of visits to the two layers using randomness independent
of the product samples. Whenever the skeleton visits \(a_i\), use the
next sample from \(q_i\), together with fresh uniform randomness, to
generate the next state in \(\mathcal B\). For
\(m=\lceil 8\pi_{\min}T\rceil\), the probability that any user needs
more than \(m\) samples from any row is low. Hence the simulation succeeds with high constant
probability once \(m = \tilde{\Omega}(1)\), which gives the condition
\(T=\tilde{\Omega}(1/\pi_{\min})\). The exceptional event depends only
on the data-independent skeleton and can be handled by returning a
fixed output.
\begin{enumerate}
    \item \textbf{Privacy:} Replacing all the product samples of one source user can change only that user's simulated trajectory. The reduction therefore preserves user-level adjacency, so an \((\eps,\del)\)-private Markov estimator induces an \((\eps,\del)\)-private product-distribution estimator; the same argument applies when \(\del=0\).
    \item \textbf{Utility:} Since \(\widehat R\) is the \(\mathcal A\)-to-\(\mathcal B\) block of the estimated transition matrix, we compute
\[
    Z=\frac{\widehat R-\eta U}{1-\eta}
\]
and project each row of \(Z\) onto the simplex. Since \(\eta\leq 1/2\) in the parameter range of the reduction, the resulting estimate \(\widehat Q\) satisfies $\|\widehat Q-Q\|_\infty \leq 4\|\widehat P-P_{\pi_{\min},Q}\|_\infty$. Thus, up to a constant loss in accuracy and the small simulation failure probability, an accurate Markov chain estimator would yield an accurate estimator for the original product-distribution instance.
\end{enumerate}
A formal version of the reduction appears
in Appendix~\ref{ssec:markov-lower}. Thus, to obtain our lower bounds, we apply the i.i.d. product-distribution lower bounds (Section~\ref{sec:kd_dist_bounds}), to obtain the following result.

\begin{theorem}[Complete version in \Cref{thm:markov-chain-estimation-approx}]
\label{thm:user-level-markov-approx-lower-informal}
Let $0 < \alpha \leq \frac{1}{384\sqrt{2}}$, $0 < \varepsilon < 1$ and $\delta \lesssim \min\left\{\frac{\varepsilon}{S}, \varepsilon^2\right\}$. For
\(T = \tilde{\Omega}\left(\frac{1}{\pi_{\min}}\right)\), every user-level
\((\eps,\del)\)-differentially private estimator based on independent
length-\(T\) trajectories that achieves error at most \(\alpha\) with
probability at least \(3/4\), uniformly over
\(\mathcal{P}(S,\pi_{\min},\gamma_{ps})\), requires
$
    N
    =\widetilde{\Omega}\!\left(
        \frac{1}{\eps}
        +\frac{S^{3/2}}{\alpha\eps\sqrt{\pi_{\min}T}}
        +\frac{S}{\pi_{\min}T\alpha^2}
    \right).
$
\end{theorem}
The corresponding result for Pure DP appears in Appendix~\ref{ssec:markov-lower}, \Cref{thm:user-level-markov-pure-lower-formal}.

\section{Product-Distribution Estimation in Infinity Norm with User-Level DP}
\label{sec:kd_dist_bounds}

We have seen in Section~\ref{sec:userlevelest} that both upper bounds and lower bounds for user-level learning of Markov chains can be reduced to the i.i.d. setting given that the length of trajectories is large enough. In this section, we study private estimation of product distributions with user-level DP.

\paragraph{Problem formulation.}
Let $\mathcal P_{d,k} := \left\{ P\in\mathbb R^{d\times k}: P_{(j,\cdot)} \in\Delta(k) \text{ for every }j\in[d] \right\}$. We associate \(P\in\mathcal P_{d,k}\) with the product distribution
\(\bigotimes_{j=1}^d P_{(j,\cdot)}\) over \([k]^d\). User \(i\) contributes a
block $X^{(i)}=(X^{(i)}_{1},\ldots,X^{(i)}_{m})\in([k]^d)^m$ where $X_{i,t}\overset{\mathrm{i.i.d.}}{\sim}\bigotimes_{j=1}^d P_{(j,\cdot)}$,
and the full dataset is \(D=(X^{(1)},\ldots, X^{(N)})\).
Datasets \(D\) and \(D'\) are user-level neighbors if there exists \(i^\star\in[N]\) such
that \(X^{(i)} = {X^{(i)}}'\) for every \(i\neq i^\star\) and the entire blocks \(X^{(i^\star)}\) and \({X^{(i^\star)}}'\) may differ arbitrarily. The user-level privacy guarantee is required to hold for every such pair of datasets.

A pure and approximate DP estimator $\mathcal{M}$ outputs a row-stochastic matrix
\({\cal M}(D)=\widehat P\in\mathcal P_{d,k}\), which induces a product distribution. For accuracy \(\alpha\in(0,1)\) and failure probability \(f\in(0,1)\), we require $\widehat P$ to be an $(\alpha,f)$-accurate estimate of $P$.

We first give upper and lower bounds for approximate user-level DP, which match up to logarithmic factors in a small-\(\del\) regime.
We begin with an estimator that exploits concentration within each user's block of \(m\) samples and a powerful  filtering technique from \citet{tsfadia2022friendlycore}. This yields the following upper bound.

\begin{theorem}[Complete version in \Cref{thm:user-level-product-marginals-priv,thm:user-level-product-marginals-util}]
\label{thm:user-level-product-marginals-main}
For \(0<\alpha,f<1\) and \(0<\eps\leq 1\), there exists a user-level \((\eps,\del)\)-differentially private
estimator such that
$
    N
    =
    \widetilde{\mathcal O}\!\left(
        \frac{1}{\eps}
        +\frac{k\sqrt d}{\alpha\eps\sqrt m}
        +\frac{k}{m\alpha^2}
    \right)
$
users suffice to achieve \(\rho(\widehat P,P)\leq\alpha\) with
high probability, uniformly over
\(P\in\mathcal P_{d,k}\).
\end{theorem}

\begin{proof}{\textbf{Sketch}}
Represent user \(i \in [N]\) by its $d \times k$ empirical
distribution matrix $Q_i(j,a) := \frac{1}{m}\sum_{t=1}^m\mathbf 1\{X^{(i)}_{t,j}=a\}$. Then \(\mathbb E[Q_i]=P\) and \(\mathbb E\|Q_i-P\|_{\mathrm F}^2\leq \frac{d}{m}\). Hence, with high probability, all the matrices \(Q_i\)'s lie in a Frobenius ball of radius \(R = \widetilde{\mathcal O}\left(\sqrt{\frac dm}\right)\) around \(P\), and therefore have pairwise distance at most \(2R\). Using this observation, we invoke the \textit{Friendly-Core} private averaging procedure \citep{tsfadia2022friendlycore} described
in the Appendix, where two matrices $A$ and $B$ are considered close if
\(\|A-B\|_{\mathrm F}\leq 2R\). The Friendly-Core averaging procedure consists of two parts: filtering and private averaging. At a high level, filtering retains only those users' empirical distribution
matrices having sufficiently many nearby
neighbors, with inclusion probabilities chosen stably under the addition or removal of one user. Additionally, it ensures that every retained point has more than half of the dataset as neighbors. Hence, the neighbor sets of retained points from two neighboring executions must intersect. Thus, even on an arbitrary input, the filter certifies that the union of the two cores has diameter \(\mathcal O(R)\).

So, if the core has size \(n_C\), changing one retained user changes its mean
by only \(\mathcal O\left(\frac{R}{n_C}\right)\). Hence, private averaging using the Gaussian mechanism suffices to prove privacy for this case. However, \(n_C\) may leak information about the data. Thus, the mechanism uses a private lower estimate of \(n_C\) to calibrate Gaussian noise without revealing the core size, while the stability of the filter yields an end-to-end DP guarantee \citep{tsfadia2022friendlycore}. Finally, each noisy row is projected onto \(\Delta(k)\).

Since the users’ empirical matrices are pairwise neighbors, the filter retains everyone with high probability and \(n_C=N\). The private count requires \(N=\widetilde{\Omega}(1/\eps)\). The core mean is the usual empirical distribution matrix \(\overline Q=\frac{1}{N}\sum_iQ_i\). The privatized empirical matrix differs from the core mean by an independent additive Gaussian noise. Thus, the private to empirical gap is given by $\max_j\|Z_{(j,\cdot)}\|_1 =\widetilde{\mathcal O}\!\left(\frac{k\sqrt d}{N\eps\sqrt m}\right)$. Concentration gives the empirical-to-population bound $\max_j\|\overline Q_{(j,\cdot)}-p_j\|_1 =\widetilde{\mathcal{O}}\!\left(\sqrt{\frac{k}{Nm}}\right)$. Projection changes the error by at most a constant factor. Making both terms at most \(\alpha\), together with the private-count requirement, gives the stated user complexity.
\end{proof}

The complete statement of the theorem and the complete proof can be found in Appendix~\ref{ssec:approx-dp-upper-bound-formal}. The upper bound contains one non-private sampling term and two
privacy-dependent terms. A corresponding result for pure DP is given by Appendix~\ref{ssec:pure-dp-upper-bound-formal}, \Cref{thm:pure-dp-product-distribution-priv,thm:pure-dp-product-distribution-util}. The next result shows that, when \(\del\) is
sufficiently small, all three are unavoidable up to logarithmic
factors.

\begin{theorem}[Complete version in \Cref{thm:high-prob-approx-final-one-infty-norm}]
\label{thm:high-prob-approx-final-one-infty-norm-main}
For $0<\alpha\le \frac{1}{48\sqrt{2}}$, $0<\eps\le 1$, and $\delta \lesssim \min\left\{\frac{\eps}{\sqrt{dk}}, \eps^2\right\}$, every
user-level \((\eps,\del)\)-differentially private estimator that
achieves error at most \(\alpha\) with probability at least \(2/3\),
uniformly over \(\mathcal P_{d,k}\), requires
$
    N
    =
    \widetilde{\Omega}\!\left(
        \frac{1}{\eps}
        +\frac{k\sqrt d}{\alpha\eps\sqrt m}
        +\frac{k}{m\alpha^2}
    \right).
$
\end{theorem}
\begin{proof}{\textbf{Sketch.}}
    We establish the three terms separately and provide references to the formal statements of each of these terms in Appendix~\ref{ssec:approx-dp-lower-bound-formal}.

    $\tilde{\Omega}\left(\frac{1}{\varepsilon}\right)$: Lemma~\ref{lem:approx-dp-few-user-packing} constructs a family of well-separated product distributions. Accuracy would allow an estimator to identify the generating member of this family, whereas group privacy prevents such identification when the number of users is too small. Adapting the packing argument of \citet{esfandiari2022tight} gives \(N\gtrsim\eps^{-1}\min\{kd,\log(1/\del)\}\), which yields the first term in the stated small-\(\del\) regime.

    $\tilde{\Omega}\left(\frac{k\sqrt d}{\alpha\eps\sqrt m}\right)$:
     Theorem~\ref{thm:main-expected-approx} proves an expected-risk lower
    bound for the mixed loss $\xi(\widehat P,P) := \left( \sum_{j=1}^d \|\widehat P_{(j,\cdot)}-P_{(j,\cdot)}\|_1^2 \right)^{1/2}$. There are two broad steps.
    \begin{enumerate}[leftmargin=2em]
        \item First, following \citet{esfandiari2022tight}, Poissonize a paired family in each $j \in [d]$ marginal:
    \[
    X^{+}_{i,j,b}\sim\Pois\!\left(m'\left(\frac1k+\theta_{j,b}\right)\right),
    X^{-}_{i,j,b}\sim\Pois\!\left(m'\left(\frac1k-\theta_{j,b}\right)\right).
    \]
    The fixed mass of each pair keeps every marginal normalized, while the category counts are independent. Conditional on a row total, expanding its counts gives independent labels from $p_j^\theta$; combining labels across the $d$ rows gives samples from $P_\theta=\bigotimes_{j=1}^d p_j^\theta$. For $m'=\widetilde{\Theta}(m)$, all user--row totals exceed $m$ with high probability. Hence, truncating the samples, applying the product-distribution estimator, and taking within-pair differences yield a solution to the Poisson problem. This user-wise reduction preserves adjacency.
    \item Second, apply the fingerprinting argument of \citet{esfandiari2022tight} and \citet{pmlr-v99-kamath19a} to all $D=d\lfloor k/2\rfloor$ pairs. Let $q = 1/k$. For adjacent databases $X$ and $X^{\sim i}$, define
    \begin{align*}
        S_{i,\ell}:=
        \frac{X_{i,\ell}^+-m'(q+\theta_\ell)}{q+\theta_\ell}
        -
        \frac{X_{i,\ell}^--m'(q-\theta_\ell)}{q-\theta_\ell},\quad
        Z_i(Y):=\sum_{\ell=1}^D
        (r^2-\theta_\ell^2)(\widehat\theta_\ell(Y)-\theta_\ell)S_{i,\ell}.
    \end{align*}
    Note that $X^{\sim i}$ replaces a pair $(X_{i,\ell}^+, X_{i,\ell}^{-})$ for each $\ell \in [D]$ by an independent copy, while $S_{i,\ell}$ still uses the original block. The fingerprinting identity with respect to $S_{i,\ell}$ and accuracy ensure $\sum_{i=1}^N\EE [Z_i(X)]$ is non-zero, while centering and independence make $\EE[Z_i(X^{\sim i})]$ zero. Under the stated $\delta$ condition, approximate privacy controls this gap.
    \end{enumerate}
    Independence makes variances add over the $D$ pairs, and $\ell_2$ aggregation across the $d$ rows yields $N=\widetilde{\Omega}\left(\frac{dk}{\alpha\eps\sqrt m}\right)$ for the expected mixed loss. A change in objective norm and a median of means trick yields $N=\widetilde{\Omega}\left(\frac{k\sqrt d}{\alpha\eps\sqrt m}\right)$.

    $\widetilde{\Omega}\left(\frac{k}{m\alpha^2}\right)$: This is the non-private lower bound. Without privacy, fixing all but one marginal reduces the problem to estimating a \(k\)-ary distribution from \(Nm\) samples. Thus, a sampling lower bound construction for a standard $k$-ary distribution \citep{canonne2020short} gives \(Nm=\Omega\left(\frac{k}{\alpha^2}\right)\).

    Since each of these is a valid lower bound for the risk, the eventual lower bound is the maximum of these three terms. This proves the
    claim.
\end{proof}

The complete statement of the theorem and the complete proof can be found in Appendix~\ref{ssec:approx-dp-lower-bound-formal}. The corresponding result for Pure DP can be found in Appendix~\ref{ssec:pure-dp-lower-bound-formal}, Theorem~\ref{thm:pure-user-level-product-lower-bound}.

\section{Conclusion and Open Problems}
In this work, we studied private learning of discrete Markov chains from multiple trajectories. We considered two privacy settings, namely, item-level and user-level privacy. In the item-level setting, we show that a simple perturbation to the transition count matrix yields a near-optimal algorithm.
In the user-level setting, we show that, in the large-$T$ regime, we can reduce the problem to user-level private product-distribution estimation under the matrix \(\ell_\infty\) loss.
As an intermediate result, we establish upper and lower bounds for user-level private product-distribution estimation under the \(\ell_\infty\) norm. Our results leave three main directions for future work:
\begin{itemize}[leftmargin=1em]
    \item \textbf{Mixing Term gap.} For item-level private learning of Markov chains, the upper and lower bounds differ by an additive gap of $\tilde{\Omega}\left(\frac{1}{\pi_{\min}T\gamma_{ps}}\right)$. Closing this gap remains open.

    \item \textbf{Small and intermediate trajectory lengths.}
    Our reductions for both upper and lower bounds apply only when trajectories are sufficiently long. Determining the optimal user complexity for small trajectory lengths remains open.

    \item \textbf{Pure-DP gap.}
    For product-distribution estimation under the matrix \(\ell_\infty\)
    loss, the privacy--accuracy interaction term has a factor-\(\sqrt{d}\)
    gap between the upper and lower bounds. Closing this gap would, through
    the reductions underlying both the upper and lower bounds, also close the
    corresponding factor-\(\sqrt{S}\) gap for private Markov-chain
    estimation in the large-$T$ regime.
\end{itemize}

\section*{AI Disclosure}
All proof ideas and technical tools were developed and verified entirely by the authors. Codex GPT Pro 5.6 Sol was used only for formatting, grammatical corrections, and refining the text.

\bibliography{references}

\newpage
\appendix
\startcontents[appendices]

\section*{Appendix}
\printcontents[appendices]{}{1}{\setcounter{tocdepth}{2}}

\newpage

\section{Additional Related Work}
\label{sec:relatedwork}
\paragraph{Statistical estimation of discrete Markov chains.}
Statistical estimation of Markov chains dates to \citet{anderson1957statistical} and \citet{billingsley1961statistical}, who studied maximum-likelihood estimation and asymptotic inference for transition probabilities. \citet{hao2018learning} studied transition-matrix estimation under smooth row-wise $f$-divergence losses, while \citet{pmlr-v98-wolfer19a,MarkovChainEstimationBernoulli} obtained nearly matching minimax bounds under maximum row-wise $\ell_1$ error, characterizing the dependence on the state-space size, minimum stationary probability, and pseudo-spectral gap. Beyond worst-case guarantees, \citet{wolfer2024empirical} developed empirical confidence intervals and instance-dependent bounds for transition-matrix and mixing-time estimation that exploit properties such as transition sparsity.

A related line of work estimates the chain's convergence properties. \citet{hsu2019mixing} developed data-dependent confidence bounds for mixing time of reversible chains, and \citet{wolfer2019mixing} extended this to general ergodic chains. \citet{wolfer2024relaxation} subsequently established the minimax sample complexity of estimating the pseudo-spectral gap to constant multiplicative accuracy in nonreversible chains and improved the associated empirical confidence bounds. Concentration inequalities developed by \citet{paulinconcenteration} provide important tools for finite-sample analyses of Markovian observations.

Other work considers estimation objectives beyond recovering the full transition matrix. \citet{han2021prediction} established optimal rates for next-state prediction under KL loss and examined the role of spectral-gap assumptions. \citet{han2018entropy} studied entropy-rate estimation for large-state-space chains, improving over the empirical estimator under suitable mixing assumptions. \citet{zhang2020spectral} developed spectral methods for low-rank transition estimation and recovery of aggregate state structure, with statistical guarantees and nearly matching minimax lower bounds.

\paragraph{Statistical estimation under user-level privacy.}
User-level differential privacy protects an individual's entire collection
of observations. \citet{dwork2010panprivate} explicitly distinguished
user-level from event-level privacy in streaming algorithms. \citet{mcmahan2018recurrent} demonstrated its applicability to language-model training, while \citet{amin2019bounding} analyzed the bias--variance trade-off induced by bounding user contributions.

\citet{liu2020discrete} developed algorithms for discrete distribution
estimation under $\ell_1$ loss that exploit user groupings to improve over
mechanisms operating only on aggregate counts. \citet{levy2021user}
introduced concentration-based methods for Euclidean mean estimation and
optimization, quantifying how additional observations per user reduce the
privacy cost. \citet{esfandiari2022tight} obtained nearly optimal
user--sample trade-offs for mean and discrete distribution estimation,
including the few-user regime and robustness to corrupted users.
Complementing these task-specific results, \citet{ghazi2021correlated}
developed a general learning framework using correlated sampling when
users hold sufficiently many observations, and \citet{ghazi2023user}
established generic guarantees for the regime of few observations per
user under pure and approximate privacy.

Subsequent work broadened the statistical models and privacy settings.
\citet{cummings2022heterogeneity} studied mean estimation with
heterogeneity in both users' distributions and sample counts, while
\citet{agarwal2025person} studied user-level
private mean estimation under bounded-moment assumptions, establishing
nearly matching upper and lower bounds under $\ell_2$ loss. In the local
privacy model, \citet{acharya2023discrete} obtained tight bounds for
discrete distribution estimation across nearly all parameter regimes,
and \citet{zhao2024local} developed mean-estimation and learning
algorithms, including squared Euclidean error guarantees under
$\ell_2$- and $\ell_\infty$-bounded support.

\paragraph{Differential privacy for stochastic matrices.}
\citet{fallin2023differential} introduce a matrix Dirichlet mechanism for privately releasing an already-specified stochastic matrix, apply this mechanism to a class of Markov chains with certain structure in their transition matrices, and bound the \emph{resulting} entry-wise and stationary-distribution perturbations. \citet{benvenuti2026modeling} extend this approach to empirical transition matrices, providing event-level privacy for the individual transition records used to construct the matrix and bounding the induced changes in its stationary distribution and convergence behavior.
In contrast to these works, our work studies the minimax estimation of an unknown transition matrix directly from Markovian trajectories under item and user-level privacy. Our guarantees account for sampling error, temporal dependence, state occupancy, and mixing, and characterize the required trajectory length or number of users rather than only the additional distortion introduced by privatizing a given empirical matrix.

\section{Definitions}
\label{app:markov-preliminaries}
Below, we provide a complete definitions and setup for Markov chains, following \citet{Levin:2017}.

\begin{definition}[Discrete-time Markov chain]
A discrete-time, time-homogeneous Markov chain on the discrete-space $[S]$ with transition matrix $P$ is a stochastic process $(X_t)_{t\ge 0}$ satisfying
\[
    \mathbb{P}\!\left(
        X_{t+1}=j
        \,\middle|\,
        X_0=x_0,\ldots,X_t=x_t
    \right)
    =
    \mathbb{P}\!\left(
        X_{t+1}=j
        \,\middle|\,
        X_t=x_t
    \right).
\]
for every $t\ge 0$ and $j,x_0,\ldots,x_t\in[S]$.
\end{definition}

As customary, we define  $P_{i,j}=\mathbb{P}(X_{t+1}=j\mid X_t=i)$. The matrix power $P^t$ is the $t$-step transition matrix, so that $(P^t)_{i,j}=\mathbb{P}(X_t=j\mid X_0=i)$. If the initial state has distribution $\mu$, then the distribution of $X_t$ is $\mu P^t$. A finite trajectory of length $T$ is denoted by $(X_0,\ldots,X_{T-1})$.

\begin{definition}[Reachability and irreducibility]
A state $j$ is said to be \emph{reachable} from a state $i$ if there exists an integer $t\ge 0$ such that $(P^t)_{i,j}>0$. Two states \emph{communicate} if each is reachable from the other. A stochastic matrix $P$ is called \emph{irreducible} if every pair of states communicates; equivalently, for every $i,j\in[S]$, there exists an integer $t\ge 1$ such that $(P^t)_{i,j}>0$.
\end{definition}

\begin{definition}[Periodicity]
The \emph{period} of a state $i$ is
\[
    \operatorname{per}(i)
    :=
    \gcd\bigl\{
        t\ge 1 : (P^t)_{i,i}>0
    \bigr\}.
\]
An irreducible Markov chain has the same period at every state. A Markov chain is called \emph{aperiodic} if this common period is equal to one.
\end{definition}

Ergodicity is the long-run regularity condition needed in our analysis: it ensures that the chain eventually forgets its initial state and approaches a common equilibrium while continuing to explore the entire state space.

\begin{definition}[Ergodicity]
A finite-state Markov chain with transition matrix $P$ is called \emph{ergodic} if there exists a probability vector $\pi$, with $\pi_j>0$ for every $j\in[S]$, such that
\[
    \lim_{t\to\infty}(P^t)_{i,j}=\pi_j
    \qquad
    \text{for every }i,j\in[S].
\]
Thus, the distribution of the chain converges to the same full-support limiting distribution regardless of its initial state.
\end{definition}

A finite-state Markov chain is \emph{ergodic} if and only if its transition matrix is both irreducible and aperiodic \cite{Levin:2017}.
Here, irreducibility prevents the state space from decomposing into inaccessible components, while aperiodicity rules out persistent cyclic behavior. To describe the equilibrium approached by an ergodic chain, we introduce distributions that are unchanged by one step of the chain.

\begin{definition}[Stationary distribution]
A stochastic vector $\pi$ is called a \emph{stationary distribution} for $P$ if $\pi P=\pi$.
\end{definition}

Every finite ergodic Markov chain admits a unique stationary distribution $\pi$, and this distribution satisfies $\pi_i>0$ for every $i\in[S]$.
We define
$\pi_{\min}:=\min_{i\in[S]}\pi_i$ and
$\pi_{\max}:=\max_{i\in[S]}\pi_i$.
If $P$ is ergodic, then for every initial distribution $\mu$,
\[
    \lim_{t\to\infty}
    d_{\mathrm{TV}}\!\left(
        \mu P^t,\pi
    \right)
    =0.
\]
In particular, if $X_0\sim\pi$, then $X_t\sim\pi$ for every $t\ge 0$, and the chain is said to be \emph{started in stationarity}.

Reversibility captures when a chain in stationarity has the same dynamics forward and backward in time, and it enables a particularly clean spectral analysis.

\begin{definition}[Time reversal and reversibility]
Let $P$ be irreducible with stationary distribution $\pi$. The \emph{time reversal} of $P$ is the stochastic matrix $P^*$ defined by
\[
    P^*_{i,j}
    :=
    \frac{\pi_j}{\pi_i}P_{j,i}.
\]
The matrix $P$ is called \emph{reversible} with respect to $\pi$ if $P^*=P$. Equivalently, $P$ is reversible if it satisfies the detailed-balance equations $\pi_iP_{i,j}=\pi_jP_{j,i}$ for every $i,j\in[S]$.
\end{definition}

The time-reversal $P^*$ also has stationary distribution $\pi$.
For functions $f,g:[S]\to\mathbb{R}$, define the weighted inner product $\langle f,g\rangle_\pi:=\sum_{i=1}^S \pi_i f(i)g(i)$ and the corresponding norm $\norm{f}_{2,\pi}:=\langle f,f\rangle_\pi^{1/2}$. A reversible transition matrix is self-adjoint with respect to this inner product and hence has real eigenvalues.
We order these eigenvalues as
$1=\lambda_1(P)\ge\lambda_2(P)\ge\cdots\ge\lambda_S(P)\ge-1$.

The eigenvalue $1$ corresponds to stationarity; for reversible chains, its separation from the next-largest eigenvalue is a fundamental spectral measure of convergence behavior.

\begin{definition}[Spectral gap]
The \emph{spectral gap} of an irreducible reversible stochastic matrix $P$ is
\[
    \gamma_{\mathrm{rev}}(P)
    :=
    1-\lambda_2(P),
\]
where $\lambda_2(P)$ denotes the second-largest eigenvalue of $P$.
\end{definition}

For nonreversible chains, the spectrum need not have the same direct interpretation. Multiplicative reversiblization provides a reversible surrogate from which an analogous gap can be defined.

\begin{definition}[Multiplicative reversiblization and pseudo-spectral gap \citep{paulinconcenteration}]
Let $P$ be an irreducible stochastic matrix with time reversal $P^*$. For every integer $k\ge 1$, the matrix $(P^*)^kP^k$ is called the $k$th \emph{multiplicative reversiblization} of $P$. The \emph{pseudo-spectral gap} of $P$ is defined by
\[
    \gamma_{\mathrm{ps}}(P)
    :=
    \sup_{k\ge 1}
    \frac{
        \gamma_{\mathrm{rev}}
        \!\left((P^*)^kP^k\right)
    }{k}.
\]
\end{definition}
For every $k\ge 1$, the matrix $(P^*)^kP^k$ is reversible with respect to $\pi$. Here, $(P^*)^k$ denotes the $k$th matrix power of the time-reversal $P^*$.

\section{Auxiliary Lemmas and Results}

\begin{lemma}[Lower tail for sums of geometric random variables]
\label{lem:geomchernoff}
Let $L_1,\dots,L_N \stackrel{\text{i.i.d.}}{\sim} \mathrm{Geom}(p)$ on $\mathbb{N}$ and $S_N = \sum_{i=1}^N L_i$. Then for any $\varepsilon \in (0,1)$,
\[
\mathbb{P}\!\left(S_N \le (1-\varepsilon)\frac{N}{p}\right)
\le \exp\!\left(-c N \varepsilon^2\right),
\]
for some universal constant $c>0$.
\end{lemma}
\begin{proof}
We apply the Chernoff method to the lower tail. For any $\lambda > 0$,
\[
\mathbb{P}(S_N \le t)
= \mathbb{P}(e^{-\lambda S_N} \ge e^{-\lambda t})
\le e^{\lambda t} \, \mathbb{E}[e^{-\lambda S_N}]
= e^{\lambda t} \left(\mathbb{E}[e^{-\lambda L_1}]\right)^N.
\]

The moment generating function at $\lambda > \ln(1 - p)$ is
\[
\mathbb{E}[e^{-\lambda L_1}]
= \sum_{\ell=1}^\infty (1-p)^{\ell-1} p\, e^{-\lambda \ell}
= \frac{p e^{-\lambda}}{1-(1-p)e^{-\lambda}}.
\]

Thus,
\[
\mathbb{P}(S_N \le t)
\le \exp\!\left(
\lambda t + N \log \frac{p e^{-\lambda}}{1-(1-p)e^{-\lambda}}
\right).
\]

Set $t = (1-\varepsilon)\frac{N}{p}$. For $\lambda \le \frac{p}{2}$, a Taylor expansion gives
\[
\log \frac{p e^{-\lambda}}{1-(1-p)e^{-\lambda}}
\leq -\frac{\lambda}{p} + \frac{\lambda^2}{p^2}.
\]

Plugging in,
\[
\mathbb{P}(S_N \le t)
\le \exp\!\left(
\lambda \frac{N}{p}(1-\varepsilon)
- \frac{N\lambda}{p}
+ \frac{N \lambda^2}{p^2}
\right)
= \exp\!\left(
- \lambda \frac{N}{p}\varepsilon
+ \frac{N \lambda^2}{p^2}
\right).
\]

Optimizing by taking $\lambda = \frac{p \varepsilon}{2}$ yields
\[
\mathbb{P}\!\left(S_N \le (1-\varepsilon)\frac{N}{p}\right)
\le \exp(-c N \varepsilon^2),
\]
for a universal constant $c>0$.
\end{proof}

\begin{lemma}[Bounded comparison under approximate DP {\citet[Equation 18]{pmlr-v99-kamath19a}}]\label{lem:approx-comparison}
Let $\cM$ be $(\eps,\delta)$-DP with $0<\eps\le 1$.  Let $x,x'$ be neighboring datasets, and let $Y=\cM(x)$ and $Y'=\cM(x')$. Let $\varphi$ be a real-valued measurable function of the output such that $|\varphi(y)|\le B$ for every possible output $y$.  Then
\[\label{eq:approx-comparison}
    \EE\varphi(Y)
    \le
    \EE\varphi(Y')
    +2\eps\,\EE|\varphi(Y')|
    +2\delta B .
\]
\end{lemma}

\begin{proof}
Let $P$ and $Q$ denote the laws of $Y$ and $Y'$.  Write $\varphi=\varphi_+-\varphi_-$ where $\varphi_+=\max\{\varphi,0\}$ and $\varphi_- = \max\{-\varphi,0\}$. Note that both $\varphi_+$ and $\varphi_-$ take values in $[0,B]$. For any nonnegative function $h$ bounded by $B$, $(\eps,\delta)$-DP implies
\begin{align*}
    \EE_P h
    =
    \int_0^B P(h\ge t)\,dt
    \le
    \int_0^B \{e^\eps Q(h\ge t)+\delta\}\,dt
    =
    e^\eps\EE_Q h+\delta B.
\end{align*}
Applying the same argument in the reverse direction gives $\EE_Q h \le e^\eps \EE_P h+\delta B$, which implies $\EE_P h \ge e^{-\eps}\EE_Q h-e^{-\eps}\delta B$.
Using these two inequalities with $h=\varphi_+$ and $h=\varphi_-$, we get
\begin{align*}
    \EE_P\varphi
    &=
    \EE_P\varphi_+ - \EE_P\varphi_- \\
    &\le
    e^\eps\EE_Q\varphi_+ +\delta B
    -e^{-\eps}\EE_Q\varphi_-+e^{-\eps}\delta B \\
    &=
    \EE_Q\varphi
    +(e^\eps-1)\EE_Q\varphi_+
    +(1-e^{-\eps})\EE_Q\varphi_-
    +(1+e^{-\eps})\delta B .
\end{align*}
Since $0<\eps\le 1$, we have $e^\eps-1\le 2\eps$, $1-e^{-\eps}\le \eps$, $1+e^{-\eps}\le 2$.
Therefore
\[
    \EE_P\varphi
    \le
    \EE_Q\varphi
    +2\eps\EE_Q|\varphi|
    +2\delta B,
\]
which is \eqref{eq:approx-comparison}.
\end{proof}

\begin{lemma}[McDiarmid's inequality
\citep{mcdiarmid1989method}]
\label{lem:mcdiarmid}
Let $Z_1,\ldots,Z_m$ be independent random variables, with
$Z_t$ taking values in a set $\mathcal{Z}_t$, and let $g:\prod_{i=1}^m \mathcal{Z}_i \to \mathbb{R}$ be a measurable function. Suppose that there exist constants $c_1,\ldots,c_m\geq 0$ such that, for every $t\in[m]$ and every $z,z'\in\prod_{i=1}^m\mathcal{Z}_i$ that differ only in their $t$-th coordinate,
\[
    |g(z)-g(z')|\leq c_t.
\]
Then, for every $u\geq 0$,
\[
    \mbb P\!\left[
            g(Z_1,\ldots,Z_m)
            -
            \mbb E[g(Z_1,\ldots,Z_m)]
        \geq u
    \right]
    \leq
    \exp\!\left(
        -\frac{2u^2}{\sum_{t=1}^m c_t^2}
    \right).
\]
\end{lemma}

The following statement follows directly from applying the previous result to
\[\left\|G(Z_1,\ldots,Z_m)- \EE G(Z_1,\ldots,Z_m)\right\|.\]

\begin{lemma}[Norm concentration via bounded differences]
\label{lem:norm-concentration-bounded-differences}
Let $Z_1,\ldots,Z_m$ be independent random variables, where $Z_t$
takes values in a measurable space $\mathcal{Z}_t$. Let
$(\mathcal{V},\|\cdot\|)$ be a normed vector space and let $G:\prod_{t=1}^m \mathcal{Z}_t \to \mathcal{V}$ be a measurable function such that $G(Z_1,\ldots,Z_m)$ is integrable.
Suppose that there exist constants $c_1,\ldots,c_m\geq 0$ such that,
for every $t\in[m]$ and every $z,z'\in\prod_{s=1}^m\mathcal{Z}_s$
that differ only in their $t$-th coordinate,
\[
    \bigl\|G(z)-G(z')\bigr\|\leq c_t.
\]
Let $\mu:=\EE G(Z_1,\ldots,Z_m)\in\mathcal{V}$. Then, for every $u\geq 0$,
\[
    \mbb P\!\left(
        \left\|G(Z_1,\ldots,Z_m)- \mu\right\|
        \geq
        \EE\!\left[
            \left\|G(Z_1,\ldots,Z_m)
            - \mu\right\|
        \right]
        +u
    \right)
    \leq
    \exp\!\left(
        -\frac{2u^2}{\sum_{t=1}^m c_t^2}
    \right).
\]
\end{lemma}

\begin{lemma}
\label{lem:finiteterm}
Let \(m,b<\infty\) be nonnegative. Then
\[
\sum_{z\in\mathbb{Z}_{\geq 0}^{2N}}
\left(\sum_{j=1}^{2N} z_j\right)
\prod_{j=1}^{2N}\frac{(mb)^{z_j}}{z_j!}
<\infty.
\]
\end{lemma}

\begin{proof}
Let \(Z_1,\ldots,Z_{2N}\) be independent
\(\operatorname{Poisson}(mb)\) random variables. Their joint probability
mass function is
\[
\mathbb{P}\bigl((Z_1,\ldots,Z_{2N})=z\bigr)
=
e^{-2Nmb}
\prod_{j=1}^{2N}\frac{(mb)^{z_j}}{z_j!}.
\]
Consequently,
\begin{align*}
\sum_{z\in\mathbb{Z}_{\geq 0}^{2N}}
\left(\sum_{j=1}^{2N}z_j\right)
\prod_{j=1}^{2N}\frac{(mb)^{z_j}}{z_j!}& =
e^{2Nmb}
\sum_{z\in\mathbb{Z}_{\geq 0}^{2N}}
\left(\sum_{j=1}^{2N}z_j\right)
\mathbb{P}\bigl((Z_1,\ldots,Z_{2N})=z\bigr) \notag\\
& =
e^{2Nmb}\,
\mathbb{E}\left[\sum_{j=1}^{2N}Z_j\right].
\end{align*}
Since each \(Z_j\) has mean \(mb\), linearity of expectation gives
\[
\mathbb{E}\left[\sum_{j=1}^{2N}Z_j\right]
=
\sum_{j=1}^{2N}\mathbb{E}[Z_j]
=
2Nmb<\infty.
\]
It follows that the original series is finite.
\end{proof}

\begin{lemma}[{\citet[Theorem~1]{canonne2019poisson}}]
\label{lem:poissontailbound}
    If $X \sim \operatorname{Pois}(\lambda)$, then for all $0 \leq t < \lambda$,
    \begin{align*}
        \mbb P\left(X \leq \lambda - t\right) \leq \exp\left({\frac{-t^2}{2\lambda}}\right).
    \end{align*}
\end{lemma}

\section{Proofs of Section~\ref{sec:ItemLevelEstimation}}

\subsection{Complete Statement and Proof of Theorem~\ref{thm:item-level-dp-upperbound-informal}}
\label{app:upperbound-item-formal}

\begin{algorithm}[t]
\caption{\textsc{PrivateTransitionEstimator}}
\label{alg:private-transition-estimator-multiple-chains}

\noindent
\textbf{\textit{Input:}}
Markov Chains $\bigl(X^{(1)}_{1:\len},\ldots,X^{(\num)}_{1:\len}\bigr)$; noise distribution \(\mathcal{D}\).

\noindent
\textbf{\textit{Operation:}}
\begin{enumerate}
    \item For all $i,j\in[S]$, obtain the transition counts $K_{i,j} = \sum_{\ell=1}^N\sum_{t=1}^{T-1} \mathbf{1}\{X^{(\ell)}_t = i, X^{(\ell)}_{t+1} = j\}$ across all chains.
    \item For every \(i,j\in[S]\), independently sample $Z_{i,j}\sim\mathcal{D}$ and compute $\widetilde{K}_{i,j} := \left[K_{i,j}+Z_{i,j}\right]_+$, where $[a]_+ = \max\{a, 0\}$.
    \item For every \(i\in[S]\), compute $\widetilde{K}_i := \sum_{k=1}^{S}\widetilde{K}_{i,k}$ and, for every \(j\in[S]\), set $\widetilde{P}_{i,j} := \frac{\widetilde{K}_{i,j}}{\widetilde{K}_i}$ if $\widetilde{K}_i > 0$ otherwise set $\widetilde{P}_{i,j} := \frac{1}{S}$.
    \item Output the estimated transition matrix $\widetilde{P} := \bigl(\widetilde{P}_{i,j}\bigr)_{i,j\in[S]}$.
\end{enumerate}
\end{algorithm}

Before providing the formal statement and the corresponding proof, we introduce some helper lemmas that will be used to prove our claim. We use the notation from Algorithm~\ref{alg:private-transition-estimator-multiple-chains}. Recall that for \(i,j\in[S]\),
\[
    K_{i,j}
    :=\sum_{u=1}^N\sum_{t=1}^{T-1}\mathbf{1}\{X^{(u)}_t=i,X^{(u)}_{t+1}=j\},
    \qquad
    K_i:=\sum_{j=1}^S K_{i,j}.
\]
Here, \(K_i\) is the number of observed departures from state \(i\).
Whenever \(K_i>0\), the empirical estimator is
\[
    \widehat P^{\mathrm{np}}_{i,j}:=\frac{K_{i,j}}{K_i}.
\]
If \(K_i=0\), its \(i\)-th row can be set to any fixed probability
vector, such as the uniform distribution.

\begin{lemma}[\cite{leskela2026robust}]
\label{thm:leskelarobust}
\label{lem:markov-chain-estimation-item}
For a universal constant $C_1$, if
\[
N \geq C_1 \max\left\{\frac{S}{\pi_{\min}T\alpha^2}\log\left(\frac{S}{f}\right) + \frac{1}{\pi_{\min}} \ln\left(\frac{1}{\pi_{\min}}\right), \frac{1}{\pi_{\min}\gamma_{ps} T}\log\left(\frac{S}{f}\right)\right\},
\]
then $P^{\mathrm{np}}$ satisfies $\mbb{P}\left(\|\widehat P^{\mathrm{np}} - P\|_{\infty} \le \alpha\right) \geq 1 - f$.
\end{lemma}

\begin{lemma}[\citet{leskela2026robust}]
\label{lem:counts}
For a universal constant $C_2$, if
\begin{align*}
    NT \geq \frac{C_2}{\pi_{\min} \gamma_{ps}} \log\left(\frac{S}{f}\right),
\end{align*}
then $\mbb P\left(\exists i \in [S], K_i \notin \left[\frac{1}{2}\pi_i NT , \frac{3}{2} \pi_i NT\right]\right) \leq f$.
\end{lemma}

\begin{lemma}
\label{lem:diffempvsnoisyemp}
    Let $(K_{i,j})_{i,j\in S}$ be a matrix with nonnegative entries such that for each row $i\in [S]$, $K_i:=\sum_{j=1}^S {K_{i,j}} > 0$, then for any $K_{i,j} \geq 0$,
    for each $i, j \in [S]$, we have that
    \begin{align*}
        \sum_{j=1}^S \left| \frac{[K_{i,j} + Z_{i,j}]_+}{\sum_{k=1}^S [K_{i,k} + Z_{i,k}]_+} - \frac{K_{i,j}}{K_i}\right| \leq \frac{2\sum_{j=1}^S|Z_{i,j}|}{K_i - \sum_{k=1}^S |Z_{i,k}|}.
    \end{align*}
\end{lemma}

\begin{proof}
    Simplifying the LHS of the inequality gives
\begin{align*}
    \sum_{j=1}^S &\left| \frac{[K_{i,j} + Z_{i,j}]_+}{\sum_{k=1}^S [K_{i,k} + Z_{i,k}]_+} - \frac{K_{i,j}}{\sum_{k=1}^S K_{i,k}}\right| \\
    &\qquad \qquad \qquad = \sum_{j=1}^S \left| \frac{[K_{i,j} + Z_{i,j}]_+\sum_{k=1}^S K_{i,k} - K_{i,j}\sum_{k=1}^S [K_{i,k} + Z_{i,k}]_+}{\sum_{k=1}^S [K_{i,k} + Z_{i,k}]_+\sum_{k=1}^S K_{i,k}}\right| \\
    &\qquad \qquad \qquad = \sum_{j=1}^S \left| \frac{\sum_{k=1}^S \left([K_{i,j} + Z_{i,j}]_+ K_{i,k} - K_{i,j} [K_{i,k} + Z_{i,k}]_+\right)}{\sum_{k=1}^S [K_{i,k} + Z_{i,k}]_+\sum_{k=1}^S K_{i,k}}\right| \\
    &\qquad \qquad \qquad = \sum_{j=1}^S \left| \frac{\sum_{k=1}^S ([K_{i,j}K_{i,k} + K_{i,k}Z_{i,j}]_+  - [K_{i,k}K_{i,j} + K_{i,j}Z_{i,k}]_+)}{\sum_{k=1}^S [K_{i,k} + Z_{i,k}]_+\sum_{k=1}^S K_{i,k}}\right| \\
    &\qquad \qquad \qquad = \frac{\sum_{j=1}^S\left|\sum_{k=1}^S ([K_{i,j}K_{i,k} + K_{i,k}Z_{i,j}]_+  - [K_{i,k}K_{i,j} + K_{i,j}Z_{i,k}]_+)\right|}{\sum_{k=1}^S [K_{i,k} + Z_{i,k}]_+\sum_{k=1}^S K_{i,k}} \\
    &\qquad \qquad \qquad \leq \frac{\sum_{j=1}^S \sum_{k=1}^S \left|[K_{i,j}K_{i,k} + K_{i,k}Z_{i,j}]_+  - [K_{i,k}K_{i,j} + K_{i,j}Z_{i,k}]_+\right|}{\sum_{k=1}^S [K_{i,k} + Z_{i,k}]_+\sum_{k=1}^S K_{i,k}} \\
    &\qquad \qquad \qquad \leq \frac{\sum_{j=1}^S \sum_{k=1}^S \left|K_{i,k}Z_{i,j} - K_{i,j}Z_{i,k}\right|}{\sum_{k=1}^S [K_{i,k} + Z_{i,k}]_+\sum_{k=1}^S K_{i,k}}\\
    &\qquad \qquad \qquad \leq \frac{\sum_{j=1}^S \sum_{k=1}^S (K_{i,k}|Z_{i,j}| + K_{i,j}| Z_{i,k}|)}{\sum_{k=1}^S [K_{i,k} + Z_{i,k}]_+\sum_{k=1}^S K_{i,k}} \\
    &\qquad \qquad \qquad \leq \frac{ (\sum_{k=1}^S K_{i,k})(\sum_{j=1}^S|Z_{i,j}|) + (\sum_{j=1}^S K_{i,j})(\sum_{k=1}^S| Z_{i,k}|)}{\sum_{k=1}^S [K_{i,k} + Z_{i,k}]_+\sum_{k=1}^S K_{i,k}} \\
    &\qquad \qquad \qquad = \frac{2\sum_{j=1}^S|Z_{i,j}|}{\sum_{k=1}^S [K_{i,k} + Z_{i,k}]_+} \leq \frac{2\sum_{j=1}^S|Z_{i,j}|}{\sum_{k=1}^S K_{i,k} - \sum_{k=1}^S |Z_{i,k}|}.
\end{align*}

This proves the claim.
\end{proof}

\begin{theorem}[Complete Privacy Guarantee of Theorem~\ref{thm:item-level-dp-upperbound-informal}]
\label{thm:pure-dp-item-level-upper-bound:privacy}
Algorithm~\ref{alg:private-transition-estimator-multiple-chains} with $\mathcal{D} = \mathrm{Lap}\left(4/\varepsilon\right)$ is $\varepsilon$ differentially private.
\end{theorem}
\begin{proof}
    Consider the count matrix $K'$ corresponding to the adjacent database. Changing a single data point in a chain can affect at most two transitions, which implies that the $\ell_1$ norm between the vectorized versions of the matrices is upper bounded by $4$. Hence, for $\mc{D} = Lap\left(\frac{4}{\varepsilon}\right)$, we get that the count matrix is $\varepsilon$-DP by the Laplace mechanism \cite{dwork2014}. Steps 2 and 3 are post-processing operations, which implies that the resultant matrix $\hat{P}^{\mathrm{np}}$ satisfies $\varepsilon$-DP.
\end{proof}

\begin{theorem}[Complete Utility Guarantee of Theorem~\ref{thm:item-level-dp-upperbound-informal}]
\label{thm:pure-dp-item-level-upper-bound:utility}
Fix \(0<\alpha,f<1\) and \(0<\eps\leq 1\). Uniformly over
\(P\in\mathcal{C}(S,\pi_{\min},\gamma_{ps})\) and arbitrary initial
distributions, Algorithm~\ref{alg:private-transition-estimator-multiple-chains} with
\(\mathcal{D}=\operatorname{Lap}(0,4/\eps)\) satisfies $\Pr\!\left(\rho(\widehat P,P)\leq\alpha\right)\geq 1-f$ for
\[
    N = {\mathcal{O}}\left(\frac{S\log\left(\frac{S}{f}\right)}{\pi_{\min}T\alpha^2} + \frac{\ln\left(\frac{1}{\pi_{\min}}\right)}{\pi_{\min}T} + \frac{\log\left(\frac{S}{f}\right)}{\gamma_{ps} \pi_{\min}T} + \frac{S +\max\left\{\sqrt{S\log\left(\frac{S}{f}\right)}, \log\left(\frac{S}{f}\right)\right\}}{\pi_{\min}T\alpha\varepsilon}\right).
\]
\end{theorem}

\begin{proof}
We start by bounding the matrix-norm distance between corresponding rows of $\hat{P}^{\mathrm{np}}$ with $\Hat{P}$. Consider an arbitrary $i^{th}$ row of the two matrices. We seek a high probability bound on the following quantity
\begin{align*}
    \sum_{j=1}^S \left| \frac{[K_{i,j} + Z_{i,j}]_+}{\sum_{k=1}^S [K_{i,k} + Z_{i,k}]_+} - \frac{K_{i,j}}{\sum_{k=1}^S K_{i,k}}\right|.
\end{align*}
Let $0 < R \leq \frac{1}{4} \pi_{\min} NT$ and $\esterr \leq 1$. Define $K_i := \sum_{j=1}^S K_{i,j}$ and consider the events
\begin{align*}
    \mc E_1 &= \left\{\sum_{j=1}^S |Z_{i,j}| \leq R \text{ for all rows }i \in [S]\right\} \\
    \mc E_2 &=  \left\{\forall i \in [S], K_i \in \left[\frac{1}{2}\pi_i NT, \frac{3}{2}\pi_i NT\right]\right\}.
\end{align*}
Conditional on the events $\event_1$ and $\event_2$, we have that $K_i \geq \frac{1}{4}\pi_{\min}NT \geq R$. Hence, we get that for all $i \in [S]$
\begin{align*}
    \sum_{j = 1}^S [K_{i,j} + Z_{i,j}]_+ \geq \sum_{j = 1}^S (K_{i,j} + Z_{i,j}) = K_i +  \sum_{j = 1}^S Z_{i,j} \geq K_i - \sum_{j = 1}^S |Z_{i,j}| \geq \frac{1}{4} \pi_{\min}T > 0.
\end{align*}

Thus, $\widetilde{P}_{i,j} := \frac{\widetilde{K}_{i,j}}{\widetilde{K}_i}$. For each row, Lemma~\ref{lem:diffempvsnoisyemp} bounds the above quantity by
\begin{align*}
    \sum_{j=1}^S \left| \frac{|K_{i,j} + Z_{i,j}|}{\sum_{k=1}^S [K_{i,k} + Z_{i,k}]_+} - \frac{K_{i,j}}{\sum_{k=1}^S K_{i,k}}\right| \leq  \frac{2R}{K_i - R} \leq \frac{2R}{\frac{1}{2}\pi_{\min}NT - R}.
\end{align*}

Let $\sigma^2 = \frac{8}{\eps^2}$. A Laplace random variable is sub-exponential with Bernstein parameters $\left(\sigma^2, \frac{\sigma}{\sqrt{2}}\right)$. Then, applying Bernstein's inequality \citep{Vershynin2026} for each row and then taking the union bound over all the rows, we obtain
\begin{align*}
    \mbb{P}\left(\exists\ i \in [S], \left|\sum_{j=1}^S |Z_{i,j}| - \frac{2S}{\eps}\right|\ge t\right)
    \le 2S\exp\left(-c\,\min\left(\frac{t^2}{S\sigma^2}, \frac{t}{\sigma}\right)\right).
\end{align*}

Hence, setting $R = \frac{2S}{\varepsilon} + t$, we get that $\mbb{P}(\overline{\mc E_1}) \leq 2S\exp\left(-c\,\min\left(\frac{t^2}{S\sigma^2}, \frac{t}{\sigma}\right)\right)$. Thus, the event $\mc E_1 \cap \mc E_2$ implies that $\norm{\hat{P}^{\mathrm{np}} - \hat{P}}_\infty \leq \alpha$. Thus, taking $NT \gtrsim \max\left\{\frac{R}{\pi_{\min}\alpha}, \frac{1}{\pi_{\min} \gamma_{ps}} \log\left(\frac{S}{f}\right) \right\}$ and using Lemma~\ref{lem:counts}, we get
\begin{align*}
    \mbb{P}\left(\norm{\hat{P}^{\mathrm{np}} - \hat{P}}_\infty \geq \alpha\right) &\leq \mbb{P}\left(\norm{\hat{P}^{\mathrm{np}} - \hat{P}}_\infty \geq \alpha \mid \mc E_1 \cap \mc E_2\right) + \mbb{P}(\overline{\mc E_1}) + \mbb{P}(\overline{\mc E_2}) \\
    &\leq 2S\exp\left(-c\,\min\left(\frac{t^2}{S\sigma^2}, \frac{t}{\sigma}\right)\right) + \frac{f}{3}.
\end{align*}
Setting $t = \frac{2\sqrt{2}}{\varepsilon}\max\left\{\sqrt{S\log\left(\frac{18S}{f}\right)}, \log\left(\frac{18S}{f}\right)\right\}$ and combining this with Lemma~\ref{lem:markov-chain-estimation-item}, for
\[
N \gtrsim \max\left\{\frac{S}{\pi_{\min}T\alpha^2}\log\left(\frac{S}{f}\right) + \frac{1}{\pi_{\min}} \ln\left(\frac{1}{\pi_{\min}}\right), \frac{1}{\pi_{\min}\gamma_{ps} T}\log\left(\frac{S}{f}\right)\right\},
\]
we obtain
\begin{align*}
    \mbb{P}\left(\norm{\hat{P}^{\mathrm{np}} - P}_\infty \geq \alpha\right) &\leq \mbb{P}\left(\norm{\hat{P} - P}_\infty \geq \alpha\right) + \mbb{P}\left(\norm{\hat{P}^{\mathrm{np}} - \hat{P}}_\infty \geq \alpha\right) \leq f.
\end{align*}
Moreover, note that the above value of $t$ implies that
\[
R = \Theta\left(\frac{S}{\varepsilon} + \frac{1}{\varepsilon}\max\left\{\sqrt{S\log\left(\frac{6S}{f}\right)}, \log\left(\frac{6S}{f}\right)\right\}\right),
\]
which gives the eventual user complexity
\[
    N = {O}\left(\frac{S\log\left(\frac{S}{f}\right)}{\pi_{\min}T\alpha^2} + \frac{\ln\left(\frac{1}{\pi_{\min}}\right)}{\pi_{\min}T} + \frac{\log\left(\frac{S}{f}\right)}{\pi_{\min}\gamma_{ps} T} + \frac{S +\max\left\{\sqrt{S\log\left(\frac{S}{f}\right)}, \log\left(\frac{S}{f}\right)\right\}}{\pi_{\min}T\alpha\varepsilon}\right).
\]
\end{proof}

\subsection{Gaussian Mechanism for Approximate DP}
\label{ssec:approxDP}

\begin{theorem}
Algorithm~\ref{alg:private-transition-estimator-multiple-chains} with $\mathcal{D} = \mathcal{N}(0, \sigma^2)$, where $\sigma = \frac{4\sqrt{\log(1/\delta)}}{\varepsilon}$ is $(\varepsilon, \delta)$ differentially private.
\end{theorem}
\begin{proof}
    Using the same argument as in Theorem~\ref{thm:pure-dp-item-level-upper-bound:privacy}
    , for an adjacent matrix $K'$, the $\norm{K - K'}_{F} \leq 2$. Hence, for $\mc{D} = \mc{N}(0, \sigma^2)$ where $\sigma = \frac{4\sqrt{\log(1/\delta)}}{\varepsilon}$, the privacy guarantee follows from the $(\varepsilon, \delta)$ privacy guarantee of the Gaussian mechanism on the count matrix $K$ and post-processing operations, so the resulting $\hat{P}^{\mathrm{np}}$ satisfies $(\eps, \del)$-DP.
\end{proof}

\begin{theorem}
Fix \(0<\alpha,f<1\) and \(0<\eps\leq 1\). Uniformly over
\(P\in\mathcal{C}(S,\pi_{\min},\gamma_{ps})\) and arbitrary initial
distributions, Algorithm~\ref{alg:private-transition-estimator-multiple-chains} with $\mathcal{D}= \mathcal{N}(0, \sigma^2)$, where $\sigma = \frac{4\sqrt{\log(1/\delta)}}{\varepsilon}$, satisfies $\Pr\!\left(\rho(\widehat P,P)\leq\alpha\right)\geq 1-f$ provided that
\[
    N = \Theta\left(\frac{S\log\left(\frac{S}{f}\right)}{\pi_{\min}T\alpha^2} + \frac{\ln\left(\frac{1}{\pi_{\min}}\right)}{\pi_{\min}T} + \frac{\log\left(\frac{S}{f}\right)}{\gamma_{ps} \pi_{\min}T} + \frac{S\sqrt{\log\left(\frac{1}{\delta}\right)}}{\varepsilon\pi_{\min}T\alpha} + \frac{\sqrt{S\log\left(\frac{1}{\delta}\right)\log\left(\frac{S}{f}\right)}}{\pi_{\min}T\alpha\varepsilon}\right).
\]
\end{theorem}

\begin{proof}
Since $|Z_{i,j}|$ has mean $\sqrt{\frac{2}{\pi}} \sigma$, we have $\mathbb E \sum_{j=1}^S |Z_{i,j}|= \sqrt{\frac{2}{\pi}} S\sigma$. Since the absolute value function is $1$-Lipschitz, it implies that $|Z_{i,j}| - \mathbb E|Z_{i,j}|$ is $2\sigma$ sub-gaussian.
Thus, by concentration of sum of sub-gaussian random variables \citep{Vershynin2026} applied to each row, followed by a union bound over all rows, we get that
\begin{align*}
    \mbb{P}\left(\exists\ i \in [S], \left|\sum_{j=1}^S |Z_{i,j}| -  S\sigma\sqrt{\frac{2}{\pi}}\right| \ge t\right) \le 2 S\exp\left(-\frac{t^2}{2 S \sigma^2}\right).
\end{align*}

As in Theorem~\ref{thm:pure-dp-item-level-upper-bound:utility}, setting $R = S\sigma\sqrt{\frac{2}{\pi}} + t$, we get that $\mbb{P}(\overline{\mc E_1}) \leq 2 S\exp\left(-\frac{t^2}{2 S \sigma^2}\right)$. Defining events $\mc E_1$ and $\mc E_2$ as in Theorem~\ref{thm:pure-dp-item-level-upper-bound:utility}, the event $\mc E_1 \cap \mc E_2$ implies that $\norm{\hat{P}^{\mathrm{np}} - \hat{P}}_\infty \leq \alpha$ when $N = \Theta\left(\frac{R}{\pi_{\min}\alpha}\right)$. Thus, taking $NT \gtrsim \max\left\{\frac{R}{\pi_{\min}\alpha}, \frac{1}{\pi_{\min} \gamma_{ps}} \log\left(\frac{S}{f}\right) \right\}$ and using Lemma~\ref{lem:counts},
\begin{align*}
    \mbb{P}\left(\norm{\hat{P}^{\mathrm{np}} - \hat{P}}_\infty \geq \alpha\right) &\leq \mbb{P}\left(\norm{\hat{P}^{\mathrm{np}} - \hat{P}}_\infty \geq \alpha \mid \mc E_1 \cap \mc E_2\right) + \mbb{P}(\overline{\mc E_1}) + \mbb{P}(\overline{\mc E_2}) \\
    &\leq 2 S\exp\left(-\frac{t^2}{2 S \sigma^2}\right) + \frac{f}{3}.
\end{align*}
Setting $t = 2\sigma\sqrt{S\log\left(\frac{6S}{f}\right)}$ and combining this with Lemma~\ref{lem:markov-chain-estimation-item}, for
\[
N \gtrsim \max\left\{\frac{S}{\pi_{\min}T\alpha^2}\log\left(\frac{S}{f}\right) + \frac{1}{\pi_{\min}} \ln\left(\frac{1}{\pi_{\min}}\right), \frac{1}{\pi_{\min}\gamma_{ps} T}\log\left(\frac{S}{f}\right)\right\},
\]
we obtain
\begin{align*}
    \mbb{P}\left(\norm{\hat{P}^{\mathrm{np}} - P}_\infty \geq \alpha\right) &\leq \mbb{P}\left(\norm{\hat{P} - P}_\infty \geq \alpha\right) + \mbb{P}\left(\norm{\hat{P}^{\mathrm{np}} - \hat{P}}_\infty \geq \alpha\right) \leq f.
\end{align*}
Moreover, note that the above value of $t$ implies that $R = \Theta\left(\frac{S\sqrt{\log(1/\delta)}}{\varepsilon} + \frac{\sqrt{S\log(1/\delta)}}{\varepsilon}\sqrt{\log\left(\frac{S}{f}\right)}\right)$, which gives the eventual sample complexity to be

\begin{align*}
    N = \Theta\left(\frac{S\log\left(\frac{S}{f}\right)}{\pi_{\min}T\alpha^2} + \frac{\ln\left(\frac{1}{\pi_{\min}}\right)}{\pi_{\min}T} + \frac{\log\left(\frac{S}{f}\right)}{\gamma_{ps} \pi_{\min}T} + \frac{S\sqrt{\log\left(\frac{1}{\delta}\right)}}{\varepsilon\pi_{\min}T\alpha} + \frac{\sqrt{S\log\left(\frac{1}{\delta}\right)\log\left(\frac{S}{f}\right)}}{\pi_{\min}T\alpha\varepsilon}\right).
\end{align*}
\end{proof}

\subsection{Complete Statement and Proof of Theorem~\ref{thm:item-level-dplowerbound-informal}}
\label{app:lowerbound-item-formal}
Let \(Y_{1:M}=(Y_1,\ldots,Y_M)\) be i.i.d.\ samples from an unknown
distribution \(p\in\Delta(k)\). An item-level
\((\varepsilon,\delta)\)-differentially private estimator outputs
\(\widehat p\in\Delta(k)\), and we seek the smallest \(M\) for which $\mbb{P}\left(d_{\mathrm{TV}}(\widehat p,p)\leq\alpha\right) \geq 1-f$ holds uniformly over \(p\in\Delta(k)\). Note that for $p, q \in \Delta(k)$, $\norm{p-q}_1 = 2d_{\mathrm{TV}}(p,q)$. The following result gives a risk lower bound for this problem. It is the specialization of \citet{esfandiari2022tight} to one sample per user, for which user-level privacy coincides with item-level privacy.

\begin{lemma}{{\citep[Theorem 10]{esfandiari2022tight}}}
\label{lem:mean_estimation}
    For every $(\varepsilon, \delta)$ item-level
    differentially private algorithm $\mathcal{P}$ satisfying $\delta = o\!\left(\frac{\varepsilon}{k^{3/2} \cdot n^{1/2} \cdot \log^2(\varepsilon^{-1} n k)}\right)$, there exists a distribution $\tilde{p}$ over $[k]$,
    such that if $X_{1:N} = (X_{1}, \cdots, X_n) \overset{\text{i.i.d.}}{\sim} \tilde{p}$, then with probability at least $1/3$,
\[
\norm{\mathcal{P}\left(X_{1:N}\right) - \tilde{p}}_1 = \Omega\!\left(\sqrt{\frac{k}{n}} +  \frac{k}{\varepsilon \cdot n \cdot \log^7(\varepsilon^{-1} \cdot n k)}
\right).
\]
\end{lemma}
The lemma lower-bounds the achievable error as a function of the sample size. Solving this bound for the number of samples needed to attain error \(\alpha\) gives the following sample-complexity lower bound.
\begin{corollary}
\label{cor:mean-estimation-sample-complexity}
Let $k \geq 2$, $0 < \alpha \leq 1$ and $0 < \varepsilon \leq 1$. Suppose that $\mathcal{P}$ is an $(\varepsilon,\delta)$ item-level
differentially private algorithm and that, for every distribution $p$
over $[k]$, when given $X_{1:N} = (X_1,\ldots,X_N)
\overset{\mathrm{i.i.d.}}{\sim}p$, it satisfies
\[
\mbb{P}\!\left(
    \left\|
        \mathcal{P}(X_{1:N})-p
    \right\|_1
    \leq \alpha
\right) > \frac{2}{3}.
\]
If $\delta = o\!\left( \frac{\varepsilon^{3/2} \alpha}{k^{2}\log^2\left(\frac{k}{\alpha\varepsilon}\right)}\right)$, then
\[
N = \Omega\!\left(
    \frac{k}{\alpha^2}
    +
    \frac{k}
    {\alpha\varepsilon
    \log^7\!\left(\frac{k}{\alpha\varepsilon}\right)}
\right).
\]
\end{corollary}

\begin{proof}
Let $N_\alpha := \frac{k}{\alpha^2} + \frac{k}{\alpha\varepsilon \log^7\left(\frac{k}{\alpha\varepsilon}\right)}$. If $N \geq N_\alpha$, the conclusion follows immediately. Thus, it
suffices to consider the case $N<N_\alpha$. Since $N_\alpha = O\!\left(\frac{k}{\alpha^2\varepsilon}\right)$, we have
$\log(\varepsilon^{-1}Nk) = O\left(\log\left(\frac{k}{\alpha\varepsilon}\right)\right)$. Consequently, the assumed condition on $\delta$ implies
\[
\delta
=
o\!\left(
    \frac{\varepsilon}
    {k^{3/2}\sqrt{N}\,
    \log^2(\varepsilon^{-1}Nk)}
\right),
\]
and hence Lemma~\ref{lem:mean_estimation} applies. It follows that there exists a distribution $\widetilde{p}$ over
$[k]$ such that, with probability at least $1/3$,
\[
\left\|
    \mathcal{P}(X_{1:N})-\widetilde{p}
\right\|_1
\geq
c\left(
    \sqrt{\frac{k}{N}}
    +
    \frac{k}
    {\varepsilon N
    \log^7(\varepsilon^{-1}Nk)}
\right)
\]
for some universal constant $c>0$. Since $\mathcal{P}$ has error at
most $\alpha$ with probability greater than $2/3$, we must have
\[
\sqrt{\frac{k}{N}}
+
\frac{k}
{\varepsilon N\log^7(\varepsilon^{-1}Nk)}
=
O(\alpha).
\]

The first term gives $N = \Omega\!\left(\frac{k}{\alpha^2}\right)$. Moreover, using
$\log(\varepsilon^{-1}Nk)=O\left(\log\left(\frac{k}{\alpha\varepsilon}\right)\right)$, the second term gives $N = \Omega\!\left(\frac{k}{\alpha\varepsilon L_\alpha^7}
\right)$. Combining the two bounds yields
\[
N
=
\Omega\!\left(
    \frac{k}{\alpha^2}
    +
    \frac{k}
    {\alpha\varepsilon
    \log^7\!\left(\frac{k}{\alpha\varepsilon}\right)}
\right).
\]
\end{proof}
To transfer this i.i.d. lower bound to transition-matrix estimation, we reduce discrete distribution estimation to estimating an \(S\)-state Markov chain with a privacy-preserving reduction. The next lemma gives the privacy and accuracy guarantees of this reduction.
\begin{lemma}
\label{lem:kary-to-markov-reduction}
Let $S\geq 2$, $0<\pi_{\min}\leq\frac{1}{S+2}$, and $NT \geq \frac{C_0}{\pi_{min}}$. For every $(\varepsilon, \delta)$ differentially private algorithm $\mathcal M$ that receives $N$ length-$T$ trajectories $\mathbf{X} = (X^{(1)}_{1:T}, \cdots, X^{(N)}_{1:T})$ from a $S$-state ergodic Markov chain $M$ with $\min_x\pi(x)=\pi_{\min}$ and satisfies
\[
\mathbb P\left(
\left\|
\mathcal M(\mathbf{X})-M
\right\|_\infty
\leq
\alpha
\right)
\geq\frac34,
\]
there exists a corresponding $(\varepsilon, \delta)$-DP algorithm $\mathcal{P}$ such that, for every $p \in \Delta(S-1)$ when $\mathcal{P}$ is given $\lfloor 2\pi_{\min}NT \rfloor$ i.i.d. samples from $p$, denoted by $Y_{1:\lfloor 2\pi_{\min}NT \rfloor}$, it satisfies
\[
\mathbb P\left(
\left\|
\mathcal P\left(Y_{1:\lfloor 2\pi_{\min}NT \rfloor}\right)-p
\right\|_1
\leq
4\alpha
\right)
\geq\frac23.
\]
\end{lemma}

\begin{proof}
Let ${\cal M}$ be the algorithm for Markov chain estimation with error bounded by $\alpha$ with probability at least $3/4$. Under the assumption $\pi_{\min} \leq \frac{1}{S + 2}$, let $k = S-1$. Consider the following reduction for any $p \in \Delta(k)$. Initialize $X_0$ by setting $X_0 = i$ with probability $\frac{1 - \pi_{\min}}{k}$ for each $i \in [k]$ and $X_0 = k+1$ with probability $\pi_{\min}$. Now, consider the following process:
For $j\leq k$,
\[ X_{i+1}|(X_i=j) =
\left\{
\begin{array}{lcl}
\ell & \mbox{w.p. } \frac{1-\pi_{\min}}{k} & \mbox{ if } \ell\leq k \\
k+1  & \mbox{w.p. } \pi_{\min} &\mbox{ if }\ell=k+1.
\end{array}
\right.
\]
For $j=k+1$
\[ X_{i+1}|(X_i=k+1) =
\left\{
\begin{array}{lll}
\ell & \mbox{w.p. } (1-\pi_{\min})p_{\ell} & \mbox{ if } \ell\leq k \\
k+1  & \mbox{w.p. } \pi_{\min} &\mbox{ if }\ell=k+1.
\end{array}
\right.
\]
Clearly the process above is a Markov Chain with transition matrix $M^{(p)}$,
\begin{align*}
    \begin{pmatrix}
    \frac{1 - \pi_{\min}}{k} & \cdots & \frac{1 - \pi_{\min}}{k} & \pi_{\min} \\
    \vdots & \ddots & \vdots & \vdots \\
    \frac{1 - \pi_{\min}}{k} & \cdots & \frac{1 - \pi_{\min}}{k} & \pi_{\min} \\
    (1 - \pi_{\min})p_1 & \cdots & (1 - \pi_{\min})p_k & \pi_{\min}
    \end{pmatrix}.
\end{align*}
Moreover, this transition matrix defines a chain with stationary distribution given by
\begin{align*}
\pi_i = \frac{(1 - \pi_{\min})^2}{k} + (1 - \pi_{\min})p_i \pi_{\min},
\quad \text{for } i \in [k],
\qquad
\pi_{k+1} = \pi_{\min}.
\end{align*}

Since $\pi_{\min} < \frac{1}{k+2}$, we have $\pi_{min}^2 - (k+2)\pi_{min} + 1 > 0$, which implies $\frac{(1 - \pi_{min})^2}{k} > \pi_{min}$. Hence, $\pi_{\min}$ is the minimum stationary distribution probability.

\noindent
Given $M$ i.i.d samples $Y_1, \cdots, Y_M \stackrel{i.i.d.}{\sim} p$,
define the following simulation $\mathcal{P}$ which estimates $p$:
\begin{enumerate}
    \item Set $t = 1$.
    \item For each $u \in [N]$:
    \begin{enumerate}
        \item Sample $X_0 \sim \left(\frac{1 - \pi_{\min}}{k} , \cdots , \frac{1 - \pi_{\min}}{k}, \pi_{\min}\right)$.
        \item For $i = 0, \dots, T-1$:
        \begin{enumerate}
            \item If $X_{i} = k+1$, then set
            \begin{align*}
                X_{i+1} =
                \begin{cases}
                    Y_t  &w.p.\ \ (1 - \pi_{\min})\\
                    k+1 &w.p.\ \ \pi_{\min}
                \end{cases},
            \end{align*}
            and set $t \gets t + 1$.
            \item If $X_i \neq k+1$, set $X_{i+1} \sim \left(\frac{1 - \pi_{\min}}{k} , \cdots , \frac{1 - \pi_{\min}}{k}, \pi_{\min}\right)$.
            \item If $t = M + 1$, then return $\left(\frac{1}{k}, \cdots, \frac{1}{k}\right)$.
        \end{enumerate}
    \end{enumerate}
    \item Run $\mc{M}(X^{1}_{0:T}, \cdots, X^{N}_{0:T})$ and let $\tilde{p}$ be the first $k$ entries of the $k+1$-st row.
    \item Return $\hat p = \arg\min_{q \in \Delta(k)} \norm{q - \frac{\tilde p}{1 - \pi_{min}}}$.
\end{enumerate}

We can observe two things:
\begin{itemize}
    \item The random variables $\mathbf{1}_{\{X_i = k+1\}}$ are i.i.d.~Bernoulli with parameter $\pi_{\min}$.
    Note that $t$ represents the number of times we hit $k+1$ (which corresponds to the number of times we use a sample from $p$ for determining the next state).
\item All Markov chains constructed are mutually independent.
\end{itemize}

Let $J$ be the $(M+1)$-th arrival time of state $k+1$.
If $L_i$ are the interarrival times, then
\begin{align*}
    J = \sum_{j=1}^{M+1} L_i,
\end{align*}
and $(L_i)_i \stackrel{i.i.d.}{\sim} \mathrm{Geom}(\pi_{\min})$.
Hence, by our reduction, the final value of $t$ is less than or equal to $J$. Therefore, the event $J \leq NT$ implies that the algorithm returns $\left(\frac{1}{k}, \cdots, \frac{1}{k}\right)$.

\paragraph{Adjacency-preserving coupling.}
Let \(Y_{1:M}=(Y_1,\ldots,Y_M)\) and \(Y'_{1:M}=(Y'_1,\ldots,Y'_M)\) be neighboring datasets that differ only in coordinate \(r\). Couple the two simulations using exactly the same randomness: the same initial states, the same Bernoulli choices made when leaving state \(k+1\), and the same random choices from every row in $[k]$. Under this coupling, the two simulations visit state \(k+1\) at the same times and therefore consume the samples \(Y_t\) and \(Y'_t\) in the same order. Indeed, whether the next state is \(k+1\) is determined only by the shared randomness and not by the value of \(Y_t\). Consequently, both simulations also reach the stopping event simultaneously.

Before sample \(r\) is consumed, the two simulated datasets agree exactly. When sample \(r\) is consumed, either the shared Bernoulli choice selects state \(k+1\), in which case the simulations continue
to agree, or it selects the data-dependent successor, in which case the next states are \(Y_r\) and \(Y'_r\). These are both states in \([k]\). Since all rows indexed by \([k]\) in \(M^{(p)}\) are identical, using shared randomness makes the two trajectories agree again at the following step. They remain identical thereafter. Thus, for every fixed realization of the simulation randomness, the two resulting multi-trajectory datasets are either identical or differ in exactly one state observation within one trajectory. They are therefore item-level neighbors.

\paragraph{Privacy guarantee.}
Let \(U\) collect all randomness used by the trajectory simulation, and let \(\mc{U}_U(Y)\) denote the resulting trajectory dataset. The stopping time \(J\) is determined entirely by \(U\): it depends on the initial-state draws and transition coins, but not on the values of \(Y_1,\ldots,Y_M\). Consequently, the event $E:=\{J>NT\}$ has the same probability under \(Y\) and \(Y'\), and the conditional distribution of \(U\) given \(E\) is also the same for both inputs.

For every fixed \(u\in E\), the coupling argument above shows that \(\mc{U}_u(Y)\) and \(\mc{U}_u(Y')\) are either identical or
item-level neighbors. Therefore, differential privacy of
\(\mc{M}\) gives
\[
\mbb{P}\left(\mc{M}(\mc{U}_u(Y))\in S\right)
\leq e^\varepsilon
\mbb{P}\left(\mc{M}(\mc{U}_u(Y'))\in S\right)+\delta.
\]
On the event \(E\), we have \(\mc{R}(Y)=\mc{M}(\mc{U}_U(Y))\) (Step 3 of the reduction). Hence,
\begin{align*}
\mbb{P}\left(
    \mc{R}(Y)\in S
    \,\middle|\, E
\right) &=
\mathbb E_{U\mid E}\!\left[
    \mbb{P}\left(
        \mc{M}(\mc{U}_U(Y))\in S
        \,\middle|\, U
    \right)
\right] \\
&\quad\leq
e^\varepsilon
\mathbb E_{U\mid E}\!\left[
    \mbb{P}\left(
        \mc{M}(\mc{U}_U(Y'))\in S
        \,\middle|\, U
    \right)
\right]
+\delta  \\
&\quad=
e^\varepsilon
\mbb{P}\left(
    \mc{R}(Y')\in S
    \,\middle|\, E
\right)
+\delta.
\end{align*}
Thus, the mechanism conditioned on \(E\) remains DP. Let \(q:=\mbb{P}(E)\), which is independent of the input dataset, and let
\(q_0=(1/k,\ldots,1/k)\) be the fixed output on \(E^c\). By the law of
total probability,
\[
\begin{aligned}
\mbb{P}\left(\mc{R}(Y)\in S\right)
&= q\mbb{P}\left(\mc{R}(Y)\in S \middle| E \right) + (1-q)\mathbf{1}_{\left(\frac{1}{k}, \cdots, \frac{1}{k}\right) \in S}\\
&\leq q\left(e^\varepsilon\mbb{P}\left(\mc{R}(Y')\in S\middle| E\right) + \delta \right) +
(1-q)\mathbf{1}_{\left(\frac{1}{k}, \cdots, \frac{1}{k}\right) \in S} \\
&\leq e^\varepsilon\left(q\mbb{P}\left(\mc{R}(Y')\in S \middle|\, E\right) +(1-q)\mathbf{1}_{\left(\frac{1}{k}, \cdots, \frac{1}{k}\right) \in S}\right) +\delta   \\
&=
e^\varepsilon \mbb{P}\left(\mc{R}(Y')\in S\right) +\delta.
\end{aligned}
\]
Hence, $\mc{R}$ is $(\varepsilon, \delta)$-DP and by post processing, $\mc{P}$ is also $(\varepsilon, \delta)$-DP. Note that the knowledge of $t = M+1$ does not need to be privatized since it occurs only when $J \leq NT$ and this event does not depend on the input dataset $Y_1, \cdots, Y_M$.

\paragraph{Utility Guarantee}
We are given that $p \in \Delta(k)$ and $\hat p = \arg\min_{q \in \Delta(k)} \norm{q - \frac{\tilde p}{1 - \pi_{min}}}_1$. Hence, by the triangle inequality and the fact that $\hat p = \arg\min_{q \in \Delta(k)} \norm{q - \frac{\tilde p}{1 - \pi_{min}}}_1$ we have
\begin{equation}
\label{eq:projectionineq}
    \norm{\hat{p} - p}_1 \leq \norm{\hat{p} - \frac{\tilde p}{1 - \pi_{min}}}_1 + \norm{\frac{\tilde p}{1 - \pi_{min}} - p}_1 \leq 2 \norm{\frac{\tilde p}{1 - \pi_{min}} - p}_1
\end{equation}

Let $X_{1:N} = (X_1, \cdots, X_N)$ and $Y_{1:M} = (Y_1, \cdots, Y_M)$. The utility guarantee is as follows:
\begin{align*}
    \mbb{P}\left(\norm{\mathcal{P}(Y_{1:M}) - p}_1 \geq \alpha \right) &= \mbb{P}\left(\norm{\mathcal{P}(Y_{1:M}) - p}_1 \geq \alpha| J > NT\right) \mbb{P}(J > NT) \\
    &+ \mbb{P}\left(\norm{\mathcal{P}(Y_{1:M}) - p}_1 \geq \alpha| J \leq NT\right) \mbb{P}(J \leq NT)\\
    &\leq \mbb{P}\left(\norm{\frac{\tilde{p}(\mathbf{X})}{1 - \pi_{\min}} - p}_1 \geq \frac{\alpha}{2} \middle|\, J > NT\right) \mbb{P}(J > NT) + \mbb{P}(J \leq NT)\\
    &= \mbb{P}\left(\norm{\tilde{p}(\mathbf{X}) - (1 - \pi_{\min})p}_1 \geq (1 - \pi_{\min})\frac{\alpha}{2} \middle|\, J > NT\right) \mbb{P}(J > NT) \\
    &\ \qquad + \mbb{P}(J \leq NT) \\
    &\leq \mbb{P}\left(\norm{\mathcal{M}(\mathbf{X})_{(k+1,\cdot)} - M^{(p)}_{(k+1,\cdot)}}_1 \geq (1 - \pi_{\min})\frac{\alpha}{2} \middle|\, J > NT\right)\mbb{P}(J > NT) \\
    &\ \qquad + \mbb{P}(J \leq NT) \\
    &\leq \mbb{P}\left(\norm{\mathcal{M}(\mathbf{X}) - M^{(p)}}_\infty \geq \frac{\alpha}{4} \middle|\, J > NT\right) \mbb{P}(J > NT) + \mbb{P}(J \leq NT).
\end{align*}

The first step follows from the law of total probability. The second step follows from~\eqref{eq:projectionineq}. The third step follows from multiplying $(1 - \pi_{min})$ on both sides of $\norm{\frac{\tilde{p}(\mathbf{X})}{1 - \pi_{\min}} - p}_1 \geq \alpha$. The fourth step follows from the fact that the expression $\norm{\tilde{p} - (1 - \pi_{\min})p}_1$ represents the error of the first $k$ entries of the last row of $\mathcal{M}(\mathbf{X}) - M^{(p)}$. The fifth step follows from the definition of $\ell_\infty$ error and the fact that $\pi_{min} < \frac{1}{2}$ since $k \geq 0$.

By our construction $J = \sum_{j=1}^{M+1} L_i$, where $L_i \sim \mathrm{Geom}(\pi_{\min})$. Setting $\varepsilon = \frac{1}{2}$ in Lemma~\ref{lem:geomchernoff}, yields
\begin{align*}
    \mathbb{P}\!\left(J \le \frac{M}{2\pi_{\min}}\right) \leq \mathbb{P}\!\left(J \le \frac{M+1}{2\pi_{\min}}\right) \le \exp(-c' M),
\end{align*}
for some universal constant $c'$. Set $NT = \frac{M}{2\pi_{\min}}$ and for $M$ large enough (such that $\exp(-c'M) \leq \frac{1}{12}$), the utility guarantee of $\mathcal{M}$ yields
\begin{align*}
    \mbb{P}\left(\norm{\mathcal{P}(Y_{1:\lfloor 2\pi_{min}NT}\rfloor) - p}_1 \geq 4\alpha \right) \leq \frac{1}{3}.
\end{align*}
This proves the claim.
\end{proof}

Applying Corollary~\ref{cor:mean-estimation-sample-complexity} to the estimator supplied by Lemma~\ref{lem:kary-to-markov-reduction} translates the i.i.d.\ sample lower bound into a lower bound on the trajectory length.

\begin{theorem}[Complete version of Theorem~\ref{thm:item-level-dplowerbound-informal}]
\label{thm:item-level-lowerbound-formal}
Let \(S\geq 3\), \(0<\alpha\leq 1/4\),
\(0<\varepsilon\leq 1\), \(0<\pi_{\min}\leq \frac{1}{(S+2)}\),
\(0<\gamma_{ps}\leq 1\), and $\delta=o\!\left(\frac{\varepsilon^{3/2}\alpha}{S^2\log^2\!\left(\frac{S}{\alpha\varepsilon}\right)}\right)$. Let \(\mathcal M\) be an item-level
\((\varepsilon,\delta)\)-differentially private estimator based on $N$ mutually independent length $T$ trajectories, where \(NT\gtrsim \frac{1}{\pi_{\min}}\). Assume that, for every
\(P\in\mathcal C(S,\pi_{\min},\gamma_{ps})\) and every set of initial distributions \(\mu_1, \cdots, \mu_N\), a collection of trajectories $\mathbf{X} = (X^{(1)}_{1:T}, \cdots, X^{(N)}_{1:T})$ where $X^{(u)}_{1:T}\sim(P,\mu_u)$ satisfies
\[
    \mbb{P}\!\left(
        \left\|\mathcal M(\mathbf{X})-P\right\|_\infty\leq\alpha
    \right)
    \geq\frac34.
\]
Then
\[
    N
    =\Omega\!\left(
        \frac{S}{\pi_{\min}T\alpha^2}
        +\frac{S}
        {\pi_{\min}T\alpha\varepsilon
         \log^7\!\left(\frac{S}{\alpha\varepsilon}\right)}\right).
\]
\end{theorem}

\begin{proof}
Lemma~\ref{lem:kary-to-markov-reduction} converts \(\mathcal M\) into
an estimator for distributions over \(S-1\) symbols using
\(\lfloor2\pi_{\min}NT\rfloor\) samples and achieving \(\ell_1\) error
at most \(4\alpha\). Applying
Corollary~\ref{cor:mean-estimation-sample-complexity} proves the claim.
\end{proof}

\begin{remark}
    Note that when $N=1$, we can use the same technique to show that
    \[T = \Omega\!\left(\frac{S}{\pi_{\min}\alpha^2} +\frac{S}{\pi_{\min}\alpha\varepsilon\log^7\!\left(\frac{S}{\alpha\varepsilon}\right)}\right).
    \]
    Moreover, the non-private lower bound of \citet{MarkovChainEstimationBernoulli} also applies in the single-chain case, improving our lower bound to
    \[T = \Omega\!\left(\frac{S}{\pi_{\min}\alpha^2} +\frac{S}{\pi_{\min}\alpha\varepsilon\log^7\!\left(\frac{S}{\alpha\varepsilon}\right)} + \frac{S\ln(S)}{\gamma_{ps}}\right),\] which is tight for $\pi_{\min} = \Omega\left(\frac{1}{S\ln(S)}\right)$.
\end{remark}

\section{Proofs of Section~\ref{sec:kd_dist_bounds}}
\label{sec:productdist}
\subsection{Complete Statement and Proof of Theorem~\ref{thm:user-level-product-marginals-main}}
\label{ssec:approx-dp-upper-bound-formal}

Our upper bound is based on the tools defined in \citet{tsfadia2022friendlycore}.

\begin{definition}[{\citet[Definition 1.1]{tsfadia2022friendlycore}}]
    For a function $f:\mathcal X^2 \rightarrow \{0,1\}$, a dataset $\mathcal D \subseteq \mathcal X$ is called
    \begin{enumerate}
        \item $f$-complete if for every $z, z' \in \mathcal D$, $f(z, z') = 1$.
        \item $f$-friendly if, for every $z,z' \in \mathcal D$, there exists a point $w \in \mathcal X$ such that $f(z,w)=f(z',w)=1$.
    \end{enumerate}
\end{definition}

\begin{definition}[\(f\)-friendly differential privacy {\citet[Definition B.2]{tsfadia2022friendlycore}}]
    An algorithm \(A\) is called \(f\)-friendly
    \((\varepsilon,\delta)\)-differentially private if, for every pair of
    neighboring databases \(\mathcal{D},\mathcal{D}'\) such that
    \(\mathcal{D}\cup\mathcal{D}'\) is \(f\)-friendly, the distributions
    \(A(\mathcal{D})\) and \(A(\mathcal{D}')\) are
    \((\varepsilon,\delta)\)-indistinguishable.
\end{definition}

\begin{definition}[\(f\)-friendly concentrated differential privacy {\cite[Definition B.2]{tsfadia2022friendlycore}}]
    An algorithm \(A\) is called \(f\)-friendly
    \((\varepsilon,\delta)\)-differentially private if, for every pair of
    neighboring databases \(\mathcal{D},\mathcal{D}'\) such that
    \(\mathcal{D}\cup\mathcal{D}'\) is \(f\)-friendly, the distributions
    \(A(\mathcal{D})\) and \(A(\mathcal{D}')\) are
    \((\rho,\delta)\) zCDP-indistinguishable. For friendly neighboring databases, the standard conversion from zCDP to approximate DP holds.
\end{definition}

\begin{algorithm}[h]
\caption{\textsc{BasicFilter ($\alpha = 0$)} \cite{tsfadia2022friendlycore}.}
\label{alg:basic-filter}

\noindent
\textbf{\textit{Input:}}
A database
\(\mathcal{D}=(x_1,\ldots,x_n)\in\mathcal{X}^{n}\),
a predicate
\(f:\mathcal{X}^{2}\to\{0,1\}\).

\noindent
\textbf{\textit{Operation:}}
\begin{enumerate}
    \item For \(i\in[n]\):
    \begin{enumerate}
        \item Let $z_i = \sum_{j=1}^{n} f(x_i,x_j)-\frac{n}{2}$.
        \item Sample \(v_i\gets\operatorname{Bern}(p_i)\), where
        \[
            p_i
            =
            \begin{cases}
                0,
                & z_i\leq 0,\\[2mm]
                1,
                & z_i\geq \frac{n}{2}\\
                \displaystyle
                \frac{2z_i}{n},
                & \text{otherwise}.
            \end{cases}
        \]
    \end{enumerate}
    \item Output \(v=(v_1,\ldots,v_n)\).
\end{enumerate}
\end{algorithm}

It has been shown that $f$-friendly DP algorithms can be converted into DP algorithms by composition with a pre-processing step known as a {\em filter}. This filtering algorithm selects a set of $f$-friendly data-points from the original dataset, on which $f$-friendly DP guarantees can be applied. The following result from \citet{tsfadia2022friendlycore} formalizes this. From a utility point of view, filtering may fail for arbitrary datasets, but one can use concentration arguments for certain i.i.d. datapoints to show that, with high probability, such datasets satisfy a notion of friendliness with high probability or contain large friendly subsets. This yields favorable utility guarantees whilst only requiring to satisfy a relaxed notion of privacy.

\begin{lemma}[{\citet[Theorem C.11, $\alpha = 0$]{tsfadia2022friendlycore}}]
\label{lem:filterparadigm}
    Let
    \(f:\mathcal{X}^2 \to \{0,1\}\), and let \(A\) be an
    \(f\)-friendly \((\varepsilon,\delta)\)-differentially private
    algorithm. Given a database
    \(\mathcal{D}=(x_1,\ldots,x_n)\in\mathcal{X}^*\), define $v = \operatorname{BasicFilter}(\mathcal{D},f) \in \{0,1\}^n$ and  $\mathcal{D}_v := (x_i)_{i\in[n]:\,v_i=1}$, and consider the algorithm $B(\mathcal{D}) := A(\mathcal{D}_v)$. Then \(B\) is $\left(2\bigl(e^\varepsilon-1\bigr), 2\delta e^{\varepsilon + 2\bigl(e^\varepsilon-1\bigr)} \right) $-differentially private.
\end{lemma}

\begin{algorithm}[h]
\caption{\textsc{FriendlyAvg} \cite{tsfadia2022friendlycore}.}
\label{alg:friendly-avg}

\noindent
\textbf{\textit{Input:}}
A database \(\mathcal{D}\in(\mathbb{R}^d)^*\), privacy parameters
\(\rho,\delta>0\), and \(r\geq 0\).

\medskip
\noindent
\textbf{\textit{Operation:}}
\begin{enumerate}
    \item Let $n = |\mathcal{D}|,\rho_1 = 0.1(1-\delta)\rho, \rho_2 = 0.9\rho$.

    \item Sample $\widehat n\sim {\cal N}\Big(n - \sqrt{\frac{\ln(1/\delta)}{\rho_1}} - 1,\frac{1}{2\rho_1}\Big)$.

    \item If $n=0$ or $\widehat{n}\leq 0$, output $\bot$ and abort.

    \item Otherwise, output $\operatorname{Avg}(\mathcal{D}) + \mathcal{N}\!\left(0,\sigma^2 I_{d\times d}\right)$, where $\sigma = \frac{2r}{\widehat{n}} \cdot \frac{1}{\sqrt{2\rho_2}}$.
\end{enumerate}
\end{algorithm}

We now define the appropriate notion of friendliness for our task. For $z \in\mathbb{R}^{d \times k}$ and a radius $r\geq 0$, define

\[
    F^{(2)}_r(z,z')
    :=
    \mathbf{1}
    \left\{
        \|z-z'\|_F\leq r
    \right\}.
\]

It is important to note that the Frobenius norm on a matrix corresponds to the $\ell_2$ norm of the vectorization of the matrix. Hence, for algorithms with matrix inputs and guarantees stated in the Frobenius norms, it simply helps to think about the operations acting on vectorized versions of the input matrices.

\begin{lemma}[{\citet[Claim B.4]{tsfadia2022friendlycore}}]
\label{lem:friendlyavgDP}
    \textsc{FriendlyAvg} is $(\rho, \delta)$ $F^{(2)}_r$-friendly zCDP.
\end{lemma}

Note that \textsc{FriendlyAvg} assumes that the value of $n$ is not public and hence computes a private estimate \(\widehat n\) that underestimates \(n\) with high probability while ensuring privacy at the same time. Then, for any two neighboring datasets $D$ and $D'$, their union being friendly implies that for any pair $z, z' \in D \cup D'$, there exists a point $w$ such that $\Vert z-w\Vert_F \leq r$ and  $\Vert z'-w\Vert_F \leq r$. This implies that $\Vert z- z'\Vert_F \leq 2r$, so the mean is $2r$-sensitive and noise is added calibrated to this sensitivity. A complete proof of this result can be found in \citet[Claim B.4]{tsfadia2022friendlycore}.

\noindent

We now present a private averaging procedure for $\mathbb{R}^{d \times k}$ matrices with utility measured in the $\norm{\cdot}_{\infty}$ norm. The idea is simply to filter the data using the $F_r^{(2)}$ filter and then privatize the output using a mechanism that yields a private estimate for \(F_r^{(2)}\)-complete database inputs.

\begin{algorithm}[h]
\caption{\textsc{JointAvg}}
\label{alg:joint-avg}

\noindent
\textbf{\textit{Input:}}
A database $D \in (\mathbb{R}^{d \times k})^*$, privacy parameters
\(\varepsilon,\delta>0\), and \(r\geq 0\).

\medskip
\noindent
\textbf{\textit{Operation:}}
\begin{enumerate}
    \item Let $v$ be the output of \textsc{BasicFilter} on \(D\) with predicate \(F^{(2)}_r\),
    and let
    \(D' = \{D_i: v_i = 1, i \in [n]\} \).

    \item Run \textsc{FriendlyAvg} on
    $D'$ with zCDP-privacy parameters $\left(\frac{\varepsilon^2}{256\log(2e^{2}/\delta)},\frac{\delta}{2e^2}\right)$ and radius \(r\) and denote its output by $\tilde{\mu}$.

    \item If $\tilde{\mu} = \bot$, set $\tilde \mu = \frac{1}{k}\mathbf{1}\mathbf{1}^\top$.

    \item Return $\tilde{\mu}$.
\end{enumerate}
\end{algorithm}

\begin{lemma}[Privacy of \textsc{JointAvg}]
\label{lem:one-core-private-averaging-privacy}
    Let $0<\varepsilon\leq 1$, $0<\delta<1$. \textsc{JointAvg} is $\left(\varepsilon, \delta\right)$-DP for any $r \geq 0$.
\end{lemma}
\begin{proof}
To prove the privacy guarantee, it suffices to establish privacy for Steps 2 and 3 of Algorithm~\ref{alg:joint-avg}. By Lemma~\ref{lem:friendlyavgDP},
\textsc{FriendlyAvg} is $F^{(2)}_{r}$-friendly $\left(\frac{\varepsilon^2}{256\log(4e^{2}/\delta)},\frac{\delta}{4e^2}\right)$-zCDP.
Applying the conversion from approximate zCDP to approximate DP implies that the mechanism is $F^{(2)}_r$-friendly $\left(\frac{\varepsilon}{4}, \frac{\delta}{2e^2}\right)$-DP.
Because $\varepsilon \leq 1$, Lemma~\ref{lem:filterparadigm} implies that $\mathsf{JointAvg}$ is  $\left(\varepsilon, \delta\right)$ DP.

\end{proof}

The privacy guarantees follow primarily from the implied Frobenius filter. The following lemma gives the utility guarantee with respect to the matrix $\norm{\cdot}_\infty$ norm given that its input is $F_r^{(2)}$-complete.

\begin{lemma}
\label{lem:one-core-private-averaging-utility}
Let $0 < \varepsilon, \delta,\beta < 1$, and $r\geq 0$. Suppose that $D=(z_1,\ldots,z_N)$ is $F^{(2)}_r$-complete and $N \geq \frac{C}{\varepsilon} \sqrt{ \log\left(\frac{1}{\delta}\right) \log\left(\frac{1}{\beta\delta}\right)}$. Then, with probability at least $1-\beta$, the output $\Tilde{\mu}$ of Algorithm~\ref{alg:joint-avg}, satisfies
\[
    \max_{j\in[d]}
    \left\|
        \widetilde\mu_{(j,\cdot)}
        -
        \frac{1}{N}\sum_{i=1}^N {z_{i}}_{(j,\cdot)}
    \right\|_1
    \leq
    C
    \frac{
        r\sqrt{\log(C/\delta)}
    }{
        \varepsilon N
    }
    \left(
        k+\sqrt{k\log\left(\frac{Cd}{\beta}\right)}
    \right),
\]
where $C>0$ is a universal constant.
\end{lemma}

\begin{proof}
Given that $D$ is $F^{(2)}_r$-complete, \textsc{BasicFilter} retains every element with probability one. Consequently, \textsc{FriendlyAvg} is applied to the full vectorized database. Let $\rho = \frac{\varepsilon^2}{256\log(2e^2/\delta)}$. The private lower estimate of the core size used by \textsc{FriendlyAvg} is
\[
    \widehat n
    =
    N
    -
    \sqrt{
        \frac{
            \log(2e^2/\delta)
        }{
            \rho_1
        }
    }
    -1
    +W,
    \qquad
    W
    \sim
    \mathcal{N}
    \left(
        0,
        \frac{1}{2\rho_1}
    \right),
\]
where $\rho_1:=0.1\left(1-\frac{\delta}{2e^2}\right)\rho$ and
$\rho_2:=0.9\rho$. A Gaussian tail bound gives $W \geq -\sqrt{\frac{\log(2/\beta)}{\rho_1}}$ with probability at least $1-\beta/2$. Hence
$\widehat n\geq N/2$ whenever
\[
    N
    \geq
    2\left(
        \sqrt{
            \frac{
                \log(2e^2/\delta)
            }{
                \rho_1
            }
        }
        +
        \sqrt{
            \frac{
                \log(2/\beta)
            }{
                \rho_1
            }
        }
        +1
    \right).
\]
Because $\rho_1\geq c\rho$ and
$\frac{1}{\sqrt\rho}=\frac{16}{\varepsilon}\sqrt{\log\left(\frac{2e^2}{\delta}\right)}$
, $N \geq \frac{C}{\varepsilon} \sqrt{ \log\left(\frac{C}{\delta}\right) \log\left(\frac{C}{\beta\delta}\right)}$ implies this condition after increasing the
universal constant $C$. Conditioned on $\widehat n\geq N/2$, \textsc{FriendlyAvg} releases the
empirical average plus an isotropic Gaussian vector $G$ in $\mathbb{R}^{dk}$. Writing this Gaussian vector blockwise as
$G=(G_1,\ldots,G_d)$, each block $G_j = \Tilde{\mu_j} - \frac{1}{B}\sum_{i=1}^Nz_{i,j}$ satisfies
$G_j\sim\mathcal{N}(0,\sigma^2I_k)$, where
\[
    \sigma =
    \frac{2r}{\widehat n}
    \sqrt{\frac{1}{2\rho_2}} \leq C\frac{r\sqrt{\log(C/\delta)}}{\varepsilon N}.
\]

For each $j$, we have
$\mathbb{E}\|G_j\|_1=k\sigma\sqrt{2/\pi}\leq k\sigma$.
Moreover, the map $x\mapsto\|x\|_1$ is $\sqrt k$-Lipschitz with
respect to the Euclidean norm. The Gaussian isoperimetric inequality for Lipschitz functions therefore
gives
\[
    \mbb{P}\left(
        \|G_j\|_1
        >
        \sigma
        \left(
            k
            +
            \sqrt{
                2k\log\left(\frac{2d}{\beta}\right)
            }
        \right)
    \right)
    \leq
    \frac{\beta}{2d}.
\]

A union bound over $j\in[d]$, together with the high probability event that $\hat{n} \geq N/2$, proves the claimed utility bound with
probability at least $1-\beta$.
\end{proof}

\begin{algorithm}[h]
\caption{User-level private estimation of product distributions with approximate DP.}
\label{alg:private-product-estimation}

\noindent
\textbf{\textit{Input:}}
A database
\(\mathcal{D}=(X^{(i)}_t)_{i\in[N],\,t\in[m]}\), where
\(X^{(i)}_t\in[k]^d\), privacy parameters \(\varepsilon,\delta>0\), and
failure probability \(f\in(0,1)\).

\medskip
\noindent
\textbf{\textit{Operation:}}
\begin{enumerate}
    \item For every user \(i\in[N]\) and product coordinate \(j\in[d]\),
    compute the empirical marginal
    \[
        \overline{P}_{i,(j,\cdot)}
        :=
        \frac{1}{m}\sum_{t=1}^m \mathbf{e}_{X^{(i)}_{t,j}}
        \in\Delta(k),
    \]
    where \(X^{(i)}_{t,j}\) is the \(j\)-th coordinate of \(X^{(i)}_t\) and
    \(e_a\in\mathbb{R}^k\) is the \(a\)-th standard basis vector.

    \item Set $r_m := \min\left\{\sqrt{2},\frac{2}{\sqrt{m}}{\left(1+\sqrt{\log\left(\frac{3Nd}{f}\right)}\right)}\right\}$. Here, \(r_m\) is set using the public database size
    \(N\).

    \item Run $\text{\textsc{JointAvg}}(\overline{P}_1,\ldots,\overline{P}_N)$ with radius $r_m\sqrt{d}$ and privacy parameters $(\varepsilon, \delta)$ and let the output be $\widetilde{P}$.

    \item For every \(j\in[d]\), compute
    \[
        \widehat P_{(j,\cdot)}
        \in
        \operatorname*{argmin}_{q\in\Delta(k)}
        \|q-\widetilde P_{(j,\cdot)}\|_1
    \] and output $\hat{P}$.
\end{enumerate}
\end{algorithm}

Intuitively, Algorithm~\ref{alg:private-product-estimation} applies the \textsc{JointAvg} mechanism to estimate the distribution matrix of a $k$-ary, $d$-dimensional distribution. The privacy guarantee follows directly from the privacy of \textsc{JointAvg}. Hence, the main remaining step is to prove that the constructed empirical distribution matrices for each user are $F^{(2)}_{r_m\sqrt{d}}$ complete with high probability.

\begin{theorem}[Complete Privacy Guarantee in Theorem~\ref{thm:user-level-product-marginals-main}]
\label{thm:user-level-product-marginals-priv}
    Algorithm~\ref{alg:private-product-estimation} is $\left(\varepsilon, \delta\right)$-DP.
\end{theorem}
\begin{proof}
    Replacing all $m$ samples of one user changes exactly one tuple $Y_i$. Lemma~\ref{lem:one-core-private-averaging-privacy} therefore gives user-level $\left(\varepsilon, \delta\right)$-differential privacy, and the simplex projection is post-processing.
\end{proof}

\begin{theorem}[Complete Utility Guarantee in Theorem~\ref{thm:user-level-product-marginals-main}]
\label{thm:user-level-product-marginals-util}
Fix \(0<\alpha,f<1\) and \(0<\eps\leq 1\). Uniformly over
\(P\in \mathcal{P}_{d,k}\), the output of Algorithm~\ref{alg:private-product-estimation}, $\hat{P}$, satisfies $\mbb{P}\left(\rho(\widehat P,P)\leq\alpha\right)\geq 1-f$ for
\[
    N
    \geq
    C\left[
        \frac{
            k+\log(d/f)
        }{
            m\alpha^2
        } +
        \frac{k\sqrt{d}}{\varepsilon\alpha\sqrt m}\log^{3/2}\left(\frac{dkm}{\alpha\varepsilon\delta f}\right) +
        \frac{1}{\varepsilon}
        \sqrt{
            \log\left(\frac{1}{\delta}\right)
            \log\left(\frac{1}{\delta f}\right)
        }
    \right].
\]
\end{theorem}

\begin{proof}
Let
\[
    \overline P_{(j,\cdot)}
    :=
    \frac{1}{N}
    \sum_{i=1}^N \overline{P}_{i,(j,\cdot)}
    =
    \frac{1}{Nm}
    \sum_{i=1}^N
    \sum_{t=1}^m \mathbf{e}_{X^{(i)}_{t,j}}.
\]
Recall that $\mathbf{e}_{X^{(i)}_{t,j}}$ corresponds to the standard basis vector in $\mathbb{R}^k$ indexed by $X^{(i)}_{t,j}$, which implies that $\overline{P}_{(j,\cdot)} \in\Delta(k)$. Since $P_{(j,\cdot)} \in\Delta(k)$ and $\widehat P_{j,\cdot}$ is an
$\ell_1$-nearest point in $\Delta(k)$ to $\widetilde P_{(j,\cdot)}$, we have $\|\widehat P_{(j,\cdot)}-\widetilde P_{(j,\cdot)}\|_1 \leq \|P_{(j,\cdot)}-\widetilde P_{(j,\cdot)}\|_1$. Consequently,
\[
\begin{split}
    \|\widehat P_{(j,\cdot)}-P_{(j,\cdot)}\|_1
    &\leq
    \|\widehat P_{(j,\cdot)}-\widetilde P_{(j,\cdot)}\|_1
    +
    \|\widetilde P_{(j,\cdot)} - P_{(j,\cdot)}\|_1 \\
    &\leq
    2\|\widetilde P_{(j,\cdot)} - P_{(j,\cdot)}\|_1 \\
    &\leq
    2\|\overline P_{(j,\cdot)} - P_{(j,\cdot)}\|_1
    +
    2\|\widetilde P_{(j,\cdot)} - \overline P_{(j,\cdot)}\|_1.
\end{split}
\]

Taking the maximum on both sides gives,
\[
\max_{j \in [d]} \|\widehat P_{(j,\cdot)}-P_{(j,\cdot)}\|_1 \leq 2 \underbrace{\max_{j \in [d]} \|\widetilde P_{(j,\cdot)} - \overline P_{(j,\cdot)}\|_1}_{T_1} + 2 \underbrace{\max_{j \in [d]} \|\overline P_{(j,\cdot)} - P_{(j,\cdot)}\|_1}_{T_2}.
\]

\paragraph{Upper bounding $T_1$ w.h.p.} To bound this term, we show that the dataset is $F^{(2)}_{r_m\sqrt{d}}$-complete with high probability and appeal to the utility guarantee in Lemma~\ref{lem:one-core-private-averaging-utility}.
Towards this, fix $i\in[N]$ and $j\in[d]$. Since $\overline{P}_{i,(j,\cdot)}$ is the empirical
distribution of $m$ independent samples from $P_{(j,\cdot)}$,
\[
\begin{split}
    \mathbb{E}\|\overline{P}_{i,(j,\cdot)}-P_{(j,\cdot)}\|_2^2
    &=
    \sum_{a=1}^k
    \operatorname{Var}(\overline{P}_{i,(j,a)}) =\frac{1}{m} \sum_{a=1}^k P_{(j,a)}(1-P_{(j,a)}) = \frac{1-\|P_{(j,\cdot)}\|_2^2}{m} \leq \frac{1}{m}.
\end{split}
\]
Hence $\mathbb{E}\|\overline{P}_{i,(j,\cdot)}-P_{(j,\cdot)}\|_2\leq \frac{1}{\sqrt m}$. Changing one of the $m$ samples changes $\overline{P}_{i,(j,\cdot)}$ in Euclidean norm by at most $\frac{\sqrt 2}{m}$. Lemma~\ref{lem:norm-concentration-bounded-differences} therefore yields, for every $t>0$,
\[
    \mbb{P}\left(
        \|\overline{P}_{i,(j,\cdot)}-P_{(j,\cdot)}\|_2
        >
        \frac{1}{\sqrt m}+t
    \right)
    \leq
    e^{-mt^2}.
\]
Taking $t=\frac{1}{m}\sqrt{\log\left(\frac{3Nd}{f}\right)}$ and applying a union bound over all $i\in[N]$ and $j\in[d]$ gives, with probability at least $1-\frac{f}{3}$,
\[
    \|\overline{P}_{i,(j,\cdot)}-P_{(j,\cdot)}\|_2
    \leq
    \frac{1}{\sqrt m}\left(1+\sqrt{\log\left(\frac{3Nd}{f}\right)}\right),
\]
for every $i\in[N]$ and $j\in[d]$. On this event, the triangle inequality gives
\[
\|\overline{P}_{i,(j,\cdot)}-\overline{P}_{i',(j,\cdot)}\|_2 \leq \frac{ 2}{\sqrt m}\left(1+\sqrt{\log\left(\frac{3Nd}{f}\right)}\right),
\]
for every $i,i'\in[N]$ and $j\in[d]$. In addition, any two
vectors in $\Delta(k)$ are at Euclidean distance at most $\sqrt 2$.
Recalling that $r_m := \min\left\{\sqrt{2},\frac{2}{\sqrt{m}}{\left(1+\sqrt{\log\left(\frac{3Nd}{f}\right)}\right)}\right\}$, it follows that the database $(Y_1,\ldots,Y_N)$ is
$F^{(2)}_{r_m\sqrt{d}}$-complete with probability at least $1 - f/3$.

Condition on the event that $(Y_1,\ldots,Y_N)$ is
$F^{(2)}_{r_m\sqrt{d}}$-complete. Taking $N \geq \frac{C}{\varepsilon}\sqrt{\log\left(\frac{C}{\delta}\right)\log\left(\frac{C}{\delta f}\right)}$ in the theorem allows
us to apply Lemma~\ref{lem:one-core-private-averaging-utility} with
$\beta=f/3$. Therefore, with conditional probability at least
$1-f/3$, the mechanism does not return $\frac{1}{k} \mathbf{1}\mathbf{1}^\top$ from $\bot$ and
\[
\begin{split}
    \max_{j\in[d]}
    \|\widetilde P_{(j,\cdot)}-\overline P_{(j,\cdot)}\|_1
    &\leq
    C
    \frac{
        r_m\sqrt{d\log(C/\delta)}
    }{
        \varepsilon N
    }
    \left(
        k+\sqrt{k\log(Cd/f)}
    \right)                                              \\
    &\leq
    C
    \frac{
        \left(
            k+\sqrt{k\log(Cd/f)}
        \right)
        \sqrt{
            d
            \log(C/\delta)
            \log(CNd/f)
        }
    }{
        \varepsilon N\sqrt m
    }.
\end{split}
\]

\paragraph{Upper bounding $T_2$ w.h.p.}
This amounts to bounding an empirical-to-population gap using concentration for the mean of bounded random vectors. $\overline{P}_{(j,\cdot)}$ is the empirical distribution of $Nm$ independent samples
from $P_{(j,\cdot)}$. As above,
$\mathbb{E}\|\overline P_{(j,\cdot)}-P_{(j,\cdot)}\|_2^2\leq \frac{1}{Nm}$, and hence $\mathbb{E} \|\overline P_{(j,\cdot)}-P_{(j,\cdot)}\|_1 \leq \sqrt{\frac{k}{Nm}}$. Changing one of the $Nm$ observations changes
$\|\overline P_{(j,\cdot)}-P_{(j,\cdot)}\|_1$ by at most $\frac{2}{Nm}$. Lemma~\ref{lem:norm-concentration-bounded-differences} gives
\[
    \mbb{P}\left(
        \|\overline P_{(j,\cdot)}-P_{(j,\cdot)}\|_1
        >
        \sqrt{\frac{k}{Nm}}+t
    \right)
    \leq
    \exp\left(
        -\frac{Nm\,t^2}{2}
    \right).
\]
Taking
$t=\sqrt{\frac{2\log(3d/f)}{Nm}}$ and applying a union bound over $j\in[d]$, we
obtain, with probability at least $1-f/3$,
\[
    \max_{j\in[d]}
    \|\overline P_{(j,\cdot)}-P_{(j,\cdot)}\|_1
    \leq
    C
    \sqrt{
        \frac{
            k+\log(Cd/f)
        }{
            Nm
        }
    }.
\]

\paragraph{Combining the bounds}
The $F^{(2)}_{r_m\sqrt{d}}$ completeness event, the global empirical-concentration event, and the private-release event each fail with probability at most
$f/3$. Thus, taking $N$ large enough to control these three events leads to an $\mathcal{O}(\alpha)$ error. The sampling term is at most a constant multiple of $\alpha$ once
\[
N \geq C \frac{k+\log(Cd/f)}{m\alpha^2}.
\]
For the private-release term,
\[
    N
    \geq
    C
    \frac{
        \left(
            k+\sqrt{k\log(Cd/f)}
        \right)
        \sqrt{
            d
            \log(C/\delta)
            \log(CNd/f)
        }
    }{
        \varepsilon\alpha\sqrt m
    }.
\]

Since $N$ is polynomial in
$d,k,m,1/\alpha,1/\varepsilon,1/\delta$, and $1/f$, we have $\log\left(\frac{CNd}{f}\right) \leq C'\log\left(\frac{dkm}{\alpha\varepsilon\delta f}\right)$ after increasing the universal constant. This proves the explicit user-complexity bound.
\end{proof}
\subsection{Complete Statement and Proof of Theorem~\ref{thm:high-prob-approx-final-one-infty-norm-main}}
\label{ssec:approx-dp-lower-bound-formal}
\begin{theorem}[Complete version of Theorem~\ref{thm:high-prob-approx-final-one-infty-norm-main}]
\label{thm:high-prob-approx-final-one-infty-norm}
There are universal constants $C^\rho_\ell, C_\delta >0$ such that the following
holds. Let $0<\alpha\le \frac{1}{48\sqrt{2}}$, $0<\eps\le 1$, $k \geq 40$ and $\delta\le \min\left\{\frac{C_\delta\eps}{\sqrt{dk}}, \varepsilon^2\right\}$. Let $\mathcal A$ be an $(\eps,\delta)$-user-level DP estimator with $N$ users, each contributing $m$ samples. Suppose that, uniformly over
\(P \in \mathcal{P}_{d,k}\), $\mathcal{A}$ outputs $\widehat{P} \in \mathcal{P}_{d,k}$ such that
\begin{equation}
\label{eq:highprobcondition}
    \PP\left(
        \rho(\widehat P,P)\le \alpha
    \right)
    \ge \frac23,
\end{equation}
then
\begin{equation}
\label{eq:highprobsamplb}
    N
    \gtrsim \frac1\eps\min \left\{kd, \log\left(\frac{1}{\delta}\right)\right\} + \frac{k\sqrt{d}}
    {\alpha\eps\sqrt m\log^{C^\rho_\ell}((\alpha\eps)^{-1}edk)} + \frac{k}{m\alpha^2}.
\end{equation}
\end{theorem}
We prove the three terms in the user-level lower bound in three separate parts.
\begin{enumerate}
    \item Lemma~\ref{lem:approx-dp-few-user-packing} proves a lower bound on the number of users by showing that for any algorithm achieving the desired utility needs $N \gtrsim  \frac1\eps\min \left\{kd, \log\left(\frac{1}{\delta}\right)\right\}$ users. This bound is proved via an adaptation of the packing argument
    in \citet{esfandiari2022tight} to our setting.
    \item Theorem~\ref{thm:main-expected-approx} derives the lower bound on the number of users required to bound, in expectation, the error defined by $\xi(\widehat P,P) := \sqrt{\sum_{j=1}^d \|\widehat P_{(j,\cdot)}-P_{(j,\cdot)}\|_1^2}$. Thus $\xi$ is an $\ell_2$ norm over product coordinates, with an $\ell_1$ error within each coordinate. Theorem~\ref{thm:main-expected-approx} proves $N
    \gtrsim \frac{dk}{\alpha\eps\sqrt m \log^2\left((\alpha\eps)^{-1}edk\right)}$. However, the utility guarantee upper-bounds a different norm of the difference between the estimated and actual matrices. Thus, the lower bound is adjusted for the utility guarantee stated in Theorem~\ref{thm:high-prob-approx-final-one-infty-norm} via a change of norm to align with the norm in the stated utility guarantee in Theorem~\ref{thm:high-prob-approx-final-one-infty-norm}. Additionally, since the user lower bound applies to an expected-utility guarantee, a median-of-means trick is used to show the lower bound for a stronger constant-probability utility condition yielding the second term of the lower bound.
    \item The final bound simply follows from the sample-complexity lower bound for achieving the desired utility for this task without any privacy considerations. This is made rigorous in the proof of Theorem~\ref{thm:high-prob-approx-final-one-infty-norm}.
\end{enumerate}

We formalize the aforementioned sketch of the proof below. Before proving the above theorem, we require two results which we prove later.

\begin{lemma}[User lower bound via packing]
\label{lem:approx-dp-few-user-packing}
Let $\varepsilon \leq 1$, $\delta \leq \varepsilon^2$, and $k \geq 40$. Suppose that $\mathcal A$ is an $(\eps,\delta)$-user-level
differentially private estimator using $N$ users, each contributing
$m$ samples, which outputs $\hat{P} \in \mathcal{P}_{d,k}$ and that for each $P \in \mathcal{P}_{d,k}$, satisfies
\[
    \mathbb{P}
    \left(
        \rho(\widehat P,P)\le\alpha
    \right)
    \ge\frac23.
\]
Then
\[
    N
    \gtrsim
    \frac1\eps\min \left\{kd, \log\left(\frac{1}{\delta}\right)\right\}.
\]
\end{lemma}

\begin{theorem}[Sample Lower Bound under approximate DP]\label{thm:main-expected-approx}
Let $0<\eps\le 1$, $0<\alpha\le \frac{\sqrt{d}}{48\sqrt{2}}$, and $\delta\le \frac{C_\delta\eps}{\sqrt{dk}}$.
Define
\[
\xi(\widehat P,P) := \sqrt{\sum_{j=1}^d \|\widehat P_{(j,\cdot)}-P_{(j,\cdot)}\|_1^2}.
\]
There exists a universal constant $C_\delta >0$ such that the following holds.   Suppose $\cA$ is an $(\eps,\delta)$-user-level differentially private estimator which outputs $\hat{P} \in \mathcal{P}_{d,k}$ satisfying
\[\sup_{P\in\cP_{d,k}} \EE_P\left[\xi(\widehat P,P)^2\right] \le \alpha^2\]
then
\begin{equation}\label{eq:main-lower-bound-approx}
    N
    \gtrsim \frac{dk}{\alpha\eps\sqrt m \log^2\left((\alpha\eps)^{-1}edk\right)}.
\end{equation}
\end{theorem}

We first show how Lemma~\ref{lem:approx-dp-few-user-packing} and Theorem~\ref{thm:main-expected-approx} imply Theorem~\ref{thm:high-prob-approx-final-one-infty-norm}.

\begin{proofof}{Theorem~\ref{thm:high-prob-approx-final-one-infty-norm}} For any $(\varepsilon, \delta)$-user-level DP estimator with $N$ users, we simply have to exhibit three instances in the same class, one yielding each term. The lower bound of $N
    \gtrsim
\frac1\eps\min \left\{kd, \log\left(\frac{1}{\delta}\right)\right\}$ directly follows by Lemma~\ref{lem:approx-dp-few-user-packing} from the accuracy guarantee and the conditions on $\varepsilon$ and $\delta$. We now prove the lower bounds due to private term and sampling term separately.

\paragraph{Sampling Term}
We use the standard fact that learning an arbitrary distribution on $[k]$ to total-variation accuracy $\eta$ with success probability
at least $2/3$ requires $\Omega(k/\eta^2)$ samples; see, for example, Theorem~1 of~\cite{canonne2020short}. Let $u_k$ denote the uniform distribution on $[k]$ and consider the subclass
\[
    \mathcal P'
    :=
    \left\{
        p\otimes u_k^{\otimes(d-1)}
        :
        p\in\Delta(k)
    \right\}
    \subseteq\mathcal P_{d,k}.
\]
Suppose that $Y_{i,t}\sim p$ independently for
$i\in[N]$ and $t\in[m]$. Independently generate
\[
    U_{i,t,j}\sim u_k,
    \qquad
    i\in[N],\ t\in[m],\ j=2,\ldots,d,
\]
and form $X_{i,t} := \left(Y_{i,t},U_{i,t,2},\ldots,U_{i,t,d}\right)$.
Then the resulting dataset has exactly the law of $N$ users, each
with $m$ independent samples from $P_p:=p\otimes u_k^{\otimes(d-1)}$.
Run the assumed estimator on this dataset and output its first
estimated marginal $\widehat p_1$. Since
\[
    2d_{\mathrm{TV}}(\widehat p_1,p) \leq \|\widehat p_1-p\|_1
    \le
    \rho(\widehat P,P_p),
\]
the assumed guarantee implies
\[
    \inf_{p\in\Delta(k)}
    \mathbb{P}_p\left(
        d_{\mathrm{TV}}(\widehat p_1,p)
        \le\frac{\alpha}{2}
    \right)
    \ge\frac23.
\]
The constant-probability sampling lower bound for learning
an arbitrary distribution on $[k]$ in total variation
implies
\[
    Nm
    \gtrsim
    \frac{k}{\alpha^2}.
\]
This proves the sampling lower bound.

\paragraph{Privacy Term} To establish the privacy term, we first apply a norm comparison and then a median-based amplification argument to convert the expected-error lower bound into a constant-probability lower bound, losing only logarithmic factors. The amplification preserves user-level privacy by operating on disjoint blocks of users.

Fix $0<\alpha\le \frac{1}{48\sqrt{2}}$. We prove this by contradiction. Assume that there exists an estimator $\mc A$ which satisfies Equation~\eqref{eq:highprobcondition} but does not satisfy not~\eqref{eq:highprobsamplb}. Since, $\xi(\hat{P}, P) \leq \sqrt{d}\rho(\hat{P}, P)$, this implies that $\mc A$ satisfies
\begin{equation}
\label{eq:hp-assumption-approx}
    \inf_{P\in\cP_{d,k}}
    \PP\left(
        \xi(\widehat P,P)\le \alpha\sqrt{d}
    \right)
    \ge \frac23.
\end{equation}

Let $R=\left\lceil C_1\log\left(e/\alpha\right)\right\rceil$, where $C_1>0$ is a universal constant to be determined.  Split the $RN$ users into $R$ disjoint blocks of size $N$.  Run $\cA$ independently on each block, obtaining candidate estimates $\widehat P^{(1)},\ldots,\widehat P^{(R)}$. Let $E_r$ be the event
\[
    E_r=\left\{\xi(\widehat P^{(r)},P)\le \frac{\alpha\sqrt{d}}{3\sqrt{2}}\right\}.
\]
The events $E_1,\ldots,E_R$ are independent and each has probability at least $2/3$ by \eqref{eq:hp-assumption-approx}.  Hence, choosing \(C_1>0\) sufficiently large, a Chernoff bound gives
\begin{equation}\label{eq:chernoff-good-blocks-approx}
    \PP\left(
    \sum_{r=1}^R \mathbf 1\{E_r\}<\frac{3R}{5}
    \right)
    \le
    \exp(-cR)
    \le
    \frac{\alpha^2}{8} .
\end{equation}

We now define the aggregation rule.  The loss $\xi$ is a metric: each coordinate error $\|p_j-q_j\|_1$ is a metric, and applying the $\ell_2$ norm across coordinates preserves the triangle inequality by Minkowski's inequality.  For a candidate $Q_r:=\widehat P^{(r)}$, define
\[
    \eta_r
    :=
    \inf\left\{t\ge0:
    \left|\{s\in[R]:\xi(Q_r,Q_s)\le t\}\right|
    \ge
    \frac R2
    \right\} .
\]
Choose $r_\star\in\arg\min_{r\in[R]}\eta_r$ and output $\widehat P^{\mathrm{med}}=Q_{r_\star}$.

We define a candidate $r$ to be good if $E_r$ holds. On the event that at least $\frac{3R}{5}$ candidates are good, $\widehat P^{\mathrm{med}}$ is close to $P$.  Indeed, if $g$ is good, then every other good candidate lies within distance at most $\frac{\sqrt{2}}{3}\alpha\sqrt{d}$ of $Q_g$ by the triangle inequality.  Since there are more than $\frac{R}{2}$ good candidates, $\eta_g\le \frac{\sqrt{2}}{3}\alpha\sqrt{d}$.  By minimality of $r_\star$, $\eta_{r_\star}\le \frac{\sqrt{2}}{3}\alpha\sqrt{d}$. Thus at least $\frac{R}{2}$ candidates lie within distance $\frac{\sqrt{2}}{3}\alpha$ of $Q_{r_\star}$. This set must intersect the set of good candidates, because the good candidates have cardinality larger than $\frac{R}{2}$. Pick a good candidate $Q_s$ in the intersection.  Then
\[
    \xi(Q_{r_\star},P)
    \le
    \xi(Q_{r_\star},Q_s)+\xi(Q_s,P)
    \le
    \frac{\alpha\sqrt{d}}{\sqrt{2}}.
\]
Therefore, with probability at least $1-\frac{\alpha^2}{8}$, $\xi(\widehat P^{\mathrm{med}},P) \le \frac{\alpha\sqrt{d}}{\sqrt{2}}$. Since the loss is uniformly bounded by $\xi^2\le 4d$, we have
\begin{equation}\label{eq:amplified-risk-approx}
    \EE_P\xi(\widehat P^{\mathrm{med}},P)^2
    \le
    \frac{\alpha^2d}{2}
    +
    4d\cdot\frac{\alpha^2}{8}\leq \alpha^2 d.
\end{equation}

The amplified estimator is still $(\eps,\delta)$-user-level DP.  This follows from parallel composition: the $R$ invocations of $\cA$ are run on disjoint sets of users, so changing one user can affect only one invocation.  The final metric median is a post-processing step. Since $\alpha \sqrt{d} \leq \frac{\sqrt{d}}{48\sqrt{2}}$, apply Theorem~\ref{thm:main-expected-approx} to $\cA^{\mathrm{med}}$, which uses $RN$ users and $m$ samples per user. From \eqref{eq:amplified-risk-approx},
\[
    RN\sqrt m
    \gtrsim \frac{k\sqrt{d}}{\alpha\eps\log^2(edk/(\alpha\eps))}.
\]
Since $R=\Theta(\log(ed/\alpha))$, this implies
\begin{equation}
\label{eq:hp-private-alpha-approx}
    N\sqrt m
    \gtrsim \frac{k\sqrt{d}}{\alpha\eps\log^{C^\rho_\ell}(edk/(\alpha\eps))}.
\end{equation}
\end{proofof}

\subsubsection{Proof of Theorem~\ref{thm:main-expected-approx}}

We now prove Theorem~\ref{thm:main-expected-approx}. In a nutshell, our proof can be divided into three broad parts.
\begin{itemize}
    \item We first define an intermediate problem called the Item Level Private Poisson Estimation Problem. We then define a local paired perturbation family around the uniform distribution whose elements serve as parameters for the Poisson Estimation Problem. Pairing is used to keep every $k$-ary marginal exactly normalized.

    \item Motivated by techniques from \citet{esfandiari2022tight}, we then prove a lower bound on the sample complexity for privately  estimating the parameters of the Poisson Estimation Problem via fingerprinting arguments. The fingerprinting identity essentially states that an accurate estimator must have a sufficient correlation with the users' likelihood scores. On the other hand, approximate item-level privacy upper bounds this correlation. The well-behaved nature of the Poisson distribution makes this computation explicit and easy to analyze.
    \item Finally, we establish a privacy-preserving reduction from user-level estimation in our ordinary product-sample model to the Poisson Estimation Problem by expanding Poisson counts into categorical samples.
\end{itemize}

We begin by defining the Poisson Estimation Problem.

\begin{definition}[Poisson Estimation Problem]
\label{def:poissonestmproblem}
    Let $q \in [0, 1]$ and $r \leq \frac{q}{2}$. For each $i\in[N]$ and $\ell \in [D]$ , draw independent counts $X_{i,\ell}^+\sim\Pois\left(m\left(q+\theta_\ell\right)\right)$ and  $X_{i,\ell}^-\sim\Pois\left(m\left(q - \theta_\ell\right)\right)$, where $|\theta_\ell| \leq r$. The $i^{th}$ data-point consists of the full block $X_i=\{X_{i,\ell}^+,X_{i,\ell}^-:\ell\in[D]\}$. Neighboring datasets differ in one such block.
    An algorithm $\cB$ solves the Poisson Estimation Problem with $\alpha$ accuracy if for every $\theta \in [-r, r]^D$, it outputs an estimate $\hat{\theta} \in [-r, r]^D$ that satisfies $\mbb E \norm{\hat{\theta} - \theta}_2^2 \leq \alpha$.
\end{definition}

We use fingerprinting techniques developed in \citet{esfandiari2022tight} and \citet{pmlr-v99-kamath19a} to prove this lower bound. It is important to note that we cannot directly use the fingerprinting results from \citet{esfandiari2022tight} since we are solving a different Poisson estimation problem. Thus, for our case, we require new computations to accommodate the new structure of the intermediate problem. However, the computations use insights proposed in \citet{esfandiari2022tight} and \citet{pmlr-v99-kamath19a}.

Note that a direct extension of the reduction in \citet{esfandiari2022tight} to multiple dimensions encounters a normalization issue due to the nature of their reduction. Their hard instance consists of independent Poisson counts with means \(m\lambda_1,\ldots,m\lambda_k\), where the vector \(\lambda\) need not sum to one. Expanding these counts into categorical observations therefore produces samples from the normalized distribution \(\lambda/s\), where \(s:=\sum_{a=1}^k\lambda_a\). To recover the original Poisson means, the reduction must estimate both
\(\lambda/s\) and the total intensity \(s\), with the latter requiring an additional private one-dimensional estimation step. For a single \(k\)-ary distribution, the resulting loss in privacy and accuracy can be absorbed into constant factors. In our setting, this normalization must be estimated separately for each of the \(d\) rows.
The resulting \(d\) normalization errors accumulate as their
\(\ell_2\) norm, introducing an additional factor of \(\sqrt d\) in the
mixed loss; moreover, since the same users contribute to every row,
these estimates do not benefit from parallel composition.

We avoid this difficulty by considering a paired Poisson family. Within
each row, the categories are grouped into pairs and their means are
perturbed in opposite directions, for example,
\[
    \lambda_{j,2a-1}
    =
    \frac{1}{k}+\theta_{j,a},
    \qquad
    \lambda_{j,2a}
    =
    \frac{1}{k}-\theta_{j,a}.
\]
Thus, the total intensity of every pair, and consequently of every row,
is fixed independently of the unknown parameters
\(\{\theta_{j,a}\}\). The realized Poisson counts remain random, but
their total mean is known, so no private row-wise normalization estimate
is needed. The hard family therefore varies only through the imbalance
within each pair. At the same time, the individual counts remain
independent Poisson variables, allowing the Poisson score identity and the associated fingerprinting argument to be applied as before. This retains the advantages of Poissonization while avoiding the extraneous \(\sqrt d\) accumulation caused by estimating \(d\) unknown
normalization factors.

We now prove the sample complexity necessary to achieve $\frac{Dr^2}{12}$ accuracy while preserving $(\varepsilon, \delta)$ item level DP for the Poisson Estimation Problem.

\begin{lemma}[Paired Poisson score identity]\label{lem:score}
Fix positive integers $m,N$, a number $q>0$, and a radius $0\le r\le \frac q2$.
For each parameter value $\theta\in[-r,r]$, let $X^+=(X_1^+,\ldots,X_n^+)$ and $ X^- = (X_1^-,\ldots,X_n^-)$ where $\{X_{i}^+\}_{i=1}^N \sim_{i.i.d} \Pois\!\left(m (q+\theta)\right)$ and $\{X_{i}^-\}_{i=1}^N \sim_{i.i.d.} \Pois\!\left(m (q-\theta)\right)$. Let $\mathbb P_\theta$ denote the joint probability distribution function of $(X^+, X^-)$. For each $\theta \in [-r, r]$, define $g(\theta) = \mathbb{E}_{\mathbb{P}_\theta}[f(X^+, X^-)]$ where $f:\mathbb{Z}_{\ge 0}^{2n}\to\mathbb{R}$ is a bounded measurable function over
$\theta\in[-r,r]$. Then, the derivative
of $g$ satisfies
\[
    g'(\theta)
    =
    \mathbb{E}_{\theta}\!\left[
    f(X^+, X^-)
    \sum_{i=1}^N
    \left\{
    \frac{X_{i}^+-m (q+\theta)}{(q+\theta)}
    -
    \frac{X_{i}^- - m (q-\theta)}{(q-\theta)}
    \right\}
    \right].
\]
\end{lemma}

\begin{proof}
Write $x^+=(x_1^+,\ldots,x_N^+)\in\mathbb{Z}_{\ge 0}^N$, $x^-=(x_1^-,\ldots,x_N^-)\in\mathbb{Z}_{\ge 0}^N$ and define $|x^+|=\sum_{i=1}^N x_i^+$, $|x^-|=\sum_{i=1}^N x_i^-$, $x^+! = \prod_{i=1}^N x_i^+!$, and $x^-! = \prod_{i=1}^N x_i^-!$. For a fixed parameter value $\theta\in [-r,r]$, the joint probability mass function of the counts is
\[
\begin{aligned}
    \mathbb{P}_{\theta}(X^+=x^+,X^-=x^-)
    &=
    \prod_{i=1}^N
    \left[
        e^{-m (q + \theta)}
        \frac{(m (q + \theta))^{x_i^+}}{x_i^+!}
    \right]
    \prod_{i=1}^N
    \left[
        e^{-m (q - \theta)}
        \frac{(m (q - \theta))^{x_i^-}}{x_i^-!}
    \right]                                                \\
    &=
    e^{-mN (q + \theta)}
    e^{-mN (q - \theta)}
    \frac{
        m^{|x^+|+|x^-|}
        (q + \theta)^{|x^+|}
        (q - \theta)^{|x^-|}
    }{
        x^+!x^-!
    } .
\end{aligned}
\]
Since $(q+\theta)+(q-\theta) = 2q$,
the exponential factor is independent of $\theta$. Therefore,
\[
    \mathbb{P}_{\theta}(X^+=x^+,X^-=x^-)
    =
    e^{-2mNq}
    \frac{
        m^{|x^+|+|x^-|}
        (q + \theta)^{|x^+|}
        (q - \theta)^{|x^-|}
    }{
        x^+!x^-!
    } .
\]
Hence
\[
\begin{aligned}
    g(\theta)
    &=
    \sum_{x^+\in\mathbb{Z}_{\ge 0}^N}
    \sum_{x^-\in\mathbb{Z}_{\ge 0}^N}
    f(x^+,x^-)
    \mathbb{P}_{\theta}(X^+=x^+,X^-=x^-)                                      \\
    &=
    e^{-2mNq}
    \sum_{x^+\in\mathbb{Z}_{\ge 0}^N}
    \sum_{x^-\in\mathbb{Z}_{\ge 0}^N}
    f(x^+,x^-)
    \frac{
        m^{|x^+|+|x^-|}
        (q + \theta)^{|x^+|}
        (q - \theta)^{|x^-|}
    }{
        x^+!x^-!
    } .
\end{aligned}
\]

We now prove the differentiability of $g(\theta)$ and derive the exact form of its derivative. First fix $(x^+,x^-)\in\mathbb{Z}_{\ge 0}^{2N}$. Differentiating
$(q + \theta)^{|x^+|}(q - \theta)^{|x^-|}$, we get:
\[
\begin{aligned}
    \frac{d}{d\theta}
    \left[
        (q + \theta)^{|x^+|}
        (q - \theta)^{|x^-|}
    \right]
    &=
    |x^+|
    (q + \theta)^{|x^+|-1}
    (q - \theta)^{|x^-|}
    -
    |x^-|
    (q + \theta)^{|x^+|}
    (q - \theta)^{|x^-|-1}                         \\
    &=
    (q + \theta)^{|x^+|}
    (q - \theta)^{|x^-|}
    \left(
        \frac{|x^+|}{(q + \theta)}
        -
        \frac{|x^-|}{(q - \theta)}
    \right).
\end{aligned}
\]
Thus, we have
\[
\begin{aligned}
    \frac{d}{d\theta}
    \mathbb{P}_{\theta}(X^+=x^+,X^-=x^-)
    &=
    \mathbb{P}_{\theta}(X^+=x^+,X^-=x^-)
    \left(
        \frac{|x^+|}{(q + \theta)}
        -
        \frac{|x^-|}{(q - \theta)}
    \right).
\end{aligned}
\]

\paragraph{Series of Derivatives Converges} Let
$\|f\|_{\infty}\le B$. Fix a compact interval
$J\subset (-q,q)$. Since $(q + \theta)$ and $(q - \theta)$ are positive
and continuous on $J$, there exist constants $0<a<b<\infty$ such that $a\le (q + \theta),(q - \theta)\le b$ for all $\theta\in J$. For $\theta\in J$, using the derivative formula above,
\[
\begin{aligned}
    \left|
    f(x^+,x^-)
    \frac{d}{d\theta}
    \mathbb{P}_{\theta}(X^+=x^+,X^-=x^-)
    \right|
    &\le
    B\,
    \mathbb{P}_{\theta}(X^+=x^+,X^-=x^-)
    \left(
        \frac{|x^+|}{a}
        +
        \frac{|x^-|}{a}
    \right)                                                     \\
    &\le
    \frac{B}{a}
    \bigl(|x^+|+|x^-|\bigr)
    \frac{
        m^{|x^+|+|x^-|}
        b^{|x^+|+|x^-|}
    }{
        x^+!x^-!
    } .
\end{aligned}
\]
The right-hand side is summable over
$(x^+,x^-)\in\mathbb{Z}_{\ge 0}^{2n}$. Indeed,
\[
\begin{aligned}
    &\sum_{x^+\in\mathbb{Z}_{\ge 0}^N}
    \sum_{x^-\in\mathbb{Z}_{\ge 0}^N}
    \bigl(|x^+|+|x^-|\bigr)
    \frac{
        (mb)^{|x^+|+|x^-|}
    }{
        x^+!x^-!
    } =
    \sum_{z\in\mathbb{Z}_{\ge 0}^{2n}}
    \left(\sum_{j=1}^{2N} z_j\right)
    \prod_{j=1}^{2N}\frac{(mb)^{z_j}}{z_j!}
    <\infty.
\end{aligned}
\]
The last inequality follows from Lemma~\ref{lem:finiteterm}. Thus, on every compact interval $J\subset(-q,q)$, the derivatives of the summands are dominated uniformly in $\theta\in J$ by a summable sequence. By the Weierstrass M-test \cite[Theorem 7.10]{rudin1976principles}, we obtain that
\[
\sum_{x^+\in\mathbb{Z}_{\ge 0}^N}
    \sum_{x^-\in\mathbb{Z}_{\ge 0}^N}
    f(x^+,x^-)
    \frac{d}{d\theta}\mathbb{P}_{\theta}(X^+=x^+,X^-=x^-)
\]
converges uniformly over $[-r, r]$. Moreover, $g(0) \leq B < \infty$. Hence, applying \citet[Theorem 7.17]{rudin1976principles} shows that $g(\theta)$ is differentiable and $g'(\theta)$ is equal to the sum above.
Therefore,
\begin{align*}
    g'(\theta)
    &=
    \sum_{x^+\in\mathbb{Z}_{\ge 0}^N}
    \sum_{x^-\in\mathbb{Z}_{\ge 0}^N}
    f(x^+,x^-)
    \frac{d}{d\theta}
    \mathbb{P}_{\theta}(X^+=x^+,X^-=x^-)                       \\
    &=
    \sum_{x^+\in\mathbb{Z}_{\ge 0}^N}
    \sum_{x^-\in\mathbb{Z}_{\ge 0}^N}
    f(x^+,x^-)
    \mathbb{P}_{\theta}(X^+=x^+,X^-=x^-)
    \left(
        \frac{|x^+|}{(q + \theta)}
        -
        \frac{|x^-|}{(q - \theta)}
    \right)                                                       \\
    &=
    \mathbb{E}_{\theta}
    \left[
        f(X^+, X^-)
        \left(
            \frac{\sum_{i=1}^N X_{i}^+}{(q + \theta)}
            -
            \frac{\sum_{i=1}^N X_{i}^-}{(q - \theta)}
        \right)
    \right] \\
    &=
    \mathbb{E}_{\theta}
    \left[
        f(X^+, X^-)
        \left\{
            \left(\frac{\sum_{i=1}^N X_{i}^+}{(q + \theta)} - Nm\right)
            -
            \left(\frac{\sum_{i=1}^N X_{i}^-}{(q - \theta)} - Nm\right)
        \right\}
    \right]\\
    &=
    \mathbb{E}_{\theta}\!\left[
    f(X^+, X^-)
    \sum_{i=1}^N
    \left\{
    \frac{X_{i}^+-m (q + \theta)}{(q + \theta)}
    -
    \frac{X_{i}^- - m (q - \theta)}{(q - \theta)}
    \right\}
    \right].
\end{align*}
Since $\theta\in[-r,r]$ and $r\le \frac q2$, one also has $(q+\theta),(q-\theta)\in \left[\frac{q}{2}, \frac{3q}{2}\right]$.
Hence, the score is integrable and all the above quantities are finite.
\end{proof}

\begin{lemma}[One-dimensional paired fingerprinting inequality]\label{lem:1d-fingerprint}
Let $\theta\sim\Unif[-r,r]$, and conditionally on $\theta$, let $X_{i}^+\sim\Pois(m(q+\theta))$ and $X_{i}^-\sim\Pois(m(q-\theta))$ independently for $i=1,\ldots,N$. Assume $r\le \frac q2$. Let $\widehat\theta=\widehat\theta(X)\in[-r,r]$ be any estimator. Then,
\begin{equation*}
    \EE\!\left[
    (\widehat\theta-\theta)^2
    +(r^2-\theta^2)(\widehat\theta-\theta)
    \sum_{i=1}^N \left\{\frac{X_{i}^+-m(q+\theta)}{q+\theta}
    -
    \frac{X_{i}^--m(q-\theta)}{q-\theta}\right\}
    \right]
    \ge
    \frac{r^2}{3} .
\end{equation*}
\end{lemma}

\begin{proof}
Let $g(\theta)=\EE[\widehat\theta\mid\theta]$. By Lemma~\ref{lem:score}, applied to the conditional expectation of the estimator over any internal randomness, we have
\[
    g'(\theta)
    =
    \EE\left[\widehat\theta\sum_{i=1}^N \left\{\frac{X_{i}^+-m(q+\theta)}{q+\theta}     -     \frac{X_{i}^--m(q-\theta)}{q-\theta}\right\}\,\middle|\,\theta\right].
\]
Since $\mathbb{E}\left[X_{i}^+-m(q+\theta)|\theta\right] = 0$ and $\mathbb{E}\left[X_{i}^--m(q-\theta)|\theta\right] = 0$ for all $i \in [N]$, we obtain,
\[
    g'(\theta)
    =
    \EE\left[(\widehat\theta-\theta)\sum_{i=1}^N \left\{\frac{X_{i}^+-m(q+\theta)}{q+\theta}     -     \frac{X_{i}^--m(q-\theta)}{q-\theta}\right\}\,\middle|\,\theta\right].
\]
Averaging over $\theta\sim\Unif[-r,r]$, we get,
\begin{align*}
    &\EE\left[(r^2-\theta^2)(\widehat\theta-\theta)\sum_{i=1}^N \left\{\frac{X_{i}^+-m(q+\theta)}{q+\theta}     -     \frac{X_{i}^--m(q-\theta)}{q-\theta}\right\}\right]\\
    &\qquad = \EE_\theta\left[(r^2 - \theta^2)g'(\theta)|\theta\right] \\
    &\qquad = \frac1{2r}\int_{-r}^r (r^2-\theta^2)g'(\theta)\,d\theta \\
    &\qquad = \frac1{2r}\left[(r^2-\theta^2)g(\theta)\right]_{-r}^{r}
    +
    \frac1{2r}\int_{-r}^r 2\theta g(\theta)\,d\theta \\
    &\qquad = 2\EE[\theta\widehat\theta].
\end{align*}
The second equality follows from the definition of expectation and the third equality follows from integration by parts. In the fourth equality, the boundary term vanishes because $r^2-\theta^2=0$ at $\theta=\pm r$.

By linearity of expectation, we get that
\begin{align*}
    &\EE\!\left[
    (\widehat\theta-\theta)^2
    +(r^2-\theta^2)(\widehat\theta-\theta)
    \sum_{i=1}^N \left\{\frac{X_{i}^+-m(q+\theta)}{q+\theta}     -     \frac{X_{i}^--m(q-\theta)}{q-\theta}\right\}
    \right]\\
    &\qquad = \EE\!\left[
    (\widehat\theta-\theta)^2\right]
     + \EE\!\left[(r^2-\theta^2)(\widehat\theta-\theta)
    \sum_{i=1}^N \left\{\frac{X_{i}^+-m(q+\theta)}{q+\theta}     -     \frac{X_{i}^--m(q-\theta)}{q-\theta}\right\}
    \right]\\
    &\qquad = \EE[(\widehat\theta-\theta)^2]
    +2\EE[\theta\widehat\theta] \\
    &\qquad = \EE[\widehat\theta^2]+\EE[\theta^2] \ge
    \EE[\theta^2] = \frac{r^2}{3} .
\end{align*}
This proves the lemma.
\end{proof}

Recall that for our reduction and the high-dimensional fingerprinting argument we will need to work on a lifted space of dimension $D>d$.

\begin{lemma}[Poisson lower bound under approximate DP]\label{lem:poisson-lb-approx}
Take $D \in \mbb N$. Let $\cB$ be $(\eps,\delta)$ DP algorithm which has accuracy $\frac{Dr^2}{12}$ and which solves the Poisson Estimation Problem defined in Definition~\ref{def:poissonestmproblem} with output $\hat{\theta}(X)$.
If $\delta\le \frac{C_\delta\eps}{\sqrt D}$, for some universal constant $C_\delta>0$, then
\begin{align*}
    N\sqrt m \gtrsim \frac{\sqrt{Dq}}{\eps r}.
\end{align*}
\end{lemma}

\begin{proof}
Fix an item $i \in [N]$. Conditioned on a parameter $\theta$, define the paired Poisson score for each hard coordinate $\ell$
\[
    S(X_{i,\ell}, \theta_\ell)
    :=
    \frac{X_{i,\ell}^+-m\left(q+\theta_\ell\right)}{q+\theta_\ell}
    -
    \frac{X_{i,\ell}^- - m\left(q - \theta_\ell\right)}{q - \theta_\ell} .
\]
Let
\[ Z_i:=\sum_{\ell=1}^D (r^2-\theta_\ell^2)(\widehat\theta_\ell(X) -\theta_\ell)S(X_{i,\ell}, \theta_\ell).
\]
The statistic $(r^2-\theta_\ell^2)(\widehat\theta_\ell(X)-\theta_\ell)S(X_{i,\ell}, \theta_\ell)$ measures the correlation between the estimation error on coordinate $\ell$ and the $i^{th}$ item's score in that direction.

 Let $X^{\sim i}$ be the dataset obtained by replacing item $i$'s block by an independent fresh draw from the distribution under the same $\theta$ as the samples in $X$. Define
\[
    \widetilde Z_i:=\sum_{\ell=1}^D (r^2-\theta_\ell^2)(\widehat\theta_\ell(X^{\sim i})-\theta_\ell)S(X_{i,\ell}, \theta_\ell).
\]
Note that the score $S_{i,\ell}$ is still computed from the original $i^{th}$ block.

Keeping the conditioning on $\theta$, if we further condition on the original data $X$ and on the resampled data $X^{\sim i}$, the two inputs $X$ and $X^{\sim i}$ are neighboring datasets. Moreover, for every possible output $u\in[-r,r]^D$ of the algorithm, we have $|r^2-\theta_\ell^2|\le r^2$ and $|u_\ell-\theta_\ell|\le 2r$. Hence, by Cauchy-Schwarz, the corresponding score statistic is bounded by
\[
    \left|
    \sum_{\ell=1}^D
    (r^2-\theta_\ell^2)(u_\ell-\theta_\ell)S(X_{i,\ell}, \theta_\ell)
    \right|
    \le
    2r^3\sum_{\ell=1}^D |S(X_{i,\ell}, \theta_\ell)| =:B_i.
\]

Since $\widehat{\theta}$ is $(\varepsilon, \delta)$-DP, applying Lemma~\ref{lem:approx-comparison} conditional on $\theta, X$ and $X^{\sim i}$ gives
\begin{equation*}
    \EE Z_i
    \le
    \EE\widetilde Z_i
    +2\eps\EE|\widetilde Z_i|
    +2\delta B_i .
\end{equation*}

Here the expectation is over the randomness of $\hat{\theta}$. Applying another expectation over $\theta, X, X^{\sim i}$ on both sides yeilds
\begin{equation}\label{eq:approx-Zi-bound-start}
    \EE Z_i
    \le
    \EE\widetilde Z_i
    +2\eps\EE|\widetilde Z_i|
    +2\delta \EE B_i .
\end{equation}
Since $r\le \frac{q}{2}$, $q\pm\theta_\ell$ for all $\ell \in [D]$ lie in $\left[\frac{q}{2},\frac{3q}{2}\right]$,
\begin{equation}\label{eq:upperboundS}
    \EE[S^2(X_{i,\ell}, \theta_\ell)\mid\theta]
    =
    m\left(\frac1{q+\theta_\ell}+\frac1{q-\theta_\ell}\right)
    \lesssim \frac{m}{q} .
\end{equation}
For the last term, $\EE S^2(X_{i,\ell}, \theta_\ell)\lesssim \frac{m}{q}$ and Jensen's inequality give
\begin{equation*}
    \EE B_i = 2r^3\sum_{\ell=1}^D \EE |S(X_{i,\ell}, \theta_\ell)|
    \le
    2r^3\sum_{\ell=1}^D \sqrt{\EE S^2(X_{i,\ell}, \theta_\ell)} \lesssim r^3D\sqrt{\frac{m}{q}}.
\end{equation*}

Hence, plugging this upper bound into \eqref{eq:approx-Zi-bound-start}, we obtain
\begin{equation}\label{eq:approx-Zi-bound-end}
    \EE Z_i
    \lesssim
    \EE\widetilde Z_i
    +\eps\EE|\widetilde Z_i|
    +\delta r^3D\sqrt{\frac{m}{q}}.
\end{equation}

\paragraph{Fingerprinting Lower Bound.} Conditional on \(\{X_{i,\ell'}:i\in[N],\,\ell'\ne\ell\}\), \(\widehat\theta_\ell(X)\) is a bounded function of \(\{X_{i,\ell}:i\in[N]\}\). Additionally, $X_{i,\ell} = \{X_{i,\ell}^+, X_{i,\ell}^-\}$ for $X_{i,\ell}^+\sim\Pois(m(q+\theta))$ and $X_{i,\ell}^-\sim\Pois(m(q-\theta))$ independently for all $i \in [N]$ and for all $\ell \in [D]$. Thus, for all $\ell \in [D]$, applying Lemma~\ref{lem:1d-fingerprint} to $\widehat{\theta}_{\ell}$ conditioned on $X_{i,\ell'}$ for all $i \in [N]$ and $\ell' \in [D] \setminus \{\ell\}$, we get that
\begin{align*}
    \EE\!\left[
    (\widehat\theta_\ell(X) - \theta_\ell)^2
    +(r^2-\theta^2)(\widehat\theta_\ell(X) - \theta_\ell)
    \sum_{i=1}^N S(X_{i,\ell}, \theta_\ell) \middle| \{X_{1,\ell'}, \cdots, X_{N,\ell'}\}_{\ell' \in [D] \setminus \{\ell\}}
    \right]
    \ge
    \frac{r^2}{3}.
\end{align*}

Taking expectations over the conditioning variables on both sides and using the law of iterated expectations, we get that for all $\ell \in [D]$,
\begin{align*}
    \EE\!\left[
    (\widehat\theta_\ell(X) - \theta_\ell)^2
    +(r^2-\theta^2)(\widehat\theta_\ell(X) - \theta_\ell)
    \sum_{i=1}^N S(X_{i,\ell}, \theta_\ell)\right]
    \ge
    \frac{r^2}{3}.
\end{align*}
Summing the above inequality over $\ell \in [D]$ and using linearity of expectation gives
\begin{align*}
    &\sum_{\ell = 1}^D\EE\!\left[
    (\widehat\theta_\ell(X) - \theta_\ell)^2
    +(r^2-\theta^2)(\widehat\theta_\ell(X) - \theta_\ell)
    \sum_{i=1}^N S(X_{i,\ell}, \theta_\ell)\right] \\
    &\qquad  = \EE\!\left[
    \sum_{\ell = 1}^D(\widehat\theta_\ell(X) - \theta_\ell)^2
    +\sum_{\ell = 1}^D (r^2-\theta^2)(\widehat\theta_\ell(X) - \theta_\ell)
    \sum_{i=1}^N S(X_{i,\ell}, \theta_\ell)\right] \\
    &\qquad  = \EE\!\left[
    \norm{\widehat\theta_\ell(X) - \theta_\ell}^2
    +\sum_{\ell = 1}^D (r^2-\theta^2)(\widehat\theta_\ell(X) - \theta_\ell)
    \sum_{i=1}^N S(X_{i,\ell}, \theta_\ell)\right] \\
    &\qquad  =
    \EE\norm{\widehat\theta_\ell(X) - \theta_\ell}^2
    + \EE\left[\sum_{i=1}^N \sum_{\ell = 1}^D (r^2-\theta^2)(\widehat\theta_\ell(X) - \theta_\ell)
     S(X_{i,\ell}, \theta_\ell)\right] \\
    &\qquad  = \EE\|\widehat\theta(X)-\theta\|_2^2
    +
    \sum_{i=1}^N \EE Z_i
    \ge
    \frac{Dr^2}{3} .
\end{align*}
Using the fact that $\cB$ is $\frac{Dr^2}{12}$ accurate, we get
\begin{equation}\label{eq:zi-lower-approx}
    \sum_{i=1}^N \EE Z_i
    \ge
    \frac{Dr^2}{4} .
\end{equation}

\paragraph{Approximate-DP upper bound.}

For each $i\in[N]$, conditional on $\theta$, the original score $S(X_{i,\ell}, \theta_\ell)$ is independent of $\widehat\theta(X^{\sim i})$ and has mean zero.  Therefore $\EE[\widetilde Z_i\mid\theta]=0$, which implies that $\EE\widetilde Z_i=0$.

Since $r\le \frac{q}{2}$, $q\pm\theta_\ell \in \left[\frac{q}{2},\frac{3q}{2}\right]$ for each $\ell \in [D]$,
\[
    \EE[S^2(X_{i,\ell}, \theta_\ell)\mid\theta]
    =
    m\left(\frac1{q+\theta_\ell}+\frac1{q-\theta_\ell}\right)
    \lesssim \frac{m}{q} .
\]
Conditioned on $\theta$, for each $\ell \in [D]$, $S(X_{i,\ell},\theta_\ell)$ is centered and independent of $S(X_{i,\ell'},\theta_{\ell'})$ for all $\ell' \in [D] \setminus \{\ell\}$.  Hence,
\begin{align*}
    \EE[\widetilde Z_i^2\mid\theta] &= \EE\left[\left(\sum_{\ell=1}^D (r^2-\theta_\ell^2)(\widehat\theta_\ell(X^{\sim i})-\theta_\ell)S(X_{i,\ell}, \theta_\ell)\right)^2 \middle| \theta \right] \\
    &= \EE\left[\sum_{\ell=1}^D (r^2-\theta_\ell^2)^2(\widehat\theta_\ell(X^{\sim i})-\theta_\ell)^2 S^2(X_{i,\ell}, \theta_\ell) \middle| \theta \right] \\
    &\leq 2r^4 \EE\left[\sum_{\ell=1}^D(\widehat\theta_\ell(X^{\sim i})-\theta_\ell)^2 S^2(X_{i,\ell}, \theta_\ell) \middle| \theta \right]\\
    &= 2r^4\sum_{\ell=1}^D\EE\left[(\widehat\theta_\ell(X^{\sim i})-\theta_\ell)^2 S^2(X_{i,\ell}, \theta_\ell) \middle| \theta \right]\\
    &= 2r^4\sum_{\ell=1}^D\EE\left[(\widehat\theta_\ell(X^{\sim i})-\theta_\ell)^2\middle| \theta \right] \EE\left[S^2(X_{i,\ell}, \theta_\ell) \middle| \theta \right]\\
    &\lesssim \frac{mr^4}{q}\sum_{\ell=1}^D\EE\left[(\widehat\theta_\ell(X^{\sim i})-\theta_\ell)^2\middle| \theta \right]\\
    &= \frac{mr^4}{q}
    \EE\left[\|\widehat\theta(X^{\sim i})-\theta\|_2^2\,\middle|\,\theta\right] .
\end{align*}
The first step follows from the definition of $\Tilde{Z}_i$. Since $S(X_{i,\ell},\theta_\ell)$ is centered and independent of $S(X_{i,\ell'},\theta_{\ell'})$ for all $\ell' \in [D] \setminus \{\ell\}$, the expectation of the cross terms vanish and the second step follows. The third step follows from the fact that $(r^2 - \theta_\ell^2)^2 \leq 2r^4$. The fourth step follows from linearity of expectation. The fifth step follows from the fact that $S(X_{i,\ell},\theta_\ell)$ is independent of $X^{\sim i}$. The penultimate step follows from \eqref{eq:upperboundS}. The last step follows from linearity of expectation and the definition of the $\ell_2$ norm.

Taking expectation with respect to $\theta \sim \Unif([-r, r])$ on both sides and using the fact that $\cB$ is $\frac{Dr^2}{12}$ accurate, we get $\EE \widetilde Z_i^2 \lesssim \frac{mDr^6}{q}$. Jensen's inequality implies that
\begin{equation}\label{eq:ztilde-moment-approx}
    \EE|\widetilde Z_i|
    \lesssim r^3\sqrt{\frac{mD}{q}}.
\end{equation}
Hence, considering \eqref{eq:approx-Zi-bound-end}, the first term is zero, and the second term is controlled by \eqref{eq:ztilde-moment-approx}.
Thus, we obtain that for every $i \in [N]$,
\[
    \EE Z_i
    \lesssim
    \eps r^3\sqrt{\frac{mD}{q}}
    +
    \delta r^3D\sqrt{\frac{m}{q}} .
\]
If $\delta\le \frac{C_\delta\eps}{\sqrt D}$ with $C_\delta>0$ sufficiently small, then $\EE Z_i \lesssim \eps r^3\sqrt{\frac{mD}{q}}$. Summing this inequality over $i \in [N]$ and comparing with \eqref{eq:zi-lower-approx} gives
\[
    \frac{Dr^2}{4}
    \lesssim N\eps r^3\sqrt{\frac{mD}{q}} .
\]
Rearranging the above expression proves the claim.
\end{proof}

Now we prove Theorem~\ref{thm:main-expected-approx} via a privacy preserving reduction from estimating the parameters of product samples to estimating $\theta$ in Lemma~\ref{lem:poisson-lb-approx}.

\begin{proofof}{Theorem~\ref{thm:main-expected-approx}}
We now prove Theorem~\ref{thm:main-expected-approx} by reducing the problem of estimating a $d$-dimensional $k$-ary distribution with $m$ samples per user to estimating $\theta$ in the paired Poisson model, in which one data point consists of \(2D\) Poisson random variables. Assume, toward contradiction, that there is an $(\eps,\delta)$-user-level DP estimator $\cA$ satisfying
\begin{equation}\label{eq:assumed-risk-approx}
    \sup_{P\in\cP_{d,k}}
    \EE_P\xi(\widehat P,P)^2
    \le
    \alpha^2,
\end{equation}
but that \eqref{eq:main-lower-bound-approx} fails.

\paragraph{Local Paired Family} Let $s=\left\lfloor\frac{k}{2}\right\rfloor$ and $D=ds$.
The hard parameter is $\theta=(\theta_{j,b})_{j\in[d],\,b\in[s]}\in[-r,r]^{d \times s}$,
where $r=24\sqrt{2}\frac{\alpha}{k\sqrt d}$. Note that $\alpha$ is small enough that $r\le \frac{1}{2k}$ holds. For $j\in[d]$ and $b\in[s]$, define
\begin{equation}\label{eq:paired-family}
        p_{j}^{\theta}(2b-1) = \frac{1}{k}+\theta_{j,b}, \quad p_{j}^{\theta}(2b) =\frac{1}{k}-\theta_{j,b}.
\end{equation}
If $k$ is odd, set the final unpaired coordinate to $p_{j}^{\theta}(k) = \frac{1}{k}$.
Each pair in \eqref{eq:paired-family} sums to $\frac{2}{k}$, so every $p_j^\theta$ is a probability vector in $\Del_k$. Let $P_\theta=\bigotimes_{j=1}^d p_j^\theta$,
and let the prior be $\theta_{j,b}\stackrel{\iid}{\sim}\Unif[-r,r]$.

Let $m'=C_0\log^4\left(\frac{edk}{\alpha\eps}\right) m$, where $C_0$ is a sufficiently large numerical constant.
The Poisson estimator receives independent counts $X_{(j,a)}^{(1)}, \cdots, X_{(j,a)}^{(N)}\stackrel{i.i.d.}{\sim}\Pois\left(m'p_{j}^\theta(a)\right)$, for each $j\in[d]$, $a\in[k]$. For each user-coordinate pair $(i,j)$, let $X^{(i)}_{(j)}=\sum_{a=1}^k X^{(i)}_{(j,a)}$.

Conditioned on $X_{(j)}^{(i)}=t$, if we form the expanded list containing $X_{(j,a)}^{(i)}$ copies of label $a$ and perform a random permutation on the resultant list, we get that $\widetilde{X}^{(i)}_1(j), \cdots, \widetilde{X}^{(i)}_t(j)$ are distributed as $t$ independent samples from $\frac{p_j^\theta}{\sum_{c=1}^k{p_j^\theta}(c)}$ \citep[Theorem 25]{esfandiari2022tight}. Since $p_j^\theta\in\Del_k$, $\widetilde{X}^{(i)}_1(j), \cdots, \widetilde{X}^{(i)}_t(j) \sim p^\theta_j$.

Using the facts above, we prepare the input for the product distribution estimation. For each $(i,j)$, if $X^{(i)}_{(j)} \ge m$, take the first $m$ labels from this randomly permuted expanded list.  Otherwise, fill in $m$ independent uniform labels from $[k]$. Let $\widetilde{X}_{\ell}^{(i)} = \left(\widetilde{X}_{\ell}^{(i)}(1), \cdots, \widetilde{X}_{\ell}^{(i)}(d)\right)$. Conditioned on $X^{(i)}_{(j)} \geq m$ for each $(i,j)$, this produces samples
$\widetilde{X}_{1}^{(i)}, \cdots, \widetilde{X}_{m}^{(i)} \sim \bigotimes_{j=1}^d p^\theta_j$ for each $i \in [N]$. On the complementary event, we don't have any such guarantee, however we later show that $X^{(i)}_{(j)} \geq m$ occurs with high probability for each user-coordinate pair $(i,j)$.
Run $\cA$ on $\left\{\widetilde{X}_{1}^{(i)}, \cdots, \widetilde{X}_{m}^{(i)}\right\}_{i=1}^N$ to obtain $\widehat p_1,\ldots,\widehat p_d$, and output
\begin{equation}\label{eq:theta-estimator-from-p}
    \widehat\theta_{j,b}
    =
    \Pi_{[-r,r]}
    \left(
    \frac{\widehat p_{j}(2b-1)-\widehat p_{j}(2b)}{2}
    \right),
\end{equation}
where $\Pi_{[-r,r]}$ denotes Euclidean projection onto $[-r,r]$.

\paragraph{Privacy Guarantee.}
The map from one user's Poisson counts $\left\{X_{(j,a)}^{(i)}: j \in [d], a \in [k] \right\}_{i=1}^N$ to the synthetic samples $\left\{\widetilde X_{1}^{(i)}, \cdots, X_{m}^{(i)}\right\}_{i=1}^N$ operates separately on each user. Thus neighboring Poisson datasets map to neighboring product-sample user-level datasets. Since $\cA$ is $(\eps,\delta)$-user-level DP and \eqref{eq:theta-estimator-from-p} is post-processing, the constructed Poisson estimator is therefore $(\eps,\delta)$ DP.

\paragraph{Utility Relation Between Instances}
Let $L_\alpha:=\log\left(\frac{edk}{\alpha\varepsilon}\right)$ and $A_0:=C_0L_\alpha^4 \geq 2$,
and let $G$ be the event that $X_{i,j}\ge m$ for every $i\in[n]$ and $j\in[d]$. Since $X_{i,j}\sim\operatorname{Pois}(A_0m)$ and $A_0\ge 2$, Lemma~\ref{lem:poissontailbound} implies that
\[
\mbb{P}(X_{i,j}<m)
\le
\mbb{P}\left(X_{i,j}\le \frac{A_0m}{2}\right)
\le
\exp\left(-\frac{A_0m}{8}\right)
\le
\exp\left(-\frac{A_0}{8}\right).
\]
A union bound therefore yields
\[
\mbb{P}(G^c)\le nd\exp\left(-\frac{A_0}{8}\right).
\]
Since, we have assumed the failure of \eqref{eq:main-lower-bound-approx}, and since $m \geq 1$, it is safe to assume that $N\le C_1\frac{dk}{\alpha\varepsilon}$. Then
\begin{align*}
\mbb{P}(G^c)
&\le
C_1\frac{d^2k}{\alpha\varepsilon}
\exp\left(-\frac{C_0L_\alpha^4}{8}\right) =
C_1\frac{\alpha^2}{d}
\frac{d^3k}{\alpha^3\varepsilon}
\exp\left(-\frac{C_0L_\alpha^4}{8}\right) \le
C_1\frac{\alpha^2}{d}
\exp\left(3L_\alpha-\frac{C_0L_\alpha^4}{8}\right).
\end{align*}
Since $L_\alpha\ge 1$, choosing $C_0$ to be a sufficiently large absolute constant gives
\[
\mbb{P}(G^c)\le \frac{\alpha^2}{4d}.
\]

On $G$, the data have exactly the same law as $N$ users with $m$ independent samples each from $P_\theta$.  Since $\xi^2(\hat{P},P) \le 4d$ is always true, combining the assumed risk guarantee \eqref{eq:assumed-risk-approx} conditioned on $G^c$ and applying the law of total probability gives
\begin{equation}\label{eq:fake-risk-approx}
    \EE
    \sum_{j=1}^d \|\widehat p_j-p_j^\theta\|_1^2
    \le
    2\alpha^2.
\end{equation}

\noindent
For each pair $(j,b)$, $\theta_{j,b} = \frac{p^\theta_{j}(2b-1) - p_{j}^\theta(2b)}{2}$.
Projection in \eqref{eq:theta-estimator-from-p} cannot increase distance to $\theta_{j,b}\in[-r,r]$, so
\[
    |\widehat\theta_{j,b}-\theta_{j,b}|
    \le
    \frac12
    \left(
    |\widehat p_{j}(2b-1)-p^\theta_{j}(2b-1)|
    +
    |\widehat p_{j}(2b)-p^\theta_{j}(2b)|
    \right).
\]
Since $|\widehat\theta_{j,b}-\theta_{j,b}| \le 2r$, the above expression implies that
\begin{align}
    \|\widehat\theta-\theta\|_2^2
    &\le
    r\sum_{j=1}^d \|\widehat p_j-p_j^\theta\|_1
    \le
    r\sqrt d
    \left(
    \sum_{j=1}^d \|\widehat p_j-p_j^\theta\|_1^2
    \right)^{1/2}. \label{eq:l1-to-theta-approx}
\end{align}
Taking expectations in \eqref{eq:l1-to-theta-approx}, applying Jensen's inequality, and using \eqref{eq:fake-risk-approx}, we get $\EE\|\widehat\theta-\theta\|_2^2 \le r\sqrt {2d}\alpha$. With $r=\frac{24 \sqrt{2}\alpha}{k\sqrt d}$, $\EE\|\widehat\theta-\theta\|_2^2 \leq \frac{Dr^2}{12}$. Hence, applying Lemma~\ref{lem:poisson-lb-approx} with $q = \frac1k$ and Poisson mean parameter $m'=C\log^4\left(\frac{edk}{\alpha\eps}\right)m$ gives $N\sqrt{m'} \gtrsim \frac{\sqrt{D/k}}{\eps r}$. Since $D=\Theta(dk)$ and  $r=\frac{24 \sqrt{2}\alpha}{k\sqrt d}$, we get
\[
    N\sqrt m
    \gtrsim \frac{dk}{\alpha\eps \log^2\left(\frac{edk}{\alpha\eps}\right)},
\]
contradicting the assumed failure of \eqref{eq:main-lower-bound-approx}.  This proves Theorem~\ref{thm:main-expected-approx}.
\end{proofof}

\subsubsection{Proof of Lemma~\ref{lem:approx-dp-few-user-packing}}
\begin{proofof}{Lemma~\ref{lem:approx-dp-few-user-packing}}
We first prove the result when $k$ is even. For odd $k$, let $k':=2\left\lfloor\frac{k}{2}\right\rfloor$. Apply the even-domain construction below on the first $k'$ symbols and assign probability zero to the remaining symbol. Since
$k'\ge k/2$, this changes only the universal constants in the final
lower bound.

\paragraph{Defining a Packing} Let $\mathcal C\subseteq\{0,1\}^k$ be a constant-weight binary code
such that every $c\in\mathcal C$ has Hamming weight $\frac{k}{2}$, every
distinct $c,c'\in\mathcal C$ satisfy $d_{\mathrm H}(c,c')\ge\frac{k}{8}$,
and $\log|\mathcal C|=\Omega(k)$. For $k\ge40$, this follows from
\cite[Lemma~6]{acharya2021differentially} by taking weight $\frac{k}{2}$. For every $c\in\mathcal C$, define $p_c\in\Delta(k)$ by
\[
    p_c(a)
    =
    \begin{cases}
        \dfrac{1+24\alpha}{k},
        & c_a=1,\\[6pt]
        \dfrac{1-24\alpha}{k},
        & c_a=0.
    \end{cases}
\]
Since every codeword has exactly $\frac{k}{2}$ ones,
\[
    \sum_{a=1}^k p_c(a)
    =
    \frac{k}{2}\cdot\frac{1+24\alpha}{k}
    +
    \frac{k}{2}\cdot\frac{1-24\alpha}{k} =1.
\]
Moreover, $p_c(a)\ge0$ because $\alpha<1/48$. For every $\mathbf c = \bigl(c^{(1)},\ldots,c^{(d)}\bigr) \in\mathcal C^d$,
define $P_{\mathbf c} := \bigotimes_{j=1}^d p_{c^{(j)}}$. Let $\mathcal V := \left\{ P_{\mathbf c}: \mathbf c\in\mathcal C^d \right\}$. Its cardinality is $W:=|\mathcal V|=|\mathcal C|^d$, and therefore
\begin{equation}
    \label{eq:packing-cardinality}
    \log W
    =
    d\log|\mathcal C|
    =
    \Omega(kd).
\end{equation}

\paragraph{Separation of the packing.}

Fix distinct $c,c'\in\mathcal C$ and set
$h:=d_{\mathrm H}(c,c')$. At every coordinate on which $c$ and
$c'$ differ, the corresponding probabilities differ by
$48\alpha/k$. Hence
\[
    d_{\mathrm{TV}}(p_c,p_{c'})
    =
    \frac12
    \sum_{a=1}^k
    |p_c(a)-p_{c'}(a)| =
    \frac12\cdot h\cdot\frac{48\alpha}{k} =
    \frac{24\alpha h}{k} \ge 3\alpha.
\]
Equivalently, $\|p_c-p_{c'}\|_1 \ge 6\alpha$. If $\mathbf c\ne\mathbf c'$, then
$c^{(j)}\ne c'^{(j)}$ for some $j\in[d]$. It follows from
the $\ell_1$ norm separation that

\begin{equation}
\label{eq:packing-separation}
    \rho(P_{\mathbf c},P_{\mathbf c'})
    =
    \max_{\ell\in[d]}
    \|p_{c^{(\ell)}}-p_{c'^{(\ell)}}\|_1
    \ge
    \|p_{c^{(j)}}-p_{c'^{(j)}}\|_1 \ge 6\alpha.
\end{equation}
Post-process the output of $\mathcal A$ by selecting the nearest
packing element:
\[
    \mathcal B(X)
    \in
    \operatorname*{arg\,min}_{i\in[W]}
    \rho(\widehat P,P_i),
\]
with ties broken according to a fixed ordering of $[W]$.
Since $\mathcal B$ is a post-processing of $\mathcal A$, it is
$(\eps,\delta)$-user-level differentially private.

Suppose now that the true distribution is $P_i$ and that $\rho(\widehat P,P_i)\le\alpha$. For every $\ell\ne i$, the triangle inequality and
\eqref{eq:packing-separation} give
\[
    \rho(\widehat P,P_\ell)
    \ge \rho(P_i,P_\ell) - \rho(\widehat P,P_i) \ge
    6\alpha-\alpha = 5\alpha >
    \rho(\widehat P,P_i).
\]
Thus, $P_i$ is the unique nearest packing element, and consequently
\begin{equation}
    \label{eq:packing-decoder-success}
    \mbb{P}_{X\sim P_i^{\otimes Nm}}
    \left[
        \mathcal B(X)=i
    \right]
    \ge\frac23.
\end{equation}

Fix an arbitrary user block $z_0\in([k]^d)^m$. For each $i\in[W]$ and $t\in\{0,\ldots,N\}$, let
$X^{(i,t)}$ be the random dataset in which the first $t$ user
blocks are drawn independently from $P_i^{\otimes m}$ and the
remaining $N-t$ user blocks are equal to $z_0$. Define
\[
    q_{i,t}
    :=
    \mbb{P}\left[
        \mathcal B(X^{(i,t)})=i
    \right].
\]

For every $t\in\{0,\ldots,N-1\}$, the datasets defining
$X^{(i,t)}$ and $X^{(i,t+1)}$ differ only in user $t+1$.
Conditioning on all the sampled user blocks, applying
$(\eps,\delta)$-DP, and then averaging gives
\[
    q_{i,t+1}
    \le
    e^\eps q_{i,t}+\delta.
\]
This can be rewritten as
\[
    q_{i,t+1}+\frac{\delta}{e^\eps-1}
    \le
    e^\eps
    \left(
        q_{i,t}+\frac{\delta}{e^\eps-1}
    \right).
\]
Iterating over $t=0,\ldots,N-1$ and using~\eqref{eq:packing-decoder-success} yields
\[
\begin{aligned}
    q_{i,0} + \frac{\delta}{e^\eps-1}
    &\ge
    e^{-N\eps}
    \left(
        q_{i,N} + \frac{\delta}{e^\eps-1}
    \right) \ge
    e^{-N\eps}
    \left(
        \frac23+\frac{\delta}{e^\eps-1}
    \right).
\end{aligned}
\]
Therefore,
\begin{equation}
    \label{eq:packing-hybrid-lower}
    q_{i,0}
    \ge
    e^{-N\eps}
    \left(
        \frac23 + \frac{\delta}{e^\varepsilon - 1}
    \right)
    -
    \frac{\delta}{e^\varepsilon - 1}.
\end{equation}
Crucially, $X^{(i,0)}$ is the same fixed dataset
$(z_0,\ldots,z_0)$ for every $i$. Since $\mathcal B$ outputs exactly
one element of $[W]$, $\sum_{i=1}^W q_{i,0}=1$. Thus, summing \eqref{eq:packing-hybrid-lower} over $i$ gives
\[
    1
    \ge
    W\left[
        e^{-N\eps}
        \left(
            \frac23 + \frac{\delta}{e^\varepsilon - 1}
        \right)
        -
        \frac{\delta}{e^\varepsilon - 1}
    \right].
\]
Rearranging gives
\[
    e^{-N\eps}
    \left(
        \frac23 + \frac{\delta}{e^\varepsilon - 1}
    \right)
    \le
    \frac1W + \frac{\delta}{e^\varepsilon - 1}.
\]
Hence, using $e^\varepsilon - 1 = \Theta(\varepsilon)$ for $\varepsilon \leq 1$ and $\delta \leq \varepsilon^2$, we obtain
\[
    N
    \ge
    \frac1\eps
    \log
    \left(
        \frac{\frac23 + \frac{\delta}{e^\varepsilon - 1}}
        {\frac1W + \frac{\delta}{e^\varepsilon - 1}}
    \right) \gtrsim \frac1\eps \min\left\{\log(W), \log\left(\frac{1}{\delta}\right)\right\}.
\]
Using~\eqref{eq:packing-separation} along with the above expression proves the claim.
\end{proofof}
\subsection{Pure DP Upper Bound}
\label{ssec:pure-dp-upper-bound-formal}
\begin{lemma}[{\citet[Theorem 43]{ghazi2023user}}]
\label{lem:userpuredpdiscretedistribution}
    Consider the problem of estimating a discrete distribution $p \in \Delta(k)$ over $[k]$ given  $n = \tilde{\Theta}\left( \frac{k}{\varepsilon} + \frac{k}{\alpha\varepsilon\sqrt{m}} + \frac{k}{\alpha^2m}\right)$ users with $m$ samples per user. Let $X$ denote the dataset sampled $\mathrm{i.i.d.}$ from $p$. For sufficiently small $\varepsilon, \alpha$, there exists an $\varepsilon$  user level DP algorithm $\mathcal{A}_{\varepsilon}$ which achieves $\norm{\mathcal{A}_{\varepsilon}(X)- p}_1 \leq \alpha$
    with probability greater than $1 - f$.
\end{lemma}

Based on the above result, we propose a (trivial) algorithm to estimate $\hat{P}$ with pure DP. Using the fact that each row of the distribution matrix corresponds to a $k$-ary distribution, Algorithm~\ref{alg:private-product-estimation-pure} simply applies the algorithm in Lemma~\ref{lem:userpuredpdiscretedistribution} to each row and obtains the required guarantee.

\begin{algorithm}[h]
\caption{User-level private estimation of product distributions with pure DP.}
\label{alg:private-product-estimation-pure}

\noindent
\textbf{\textit{Input:}}
A database
\(\mathcal{D}=(X_{i,t})_{i\in[N],\,t\in[m]}\), where
\(X_{i,t}\in[k]^d\), privacy parameters \(\varepsilon,\delta>0\), and
failure probability \(f\in(0,1)\).

\medskip
\noindent
\textbf{\textit{Operation:}}
\begin{enumerate}
    \item For each product coordinate \(j\in[d]\), let $\mathcal{X}_j$ represent the user level data $\{X_{i,t,j}\}_{i \in [N], t \in [m]}$, where \(X_{i,t,j}\) is the \(j^{th}\) coordinate of \(X_{i,t}\).

    \item For each $j \in [d]$ set $\hat{P}_{j,\cdot} = \mathcal{A}_{\frac{\varepsilon}{d}}(\mathcal{X}_j)$ with failure probability $\frac{f}{d}$ and output $\hat{P}$.
\end{enumerate}
\end{algorithm}

\begin{theorem}
\label{thm:pure-dp-product-distribution-priv}
    Algorithm~\ref{alg:private-product-estimation-pure} is $\eps$-DP.
\end{theorem}
\begin{proof}
    We use $d$ instances of $\mathcal{A}_{\frac{\varepsilon}{d}}$ each satisfying ${\varepsilon}/{d}$-user level DP. By simple composition \citep{dwork2014}, the algorithm is $\varepsilon$ user-level DP.
\end{proof}

\begin{theorem}
\label{thm:pure-dp-product-distribution-util}
    Fix \(0<f<1\) and let $\alpha$ and $\varepsilon$ be sufficiently small, as required by Lemma~\ref{lem:userpuredpdiscretedistribution}. The output of Algorithm~\ref{alg:private-product-estimation-pure}, $\hat{P}$, satisfies $\Pr\!\left(\rho(\widehat P,P)\leq\alpha\right)\geq 1-f$ uniformly over \(P\in \mathcal{P}_{d,k}\) for
    \begin{align*}
        N = \tilde{\Theta}\left(\frac{kd}{\eps} + \frac{kd}{\alpha \varepsilon \sqrt{m}} + \frac{k}{m\alpha^2}\right).
    \end{align*}
\end{theorem}

\begin{proof}
Because $P$ is a product distribution, each coordinate represents samples from an independent $k$-ary distribution. Hence, $N$ samples from $P$ yield $N$ samples from $P_{j,\cdot}$ for every $j \in [d]$. Lemma~\ref{lem:userpuredpdiscretedistribution} implies that, with $N = \tilde{\Theta}\left( \frac{kd}{\varepsilon} + \frac{kd}{\alpha\varepsilon\sqrt{m}} + \frac{k}{\alpha^2 m}\right)$ samples for every $j \in [d]$,
\begin{align*}
    \mbb{P} \left(\norm{\hat{P}_{j,\cdot} - P_{j,\cdot}}_1 \geq \alpha\right) \leq \frac{f}{d}.
\end{align*}
A union bound over the $d$ coordinates shows that with probability greater than $1 - f$, $\rho({\hat{P}, P}) \leq \alpha$. This concludes the proof.
\end{proof}

\subsection{Pure DP Lower Bound}
\label{ssec:pure-dp-lower-bound-formal}
Following~\cite{ghazi2023user}, we use the following notation for the lower-bound argument. A task $\mathcal T$ has a finite output
space $\mathcal O_{\mathcal T}$ and, for every valid data-generating
distribution $D$, a set $\Psi_{\mathcal T}(D)\subseteq\mathcal O_{\mathcal T}$ of correct outputs. An algorithm $\mathcal B$ is said to be useful
for $\mathcal T$ if, for every valid $D$,
\[
    \mbb{P}_{X\sim D^{\otimes nm}}
    \left(
        \mathcal B(X)\in\Psi_{\mathcal T}(D)
    \right)
    \ge \frac23,
\]
where the probability includes both the randomness of the samples and the internal randomness of $\mathcal B$. We let $n_m^{\mathcal T}(\eps)$ denote the smallest number of users
for which there exists an $\eps$-user-level differentially private
algorithm that is useful for $\mathcal T$.

\begin{lemma}[User-level DP Fano's inequality
{\cite[Lemma~39]{ghazi2023user}}]
\label{lem:pureuserdpfano}
Let $\mathcal T$ be any finite-output task. Suppose that there exist
valid distributions $D_1,\ldots,D_W$ such that, for all distinct
$i,j\in[W]$, $\Psi_{\mathcal T}(D_i) \cap \Psi_{\mathcal T}(D_j) = \varnothing$ and $D_{\mathrm{KL}}(D_i\Vert D_j)\le\beta$.
Then
\[
    n_m^{\mathcal T}(\eps)
    =
    \Omega\left(
        \frac{\log W}{\eps}
        +
        \frac{\log W}{\eps\sqrt{m\beta}}
        +
        \frac{\log W}{m\beta}
    \right).
\]
\end{lemma}

\begin{theorem}
\label{thm:pure-user-level-product-lower-bound}
Let $k \geq 40$ and $\alpha < 1/48$. Let \(\mathcal{A}\) be an \(\varepsilon\)-user-level differentially private estimator with $N$ users, each contributing $m$ samples. Suppose that, uniformly over
\(P \in \mathcal{P}_{d,k}\),
$\mathcal{A}$ outputs $\widehat{P} \in \mathcal{P}_{d,k}$ such that
\[
\mbb{P}\left(
\rho(\hat{P}, P)
\leq \alpha
\right)
\geq \frac{2}{3}.
\]
Then
\[
N
=
\Omega\left(
\frac{kd}{\varepsilon}
+
\frac{k\sqrt{d}}{\varepsilon\alpha\sqrt{m}}
+
\frac{k}{m\alpha^2}
\right).
\]
\end{theorem}

\begin{proof}
We first prove the result when $k$ is even. For odd $k$, let $k':=2\left\lfloor\frac{k}{2}\right\rfloor$. Apply the even-domain construction below on the first $k'$ symbols and assign probability zero to the remaining symbol. Since
$k'\ge k/2$, this changes only the universal constants in the final
lower bound.

\paragraph{Defining a Packing} Let $\mathcal C\subseteq\{0,1\}^k$ be a constant-weight binary code
such that every $c\in\mathcal C$ has Hamming weight $\frac{k}{2}$, every
distinct $c,c'\in\mathcal C$ satisfy $d_{\mathrm H}(c,c')\ge\frac{k}{8}$,
and $\log|\mathcal C|=\Omega(k)$. For $k_0\ge40$, this follows from
\cite[Lemma~6]{acharya2021differentially} by taking weight $\frac{k}{2}$. For every $c\in\mathcal C$, define $p_c\in\Delta(k)$ by
\[
    p_c(a)
    =
    \begin{cases}
        \dfrac{1+24\alpha}{k},
        & c_a=1,\\[6pt]
        \dfrac{1-24\alpha}{k},
        & c_a=0,
    \end{cases}
\]
where $p_c(a)$ denotes its $a$-th co-ordinate.
Since every codeword has exactly $\frac{k}{2}$ ones,
\[
    \sum_{a=1}^k p_c(a)
    =
    \frac{k}{2}\cdot\frac{1+24\alpha}{k}
    +
    \frac{k}{2}\cdot\frac{1-24\alpha}{k} =1.
\]
Moreover, $p_c(a)\ge0$ because $\alpha<1/48$. For every $\mathbf c = \bigl(c^{(1)},\ldots,c^{(d)}\bigr) \in\mathcal C^d$,
define $P_{\mathbf c} := \bigotimes_{j=1}^d p_{c^{(j)}}$. Let $\mathcal V := \left\{ P_{\mathbf c}: \mathbf c\in\mathcal C^d \right\}$. Its cardinality is $W:=|\mathcal V|=|\mathcal C|^d$, and therefore
\begin{equation}
    \label{eq:pure-packing-cardinality}
    \log W
    =
    d\log|\mathcal C|
    =
    \Omega(kd).
\end{equation}

\paragraph{Separation of the packing.}

Fix distinct $c,c'\in\mathcal C$ and set
$h:=d_{\mathrm H}(c,c')$. At every coordinate on which $c$ and
$c'$ differ, the corresponding probabilities differ by
$48\alpha/k$. Hence
\[
    d_{\mathrm{TV}}(p_c,p_{c'})
    =
    \frac12
    \sum_{a=1}^k
    |p_c(a)-p_{c'}(a)| =
    \frac12\cdot h\cdot\frac{48\alpha}{k} =
    \frac{24\alpha h}{k} \ge 3\alpha,
\]
where the last step follows since $h\geq \frac{k}{8}$. Equivalently, $\|p_c-p_{c'}\|_1 \ge 6\alpha$. If $\mathbf c\ne\mathbf c'$, then
$c^{(j)}\ne c'^{(j)}$ for some $j\in[d]$. It follows from
the $\ell_1$ norm separation that

\begin{equation}
\label{eq:product-packing-separation}
    \rho(P_{\mathbf c},P_{\mathbf c'})
    =
    \max_{\ell\in[d]}
    \|p_{c^{(\ell)}}-p_{c'^{(\ell)}}\|_1
    \ge
    \|p_{c^{(j)}}-p_{c'^{(j)}}\|_1 \ge 6\alpha.
\end{equation}

\paragraph{Reduction to a finite distribution-identification task.}

Enumerate the packing as $\mathcal V=\{P_1,\ldots,P_W\}$. Define a finite-output task $\mathcal T_{\mathcal V}$ whose valid
data-generating distributions are $P_1,\ldots,P_W$, whose output
space is $[W]$, and whose correct-output sets are $\Psi_{\mathcal T_{\mathcal V}}(P_i) := \{i\}$ for all $i\in[W]$.
In particular, the correct-output sets are pairwise disjoint. We then construct a useful algorithm for
$\mathcal T_{\mathcal V}$ from the assumed estimator $\mathcal A$.
Given a dataset $X$, run $\mathcal A$ and let $\widehat P$ be its
output. Define
\[
    \mathcal B(X)
    \in
    \operatorname*{arg\,min}_{i\in[W]}
    \rho(\widehat P,P_i),
\]
where ties are broken according to a fixed ordering of $[W]$. Since $\mathcal B$ is a deterministic post-processing of
$\mathcal A$, it is $\eps$-user-level differentially private.

Suppose now that the true distribution is $P_i$ and that $\rho(\widehat P,P_i)\le\alpha$. For every $\ell\ne i$, the triangle inequality and
\eqref{eq:product-packing-separation} give
\[
    \rho(\widehat P,P_\ell)
    \ge \rho(P_i,P_\ell) - \rho(\widehat P,P_i) \ge
    6\alpha-\alpha = 5\alpha >
    \rho(\widehat P,P_i).
\]
Thus $P_i$ is the unique nearest packing element, and $\mathcal B(X)=i$. The assumed utility guarantee for $\mathcal A$ therefore implies
that, for every $i\in[W]$,
\[
    \mbb{P}_{X\sim P_i^{\otimes Nm}}
    \left(
        \mathcal B(X)=i
    \right) = \mbb{P}_{X\sim P_i^{\otimes Nm}}
    \left(
        \mathcal B(X)
        \in
        \Psi_{\mathcal T_{\mathcal V}}(P_i)
    \right)
    \ge\frac23.
\]
Hence $\mathcal B$ is useful for
$\mathcal T_{\mathcal V}$, and consequently $N \ge n_m^{\mathcal T_{\mathcal V}}(\eps)$.

\paragraph{Pairwise KL-divergence bound.}
Set $x:=24\alpha$. Since $\alpha<1/48$, we have $0<x<1/2$. Fix distinct $c,c'\in\mathcal C$ and set
$h:=d_{\mathrm H}(c,c')$. Because both codewords have Hamming
weight $k/2$, exactly $h/2$ of their differing coordinates satisfy
$c_a=1,c'_a=0$ and exactly $h/2$ satisfy
 $c_a=0,c'_a=1$.
{This can be seen by the following argument. Take $|\{a: c_a = 1, c'_a = 0\}| = h_1$ and $|\{a: c_a = 0, c'_a = 1\}| = h_2$, then it would mean that if $c$ has Hamming weight $k/2$, then $c'$ must have a Hamming weight $ \geq h_2 + \frac{k}{2} - h_1$. On the other hand, if $c'$ has Hamming weight $\frac{k}{2}$, then we get that $c$ must have a Hamming weight of at least $h_1 + \frac{k}{2} - h_2$. Now, since both vectors have Hamming weight $\frac{k}{2}$, we get that $h_1 \leq h_2$ and $h_2 \leq h_1$, proving that $h_1 = h_2$}. Therefore,
\[
    D_{\mathrm{KL}}(p_c\Vert p_{c'})
    =
    \frac h2
    \left[
        \frac{1+x}{k}
        \log\left(\frac{1+x}{1-x}\right)
        +
        \frac{1-x}{k}
        \log\left(\frac{1-x}{1+x}\right)
    \right] =
    \frac{hx}{k}
    \log\left(\frac{1+x}{1-x}\right).
\]
For $0\le x\le1/2$, $\log\left(\frac{1+x}{1-x}\right) \le4x$. It follows that
\[
    D_{\mathrm{KL}}(p_c\Vert p_{c'}) \le
    \frac{4hx^2}{k} \le
    4x^2 =
    2304\alpha^2.
\]
By additivity of KL divergence under products, for every $\mathbf c,\mathbf c'\in\mathcal C^d$,
\[
    D_{\mathrm{KL}}
    \left(
        P_{\mathbf c}
        \,\middle\|\,
        P_{\mathbf c'}
    \right)
    =
    \sum_{j=1}^d
    D_{\mathrm{KL}}
    \left(
        p_{c^{(j)}}
        \,\middle\|\,
        p_{c'^{(j)}}
    \right) \le
    2304d\alpha^2.
\]
Thus, Lemma~\ref{lem:pureuserdpfano} applies to the finite task
$\mathcal T_{\mathcal V}$ with $\beta:=2304d\alpha^2$.
Combining
\eqref{eq:pure-packing-cardinality},
Lemma~\ref{lem:pureuserdpfano}, and
the fact that $N \ge n_m^{\mathcal T_{\mathcal V}}(\eps)$, we obtain
\[
    N
    \gtrsim
        \frac{kd}{\eps}
        +
        \frac{k\sqrt d}
        {\eps\alpha\sqrt m}
        +
        \frac{k}
        {m\alpha^2}.
\]
This proves the result.
\end{proof}

\section{A Simple Extension of the Item-Level Algorithm to User-Level Privacy}
\label{sec:itemtousertrivial}

In this section, the user complexity bounds follow by applying the item-level algorithm to the user-level case.

\begin{theorem}
\label{thm:pure-dp-user-lowT-level-upper-bound:privacy}
Algorithm~\ref{alg:private-transition-estimator-multiple-chains} with $\mathcal{D} = \mathrm{Lap}\left(\frac{4T}{\varepsilon}\right)$  is $\varepsilon$ user-level differentially private.
\end{theorem}
\begin{proof}
    Consider the count matrix $K'$ corresponding to the adjacent database. Changing one chain can affect at most 2T transitions, which implies that the $\ell_1$ norm between the vectorized versions of the matrices is upper bounded by $4T$. Hence, for $\mc{D} = Lap\left(\frac{4T}{\varepsilon}\right)$, we get that the count matrix is $\varepsilon$-DP by the Laplace mechanism \cite{dwork2014}. Steps 2 and 3 are simply post-processing, which implies that the resulting matrix $\tilde{P}$ satisfies $\varepsilon$-DP.
\end{proof}

\begin{theorem}
\label{thm:pure-dp-user-lowT-level-upper-bound:utility}
Fix \(0<\alpha,f<1\) and \(0<\eps\leq 1\). Uniformly over
\(P\in\mathcal{C}(S,\pi_{\min},\gamma_{ps})\) and arbitrary initial
distributions, Algorithm~\ref{alg:private-transition-estimator-multiple-chains} with
\(\mathcal{D}=\operatorname{Lap}(0,4T/\eps)\) satisfies $\mbb{P}\left(\rho(\widehat P,P)\leq\alpha\right)\geq 1-f$ for
\[
    N = {O}\left(\frac{S\log\left(\frac{S}{f}\right)}{\pi_{\min}T\alpha^2} + \frac{\ln\left(\frac{1}{\pi_{\min}}\right)}{\pi_{\min}T} + \frac{\log\left(\frac{S}{f}\right)}{\pi_{\min}T\gamma_{ps}} + \frac{S +\max\left\{\sqrt{S\log\left(\frac{S}{f}\right)}, \log\left(\frac{S}{f}\right)\right\}}{\pi_{\min}\alpha\varepsilon}\right).
\]
\end{theorem}

\begin{proof}
Proceeding exactly as in the item level analysis, we look at the matrix-norm distance between corresponding rows of $\Tilde{P}$ with $\Hat{P}$. Consider the arbitrary $i^{th}$ row of the two matrices. We seek a high probability bound on the following quantity
\begin{align*}
    \sum_{j=1}^S \left| \frac{[K_{i,j} + Z_{i,j}]_+}{\sum_{k=1}^S [K_{i,k} + Z_{i,k}]_+} - \frac{K_{i,j}}{\sum_{k=1}^S K_{i,k}}\right|.
\end{align*}

Let $0 < R \leq \frac{1}{2} \pi_{\min} N$ and $\esterr \leq 1$. Define $K_i := \sum_{j=1}^S K_{i,j}$ and consider the events
\begin{align*}
    \mc E_1 &= \left\{\sum_{j=1}^S |Z_{i,j}| \leq R \text{ for all rows }i \in [S]\right\} \\
    \mc E_2 &=  \left\{\forall i \in [S], K_i \in \left[\frac{1}{2}\pi_i NT, \frac{3}{2}\pi_i NT\right]\right\}.
\end{align*}

Conditional on the events $\event_1$ and $\event_2$, we have that $K_i \geq \frac{1}{4}\pi_{\min}NT \geq R$. Hence, we get that for all $i \in [S]$
\begin{align*}
    \sum_{j = 1}^S [K_{i,j} + Z_{i,j}]_+ \geq \sum_{j = 1}^S (K_{i,j} + Z_{i,j}) = K_i +  \sum_{j = 1}^S Z_{i,j} \geq K_i - \sum_{j = 1}^S |Z_{i,j}| \geq \frac{1}{4} \pi_{\min}T > 0.
\end{align*}

Thus, $\widetilde{P}_{i,j} := \frac{\widetilde{K}_{i,j}}{\widetilde{K}_i}$. For each row, Lemma~\ref{lem:diffempvsnoisyemp} bounds the above quantity by
\begin{align*}
    \sum_{j=1}^S \left| \frac{[K_{i,j} + Z_{i,j}]_+}{\sum_{k=1}^S [K_{i,k} + Z_{i,k}]_+} - \frac{K_{i,j}}{\sum_{k=1}^S K_{i,k}}\right| \leq  \frac{2R}{K_i - R} \leq \frac{2R}{\frac{1}{2}\pi_{\min}NT - R}.
\end{align*}

Let $\sigma^2 = \frac{8T^2}{\eps^2}$. A Laplace random variable is sub-exponential with Bernstein parameters $\left(\sigma^2, \frac{\sigma}{\sqrt{2}}\right)$. Then, by applying Bernstein's inequality \citep{Vershynin2026} to each row and then taking the union bound over all the rows, we get that
\begin{align*}
    \mbb{P}\left(\exists\ i \in [S], \left|\sum_{j=1}^S |Z_{i,j}| - \frac{2ST}{\eps}\right|\ge t\right)
    \le 2S\exp\left(-c\,\min\left(\frac{t^2}{S\sigma^2}, \frac{t}{\sigma}\right)\right).
\end{align*}

Hence, setting $R = \frac{2ST}{\varepsilon} + t$, we get that $\mbb{P}(\overline{\mc E_1}) \leq 2S\exp\left(-c\,\min\left(\frac{t^2}{S\sigma^2}, \frac{t}{\sigma}\right)\right)$. Thus, the event $\mc E_1 \cap \mc E_2$ implies that $\norm{\tilde{P} - \hat{P}}_\infty \leq \alpha$ when $NT = \Theta\left(\frac{R}{\pi_{\min}\alpha} + \frac{1}{\pi_{\min}\gamma_{ps}}\log\left(\frac{S}{f}\right)\right)$. Hence, by Lemma~\ref{lem:counts}
\begin{align*}
    \mbb{P}\left(\norm{\tilde{P} - \hat{P}}_\infty \geq \alpha\right) &\leq \mbb{P}\left(\norm{\tilde{P} - \hat{P}}_\infty \geq \alpha \mid \mc E_1 \cap \mc E_2\right) + \mbb{P}(\overline{\mc E_1}) + \mbb{P}(\overline{\mc E_2}) \\
    &\leq 2S\exp\left(-c\,\min\left(\frac{t^2}{S\sigma^2}, \frac{t}{\sigma}\right)\right) + \frac{f}{3}.
\end{align*}

Setting $t = \frac{2T\sqrt{2}}{\varepsilon}\max\left\{\sqrt{S\log\left(\frac{6S}{f}\right)}, \log\left(\frac{6S}{f}\right)\right\}$ and applying Lemma~\ref{thm:leskelarobust} with
\[
N \geq C_1 \max\left\{\frac{S}{\pi_{\min}T\alpha^2}\log\left(\frac{S}{f}\right) + \frac{\ln\left(\frac{1}{\pi_{\min}}\right)}{\pi_{\min}T}, \frac{1}{\pi_{\min}\gamma_{ps} T}\log\left(\frac{S}{f}\right)\right\},
\]
we obtain
\begin{align*}
    \mbb{P}\left(\norm{\tilde{P} - P}_\infty \geq \alpha\right) &\leq \mbb{P}\left(\norm{\hat{P} - P}_\infty \geq \alpha\right) + \mbb{P}\left(\norm{\tilde{P} - \hat{P}}_\infty \geq \alpha\right) \leq f.
\end{align*}

The above value of $t$ implies that $R = \Theta\left(\frac{ST}{\varepsilon} + \frac{T}{\varepsilon}\max\left\{\sqrt{S\log\left(\frac{6S}{f}\right)}, \log\left(\frac{6S}{f}\right)\right\}\right)$, which yields the following sample complexity,

\[
    N = {O}\left(\frac{S\log\left(\frac{S}{f}\right)}{\pi_{\min}T\alpha^2} + \frac{\ln\left(\frac{1}{\pi_{\min}}\right)}{\pi_{\min}T} + \frac{\log\left(\frac{S}{f}\right)}{\pi_{\min}T\gamma_{ps}} + \frac{S +\max\left\{\sqrt{S\log\left(\frac{S}{f}\right)}, \log\left(\frac{S}{f}\right)\right\}}{\pi_{\min}\alpha\varepsilon}\right).
\]
\end{proof}

It is important to note that every $\varepsilon$-DP algorithm is $(\varepsilon, \delta)$-DP for any $0 \leq \delta < 1$. However, for completeness, we also give the analysis for an algorithm based on the Gaussian mechanism below.

\begin{theorem}
Algorithm~\ref{alg:private-transition-estimator-multiple-chains} with $\mathcal{D} = \mathcal{N}(0, \sigma^2)$, where $\sigma = \frac{4T\sqrt{\log(1/\delta)}}{\varepsilon}$ is $(\varepsilon, \delta)$ differentially private.
\end{theorem}
\begin{proof}
    Using exact same argument as in Theorem~\ref{thm:pure-dp-item-level-upper-bound:privacy}
    , for an adjacent matrix $K'$, the $\norm{K - K'}_{F} \leq 2T$. Hence, for $\mc{D} = \mc{N}(0, \sigma^2)$ where $\sigma = \frac{4T\sqrt{\log(1/\delta)}}{\varepsilon}$, the privacy guarantee follows from the $(\varepsilon, \delta)$ privacy guarantee of the Gaussian mechanism on the count matrix $K$ and Steps 2 and 3 are post-processing operations, so the resulting $\Tilde{P}$ satisfies $(\eps, \del)$-DP.
\end{proof}

\begin{theorem}
Fix \(0<\alpha,f<1\) and \(0<\eps\leq 1\). Uniformly over
\(P\in\mathcal{C}(S,\pi_{\min},\gamma_{ps})\) and arbitrary initial
distributions, Algorithm~\ref{alg:private-transition-estimator-multiple-chains} with $\mathcal{D}$ being $\mathcal{N}(0, \sigma^2)$ where $\sigma = \frac{4T\sqrt{\log(1/\delta)}}{\varepsilon}$ satisfies $\mbb{P}\left(\rho(\widehat P,P)\leq\alpha\right)\geq 1-f$ for
\[
    N = O\left(\frac{S\log\left(\frac{S}{f}\right)}{\pi_{\min}T\alpha^2} + \frac{\ln\left(\frac{1}{\pi_{\min}}\right)}{\pi_{\min}T} + \frac{\log\left(\frac{S}{f}\right)}{\pi_{\min}T\gamma_{ps}} + \frac{S\sqrt{\log\left(\frac{1}{\delta}\right)}}{\varepsilon\pi_{\min}\alpha} + \frac{\sqrt{S\log\left(\frac{1}{\delta}\right)\log\left(\frac{S}{f}\right)}}{\pi_{\min}\alpha\varepsilon}\right).
\]
\end{theorem}

\begin{proof}
Since $|Z_{i,j}|$ has mean $\sqrt{\frac{2}{\pi}} \sigma$, we have $\mathbb E \sum_{j=1}^S |Z_{i,j}|= \sqrt{\frac{2}{\pi}} S\sigma$. Since the absolute value function is $1$-Lipschitz, it implies that $|Z_{i,j}| - \mathbb E|Z_{i,j}|$ is $2\sigma$ sub-gaussian.
Thus, by applying the concentration inequality for sums of sub-Gaussian random variables \citep{Vershynin2026} to each row and then taking a union bound over all rows, we get that
\begin{align*}
    \mbb{P}\left(\exists\ i \in [S], \left|\sum_{j=1}^S |Z_{i,j}| -  S\sigma\sqrt{\frac{2}{\pi}}\right| \ge t\right) \le 2 S\exp\left(-\frac{t^2}{2 S \sigma^2}\right).
\end{align*}

As in Theorem~\ref{thm:pure-dp-user-lowT-level-upper-bound:utility}, setting $R = S\sigma\sqrt{\frac{2}{\pi}} + t$, we get that $\mbb{P}(\overline{\mc E_1}) \leq 2 S\exp\left(-\frac{t^2}{2 S \sigma^2}\right)$. Thus, the event $\mc E_1 \cap \mc E_2$ implies that $\norm{\tilde{P} - \hat{P}}_\infty \leq \alpha$ when $NT = \Theta\left(\frac{R}{\pi_{\min}\alpha}\right)$. Taking $NT = \Theta\left(\frac{R}{\pi_{\min}\alpha}\right)$ and applying Lemma~\ref{lem:counts} we get
\begin{align*}
    \mbb{P}\left(\norm{\tilde{P} - \hat{P}}_\infty \geq \alpha\right) &\leq \mbb{P}\left(\norm{\tilde{P} - \hat{P}}_\infty \geq \alpha \mid \mc E_1 \cap \mc E_2\right) + \mbb{P}(\overline{\mc E_1}) + \mbb{P}(\overline{\mc E_2}) \\
    &\leq 2 S\exp\left(-\frac{t^2}{2 S \sigma^2}\right) + \frac{f}{3}.
\end{align*}

Setting $t = 2\sigma\sqrt{S\log\left(\frac{6S}{f}\right)}$ and combining this with Lemma~\ref{thm:leskelarobust}, for
\[
N \geq C_1 \max\left\{\frac{S}{\pi_{\min}T\alpha^2}\log\left(\frac{S}{f}\right) + \frac{\ln\left(\frac{1}{\pi_{\min}}\right)}{\pi_{\min}T}, \frac{1}{\pi_{\min}\gamma_{ps} T}\log\left(\frac{S}{f}\right)\right\},
\]
we obtain
\begin{align*}
    \mbb{P}\left(\norm{\tilde{P} - P}_\infty \geq \alpha\right) &\leq \mbb{P}\left(\norm{\hat{P} - P}_\infty \geq \alpha\right) + \mbb{P}\left(\norm{\tilde{P} - \hat{P}}_\infty \geq \alpha\right) \leq f.
\end{align*}

Moreover, note that the above value of $t$ implies that
\[
R = \Theta\left(\frac{ST\sqrt{\log(1/\delta)}}{\varepsilon} + \frac{T\sqrt{S\log(1/\delta)}}{\varepsilon}\sqrt{\log\left(\frac{S}{f}\right)}\right),
\]
which yields the following sample complexity,
\[
    N = {O}\left(\frac{S\log\left(\frac{S}{f}\right)}{\pi_{\min}T\alpha^2} + \frac{\ln\left(\frac{1}{\pi_{\min}}\right)}{\pi_{\min}T} + \frac{\log\left(\frac{S}{f}\right)}{\pi_{\min}T\gamma_{ps}} + \frac{S\sqrt{\log\left(\frac{1}{\delta}\right)}}{\varepsilon\pi_{\min}\alpha} + \frac{\sqrt{S\log\left(\frac{1}{\delta}\right)\log\left(\frac{S}{f}\right)}}{\pi_{\min}\alpha\varepsilon}\right).
\]
\end{proof}

\section{Proofs for Section~\ref{sec:userlevelest}}
\label{ssec:markov-upper}

The following algorithm converts each state's collection to exactly \(m_0\)
entries by truncating excess transitions and padding a deficient
collection with independent uniform draws. It then aligns one entry from
each state into each of \(m_0\) observations in \([S]^S\). With high probability, this gives a
fixed-size instance of the product-distribution problem with \(d=k=S\),
with target marginals \(\tranmat_{(1,\cdot)},\ldots,\tranmat_{(S,\cdot)}\). After coupling
the reduced data to this ideal product sample (Step 4 of Algorithm~\ref{alg:markov-to-product-reduction}), we may invoke the estimators
of Appendix~\ref{sec:productdist} as a black box. Their loss is exactly \(\rho(\widehat{\tranmat}, \tranmat)\), so their accuracy guarantees
translate directly to the transition matrix. Finally, the preprocessing
is performed separately for each trajectory, and hence replacing one
user changes only one output block, preserving user-level adjacency.

\begin{algorithm}[h]
\caption{\(\operatorname{PadAndTruncate}_{m}\)}
\label{alg:pad-and-truncate}
\begin{minipage}{0.94\linewidth}

\textbf{Input:} A multiset \(\mathcal S\in[k]^*\) and a target size
\(m\in\mathbb N\).

\textbf{Output:} A multiset
\(\widehat{\mathcal S}\in[k]^m\).
\begin{enumerate}
    \item If \(\lvert\mathcal S\rvert>m\), select \(m\) elements
    uniformly at random from \(\mathcal S\), without replacement, and
    return the resulting multiset.

    \item Otherwise, independently sample $Y_1,\ldots,Y_{m-\lvert\mathcal S\rvert} \sim \operatorname{Unif}([k])$,
    and return
    \[
    \widehat{\mathcal S} = \mathcal S \uplus \{Y_1,\ldots,Y_{m-\lvert\mathcal S\rvert}\},
    \]
    where $\uplus$ denotes multiset union.
\end{enumerate}
\end{minipage}
\end{algorithm}

\begin{algorithm}[h]
\caption{Reduction to Product-Distribution Estimation}
\label{alg:markov-to-product-reduction}

\textbf{Input:} Trajectories \(\{(X_{i,1},\ldots,X_{i,T})\}_{i=1}^N\), number of states $S$, a lower bound \(\pi_{\min}\) on the stationary probabilities, and an \((\varepsilon, \delta)\)-DP $S$ dimensional $S$ ary distribution estimator $\mathcal A$.
\begin{enumerate}
    \item Set $m_0 = \left\lfloor \frac{\pi_{\min}(T-1)}{2} \right\rfloor$.
    \item For every \(i\in[N]\) and \(j\in[S]\), create the multiset
    \(\mathcal S_j^i \gets \{X_{i,t+1}:X_{i,t}=j,\ t\in[T-1]\}\).
    \item For every \(i\in[N]\) and \(j\in[S]\), set
        \(\widehat{\mathcal S_j^i} \gets \operatorname{PadAndTruncate}_{m_0}(\mathcal S_j^i)\).
    \item For every \(i\in[N]\), create points
    \(\widetilde X_1^{(i)},\ldots,\widetilde X_{m_0}^{(i)}\in[S]^S\) as follows: for every \(\ell\in[m_0]\) and \(j\in[S]\), set
    \((\widetilde X_\ell^{(i)})_j \gets (\widehat{\mathcal S_j^i})_\ell\).
    \item For every \(i\in[N]\), set
    \(\widetilde X^{(i)} \gets (\widetilde X_1^{(i)},\ldots,\widetilde X_{m_0}^{(i)})\).
    \item Return \(\mathcal A(\widetilde X^{(1)},\ldots,\widetilde X^{(N)})\).
\end{enumerate}
\end{algorithm}

The following lemma formalizes the high-probability
event used above; on this event, every collection already contains at
least \(m_0\) transitions, so padding is never invoked. Combining this reduction with the product-distribution guarantees of Appendix~\ref{sec:productdist} yields the following upper bounds.

\begin{lemma}
\label{lem:countslowerbound}
    Let \(\{(X_{i,1},\ldots,X_{i,T})\}_{i=1}^N\)  be $N$ trajectories, where each $X_{i,1},\ldots,X_{i,T}$ is a Markov chain with a minimum stationary probability $\pi_{\min}$. For every \(i\in[N]\) and \(j\in[S]\), define the multiset \(\mathcal S_j^i \gets \{X_{i,t+1}:X_{i,t}=j,\ t\in[T-1]\}\). Then for $T \gtrsim \frac{1}{\gamma_{ps}\pi_{\min}}\log\left(\frac{N\sqrt{S}}{\sqrt{\pi_{\min}}f}\right)$,
    \[
    \mbb{P}\left(\bigcap_{i=1}^N \bigcap_{j=1}^S \left\{|S_j^i| \geq \frac{1}{2}\pi_{j}(T-1)\right\}\right) \geq 1 - f.
    \]
\end{lemma}
\begin{proof}
    Let $\event: \bigcup_{i=1}^N \bigcup_{j=1}^S \left\{|S_j^i| \leq \frac{1}{2}\pi_{j} (T - 1)\right\}$. For each $i \in [N]$, the same chain governs each trajectory, and $|S_j^i|$ equals to the number of occurrences of the element $j$. Hence, by Lemma~\ref{lem:counts}, we get that for each $j \in [S]$,
    \begin{align*}
        \mbb{P}_{\pi}\left(|S_j^i| \leq \frac{1}{2}\pi_{j} (T - 1)\right) &\leq \exp\!\left(-\frac{\gamma_{\mathrm{ps}} \left(\frac{1}{2} (T - 1) \pi_j \right)^2}{8\left(T - 1+ \frac{1}{\gamma_{\mathrm{ps}}}\right)\pi_j (1 - \pi_j) + 10 (T - 1) \pi_j}\right) \\
        &\leq \exp\!\left(-\frac{\gamma_{\mathrm{ps}} \left(\frac{1}{2} (T - 1) \pi_{\min} \right)^2}{8\left(T - 1+ \frac{1}{\gamma_{\mathrm{ps}}}\right)\pi_{\min} (1 - \pi_{\min}) + 10 (T - 1) \pi_{\min}}\right).
    \end{align*}
    By a union bound over all $S$ states, we get
    \begin{align*}
        \mbb{P}_{\pi}\left(\bigcup_{j=1}^S \left\{|S_j^i| \leq \frac{1}{2}\pi_{j} (T - 1)\right\}\right) \leq S\exp\!\left(-\frac{\gamma_{\mathrm{ps}} \left(\frac{1}{2} (T - 1) \pi_{\min} \right)^2}{8\left(T - 1 + \frac{1}{\gamma_{\mathrm{ps}}}\right)\pi_{\min} (1 - \pi_{\min}) + 10 (T - 1) \pi_{\min}}\right).
    \end{align*}
    Now, for each user $i$, trajectory $i$ has initial distribution $\mu_i$, then after accounting for the nonstationary initial distribution \cite{paulinconcenteration}, we get
    \begin{align*}
    \mbb{P}\left(\bigcup_{j=1}^S \left\{|S_j^i| \leq \frac{1}{2}\pi_{j} (T - 1)\right\}\right) \leq \frac{\sqrt{S}}{\sqrt{\pi_{\min}}}\exp\!\left(-\frac{\gamma_{\mathrm{ps}} \left(\frac{1}{2} (T - 1) \pi_{\min} \right)^2}{16\left(T - 1 + \frac{1}{\gamma_{\mathrm{ps}}}\right)\pi_{\min} (1 - \pi_{\min}) + 20 (T - 1) \pi_{\min}}\right).
    \end{align*}
    Hence, by a union bound over all users, we get
    \begin{align*}
        \mbb{P}\left(\event\right) \leq \frac{N\sqrt{S}}{\sqrt{\pi_{\min}}}\exp\!\left(-\frac{\gamma_{\mathrm{ps}} \left(\frac{1}{2} (T - 1) \pi_{\min} \right)^2}{16\left(T - 1+ \frac{1}{\gamma_{\mathrm{ps}}}\right)\pi_{\min} (1 - \pi_{\min}) + 20 (T - 1) \pi_{\min}}\right).
    \end{align*}
    Setting $T \geq \frac{C}{\gamma_{ps}\pi_{\min}}\log\left(\frac{N\sqrt{S}}{\sqrt{\pi_{\min}}f}\right)$ for a sufficiently large $C$ concludes the proof.
\end{proof}

\subsection{Complete Statement and Proof of Theorem~\ref{thm:user-level-markov-approx-upper-informal}}
Theorem~\ref{thm:user-level-product-marginals-util} showed the existence of an $(\varepsilon, \delta)$ user-level DP algorithm that receives a set of $N$ users each having $m$ points sampled from a $d$ fold $k$-ary distribution, and achieves the desired accuracy with high probability using $N = \tilde{\Theta}\left(\frac{1}{\varepsilon} + \frac{k}{m\alpha^2} + \frac{k\sqrt{d}}{\alpha \varepsilon \sqrt{m}}\right)$ users.

\begin{theorem}[Complete Privacy Guarantee in Theorem~\ref{thm:user-level-markov-approx-upper-informal}]
\label{thm:user-level-markov-approx-upper-formal-privacy}
    Algorithm~\ref{alg:markov-to-product-reduction} with $\mathcal{A}$ instantiated as Algorithm~\ref{alg:private-product-estimation} is $(\eps,\delta)$-DP.
\end{theorem}
\begin{proof}
User-level DP is defined with respect to the worst-case adjacency relation in which the entire dataset of one user is added or replaced. Since, the careful partition, reordering of datapoints, and the $\mathrm{PadAndTruncate}$ operation makes changes to the dataset separately for each user and not across users, it does not change the notion of adjacency for user-level DP as input to $\mathcal{A}$. Hence, the user-level DP guarantee of $\mathcal{A}$ applies and the resulting algorithm is $(\varepsilon, \delta)$ user-level DP.
\end{proof}

\begin{theorem}[Complete
Utility Guarantee in Theorem~\ref{thm:user-level-markov-approx-upper-informal}]
\label{thm:user-level-markov-approximate}
Fix \(0<\alpha,f<1\), \(0<\eps\leq 1\). Uniformly over
\(P\in\mathcal{C}(S,\pi_{\min},\gamma_{ps})\) and arbitrary initial
distributions, if $T \gtrsim \frac{C}{\gamma_{ps}\pi_{\min}}\log\left(\frac{N\sqrt{S}}{\sqrt{\pi_{\min}}f}\right)$, then the output of Algorithm~\ref{alg:markov-to-product-reduction} with $\mathcal{A}$ instantiated as Algorithm~\ref{alg:private-product-estimation}, $\hat{P}$, satisfies $\mbb{P}\left(\rho(\widehat P,P)\leq\alpha\right)\geq 1-f$ for
    \begin{align*}
        N
    \gtrsim
        \frac{
            S+\log(S/f)
        }{
            \pi_{\min}T\alpha^2
        } +
        \frac{S^{3/2}}{\varepsilon\alpha\sqrt {\pi_{\min}T}}\log^{3/2}\left(\frac{S\pi_{\min}T}{\alpha\varepsilon\delta f}\right) +
        \frac{1}{\varepsilon}
        \sqrt{
            \log\left(\frac{1}{\delta}\right)
            \log\left(\frac{1}{\delta f}\right)
        }.
    \end{align*}
\end{theorem}
\begin{proof}
By the Markov property of the chain, for each $j \in [S]$ the set $\mc S_j^i$ contains i.i.d. samples from $P_{j,:}$.
Under the event $\mathcal{E}$, we would have access to $\geq \frac{1}{2}\pi_{\min}(T-1)$ i.i.d. samples per user; hence, the $\mathrm{PadAndTruncate}$ algorithm would return $\geq \frac{1}{3}\pi_{\min}(T-1)$ i.i.d. samples from $P_{j,:}$ for each $j \in [S]$. Conditional on $\event$, Step 4 makes $\tilde{X}^{(i)}$ consist of $m_0 \geq \frac{1}{3}\pi_{\min}(T-1)$ i.i.d. samples from a $d$-dimensional $k$-ary distribution whose distribution matrix is the transition matrix of the Markov chain. Theorem~\ref{thm:user-level-product-marginals-util} therefore implies that when
\begin{align*}
        N
    \gtrsim
        \frac{
            S+\log(S/f)
        }{
            \pi_{\min}T\alpha^2
        } +
        \frac{S^{3/2}}{\varepsilon\alpha\sqrt {\pi_{\min}T}}\log^{3/2}\left(\frac{S\pi_{\min}T}{\alpha\varepsilon\delta f}\right) +
        \frac{1}{\varepsilon}
        \sqrt{
            \log\left(\frac{1}{\delta}\right)
            \log\left(\frac{1}{\delta f}\right)
        },
\end{align*}
we have
\begin{align*}
    \mbb{P} \left(\norm{\hat{P} - P}_\infty \geq \alpha | \bigcap_{i=1}^N \bigcap_{j=1}^S \left\{|S_j^i| \geq \frac{1}{2}\pi_{\min}(T-1)\right\} \right) \leq \frac{f}{2}.
\end{align*}
Using $\mbb{P}(A) = \mbb{P}(A|E) \mbb{P}(E) + \mbb{P}(A | \overline{E}) \mbb{P}(\overline{E})$, we obtain the following upper bound:
\begin{align*}
    \mbb{P} \left(\norm{\hat{P} - P}_\infty \geq \alpha \right) \leq \frac{f}{2} + \mbb{P}(\overline{\event})
\end{align*}
Given that $T \geq \frac{C}{\gamma_{ps}\pi_{\min}}\log\left(\frac{N\sqrt{S}}{\sqrt{\pi_{\min}}f}\right)$, Lemma~\ref{lem:countslowerbound} implies that
\begin{align*}
    \mbb{P} \left(\norm{\hat{P} - P}_\infty \geq \alpha \right) \leq f.
\end{align*}
This concludes the proof.
\end{proof}

\subsection{Pure DP Upper Bounds}
\label{ssec:pure-dp-upper-bounds}
\begin{theorem}
\label{thm:user-level-markov-pure-upper-formal-privacy}
    Algorithm~\ref{alg:markov-to-product-reduction} with $\mathcal{A}$ instantiated as Algorithm~\ref{alg:private-product-estimation-pure} is $\eps$-DP.
\end{theorem}
\begin{proof}
User-level DP is defined with respect to the worst-case adjacency relation in which the entire dataset of one user is added or replaced. Since, the careful partition, reordering of datapoints, and the $\mathrm{PadAndTruncate}$ operation makes changes to the dataset separately for each user and not across users, it does not change the notion of adjacency for user-level DP with respect to input of $\mathcal{A}$. Hence, the $\varepsilon$ user-level DP guarantee of $\mathcal{A}$ holds on the modified dataset.
\end{proof}

\begin{theorem}
\label{thm:user-level-markov-pure}
Fix \(0<\alpha,f<1\), \(0<\eps\leq 1\). Uniformly over
\(P\in\mathcal{C}(S,\pi_{\min},\gamma_{ps})\) and arbitrary initial
distributions, if $T \gtrsim \frac{C}{\gamma_{ps}\pi_{\min}}\log\left(\frac{N\sqrt{S}}{\sqrt{\pi_{\min}}f}\right)$, then the output of Algorithm~\ref{alg:markov-to-product-reduction} with $\mathcal{A}$ instantiated as Algorithm~\ref{alg:private-product-estimation-pure}, $\hat{P}$, satisfies $\mbb{P}\left(\rho(\widehat P,P)\leq\alpha\right)\geq 1-f$ for
    \begin{align*}
        N = \tilde{\Theta}\left(\frac{S^2}{\eps} + \frac{S^2}{\alpha \varepsilon \sqrt{\pi_{\min}T}} + \frac{S}{\alpha^2 \pi_{\min}T}\right).
    \end{align*}
\end{theorem}

\begin{proof}
By the Markov property of the chain, for each $j \in [S]$ the set $\mc S_j^i$ contains i.i.d. samples from $P_{j,:}$.
Under the event $\mathcal{E}$, we would have access to $\geq \frac{1}{2}\pi_{\min}(T-1)$ i.i.d. samples per user; hence, the $\mathrm{PadAndTruncate}$ algorithm would return $\geq \frac{1}{3}\pi_{\min}(T-1)$ i.i.d. samples from $P_{j,:}$ for each $j \in [S]$. Conditional on $\mathcal E$, Step 4 makes $\tilde{X}^{(i)}$ consist of $m_0 \geq \frac{1}{3}\pi_{\min}(T-1)$ i.i.d. samples from a $d$-dimensional $k$-ary distribution whose distribution matrix is the transition matrix of the Markov chain. Theorem~\ref{thm:pure-dp-product-distribution-util}, therefore implies that, when
\begin{align*}
        N = \tilde{\mathcal O}\left(\frac{S^2}{\eps} + \frac{S^2}{\alpha \varepsilon \sqrt{\pi_{\min}T}} + \frac{S}{\alpha^2 \pi_{\min}T}\right),
\end{align*}
we have
\begin{align*}
    \mbb{P} \left(\norm{\hat{P} - P}_\infty \geq \alpha | \bigcap_{i=1}^N \bigcap_{j=1}^S \left\{|S_j^i| \geq \frac{1}{2}\pi_{\min}(T-1)\right\} \right) \leq \frac{f}{2}.
\end{align*}
Using $\mbb{P}(A) = \mbb{P}(A|E) \mbb{P}(E) + \mbb{P}(A | \overline{E}) \mbb{P}(\overline{E})$, we obtain the following upper bound:
\begin{align*}
    \mbb{P} \left(\norm{\hat{P} - P}_\infty \geq \alpha \right) \leq \frac{f}{2} + \mbb{P}(\overline{\event}).
\end{align*}
Given that $T \geq \frac{C}{\gamma_{ps}\pi_{\min}}\log\left(\frac{N\sqrt{S}}{\sqrt{\pi_{\min}}f}\right)$, Lemma~\ref{lem:countslowerbound} implies that
\begin{align*}
    \mbb{P} \left(\norm{\hat{P} - P}_\infty \geq \alpha \right) \leq 2f.
\end{align*}
This concludes the proof.
\end{proof}

\subsection{Sample Complexity Lower Bounds}
\label{ssec:markov-lower}
\begin{theorem}
\label{thm:product-to-markov-reduction}
Let $S\geq 2$, $N,T\geq 1$, $0 \leq \delta < 1$ and $0<\pi_{\min}\leq\frac{1}{3S}$.
For every $(\varepsilon,\delta)$-user-level DP algorithm $\mathcal M$ that receives $N$ independent length-$T$
trajectories $\mathbf{X} = (X^{(1)}_{1:T}, \cdots, X^{(N)}_{1:T})$ from a $2S$-state ergodic Markov chain $M$ with $\min_x\pi(x)=\pi_{\min}$, and satisfies
\[
\mathbb P\left[
\left\|
\mathcal M(\mathbf X)-M
\right\|_\infty
\geq
{\alpha}
\right]
\leq\frac14,
\]
there exists a corresponding $(\varepsilon, \delta)$-DP algorithm $\mathcal{A}$ such that, for every $Q=(q_1, \cdots, q_S)^\top \in \mathcal{P}_{S,S}$, when given a dataset of $N$ users, each with $m:=\left\lceil 8\pi_{\min} T\right\rceil \geq C_0\log(NS)$ i.i.d. samples from $\bigotimes_{i=1}^S q_i$, $\mathbf{Y}$, it satisfies
\[
\mathbb P\left[
\left\|\mathcal A(\mathbf Y)-\overline Q\right\|_\infty
\geq \alpha
\right]
\leq \frac13.
\]
\end{theorem}

\begin{proof}
Proceeding similarly to the proof of Lemma~\ref{app:lowerbound-item-formal}, we construct a family of Markov chains which embeds the unknown
distribution matrix $Q$ in a block of its transition matrix.

\paragraph{Construction of the Markov chain.}
Set $\eta:=\frac{S\pi_{\min}}{1-S\pi_{\min}}$ and let $U:=\frac1S\mathbf 1\mathbf 1^\top$ denote the $S\times S$ uniform stochastic matrix. For any $Q\in\mathcal Q$, define
\[
P_{\pi_{\min},Q}
=
\begin{pmatrix}
0 & (1-\eta)Q+\eta U\\[1mm]
\eta U & (1-\eta)I_S
\end{pmatrix},
\]
Define $\mathcal A=\{a_1,\ldots,a_S\}$ and $\mathcal B=\{b_1,\ldots,b_S\}$, where the first $S$ states are mapped to $\mathcal{A}$ and the remaining states are mapped to $\mathcal{B}$. Equivalently, the transitions are as follows:
\begin{itemize}
    \item From $a_i$, move to $b_j$ with probability $(1-\eta)Q_{ij}+\frac{\eta}{S}$;
    \item From $b_j$, move to a uniformly random state in $\mathcal A$ with total probability $\eta$;
    \item With probability $1-\eta$, remain at $b_j$.
\end{itemize}

Since $\pi_{\min}<1/(2S)$, we have $0<\eta<1$.

\paragraph{Stationary distribution and ergodicity.}
Let $\pi$ be the stationary distribution of $P_{\pi_{\min}, Q}$. For $j\in[S]$, let $c_j:=\sum_{i=1}^S Q_{ij}$ be the $j$-th column sum of $Q$. We show that $\pi(a_i)=\pi_{\min}$ for every $i\in[S]$, and $\pi(b_j) = \pi_{\min}+\frac{1-2S\pi_{\min}}{S}c_j$ for $j \in [S]$. Since the matrix $Q$ is row-stochastic, $\sum_j c_j = \sum_{j} \sum_{i} Q_{i,j} = S$. Hence,
\[
\sum_{i=1}^S\pi(a_i)+\sum_{j=1}^S\pi(b_j)
=
S\pi_{\min}+S\pi_{\min}+1-2S\pi_{\min}
=
1.
\]
For every $i\in[S]$,
\[
\pi({a_i}) = \sum_{j=1}^S\pi(b_j)P_{\pi_{\min},Q}(b_j,a_i)
=
\frac{\eta}{S}\sum_{j=1}^S\pi(b_j)
=
\frac{\eta(1-S\pi_{\min})}{S}
=
\pi_{\min}.
\]
For every $j\in[S]$,
\begin{align*}
\pi(b_j) = &\sum_{i=1}^S\pi(a_i)P_{\pi_{\min},Q}(a_i,b_j)
+
\pi(b_j)P_{\pi_{\min},Q}(b_j,b_j) \\
&= \sum_{i=1}^S \pi_{\min} \left((1 - \eta)Q_{i,j} + \frac{\eta}{S}\right) + \sum_{k=1}^S \pi(b_k)P_{\pi_{\min}, Q}(b_k, b_j)\\
&= \pi_{\min}\left((1-\eta)c_j+\eta\right)
+
(1-\eta)\left(\pi_{\min}+\frac{1-2S\pi_{\min}}{S}c_j\right)\\
&= \pi_{\min} + \pi_{\min} (1-\eta)c_j + (1 - \eta)\left(\frac{1  - 2S\pi_{\min}}{S}\right)c_j \\
&= \pi_{\min} + (1 - \eta)c_j \left(\frac{1 - S\pi_{\min}}{S}\right) \\
&= \pi(b_j).
\end{align*}
The last equality follows from the fact that $\eta:= \frac{S\pi_{\min}}{1 - S\pi_{\min}}$. Thus, $\pi$ is stationary.

Since $1-2S\pi_{\min}>0$ and $c_j\geq0$, $\pi(b_j)\geq\pi_{\min}$. Therefore, $\min_{x\in\mathcal A\cup\mathcal B}\pi(x)=\pi_{\min}$. Furthermore, $(1-\eta)Q_{ij}+\frac{\eta}{S}>0$ for every $i,j$. Hence every state in $\mathcal A$ can reach every state in $\mathcal B$, and every state in $\mathcal B$ can reach every state in $\mathcal A$. The chain is therefore irreducible. Since $P_{\pi_{\min},Q}(b_j,b_j)=1-\eta>0$, the chain is also aperiodic. Hence, $P_{\pi_{\min},Q}$ is ergodic.

\paragraph{Reduction to estimating product-distribution samples.}
Suppose that each source user $u$ receives $Y^{(u)}_1,\ldots,Y^{(u)}_m \stackrel{\mathrm{i.i.d.}}{\sim}\mathsf D_Q$, where $Y^{(u)}_r = \left( Y^{(u)}_{r,1},\ldots,Y^{(u)}_{r,S} \right)$. Thus, for every $i\in[S]$, $Y^{(u)}_{1,i},\ldots,Y^{(u)}_{m,i} \stackrel{\mathrm{i.i.d.}}{\sim}q_i$. We construct one length-$T$ trajectory for each source user. Note that any transition from $\mathcal{A}$ to $\mathcal{B}$ requires a sample from $D_Q$. Consider the following two-stage algorithm for defining $\mathcal{M}$ based on $\mathcal{A}$.

\begin{enumerate}
    \item \textbf{Stage 1 Begins:} For all $i \in [S]$ and $u \in [N]$, let $c^{(u)}_i = 0$.
    \item For each $u\in[N]$, generate $L^{(u)}_0,\ldots,L^{(u)}_T \in \{B,A_1,\ldots,A_S\}$ as follows:
    \begin{enumerate}
        \item Set $L^{(u)}_0=B$.
        \item For $t=0,\ldots,T-1$:
        \begin{enumerate}
            \item If $L^{(u)}_t=A_i$, set
            $L^{(u)}_{t+1}=B$ deterministically and set $c^{(u)}_i \leftarrow c^{(u)}_i + 1$.
            \item If $L^{(u)}_t=B$, set
            \[
            L^{(u)}_{t+1}
            =
            \begin{cases}
                B & \text{with probability }1-\eta,\\
                A_i & \text{with probability }\eta/S,
                \quad i\in[S];
            \end{cases}
            \]
        \end{enumerate}
        \item If $c_i^{(u)} > m$, return $U$.
    \end{enumerate}
    \item \textbf{Stage 2 Begins:} For each $u$, generate the labeled trajectory $X^{(u)}_0,\ldots,X^{(u)}_T \in\mathcal A\cup\mathcal B$ as follows.
    \begin{enumerate}
        \item Initially sample $J_0^{(u)}\sim\operatorname{Unif}([S])$ and set $X^{(u)}_0=b_{J_0^{(u)}}$.
        \item Set $r^{(u)}_{i}=1$ for every $i\in[S]$.
        \item For $t=0,\ldots,T-1$:
        \begin{enumerate}
            \item If $L^{(u)}_t=A_i$, then
            $X^{(u)}_t=a_i$. Let $Z=Y^{(u)}_{r^{(u)}_i,i}$ be the next unused sample from $q_i$. Independently, set
            \[
            X^{(u)}_{t+1}
            =
            \begin{cases}
                b_Z & \text{with probability }1-\eta,\\
                b_J,\quad J\sim\operatorname{Unif}([S])
                & \text{with probability }\eta.
            \end{cases}
            \]

            \item If $L^{(u)}_t=B$ and $L^{(u)}_{t+1}=B$, then
            $X^{(u)}_t=b_j$ for some $j$, and we set $X^{(u)}_{t+1}=b_j$.

            \item If $L^{(u)}_t=B$ and $L^{(u)}_{t+1}=A_i$, set $X^{(u)}_{t+1}=a_i$.
        \end{enumerate}
    \end{enumerate}
    \item \textbf{Estimating $Q$ from simulated samples:} Run $\widehat P = \mathcal M\left(X^{(1)}_0,\ldots,X^{(1)}_T,\ldots,X^{(N)}_0,\ldots,X^{(N)}_T\right)$.
    \item Let $\widehat R$ denote the $\mathcal A$-to-$\mathcal B$ block of $\widehat P$ (the block corresponding to the first $S$ rows and last $S$ columns) and define
    \[
    Z
    =
    \frac{\widehat R-\eta U}{1-\eta}.
    \]
    \item For every $i\in[S]$, set $\widehat Q_{i,\cdot} \in \arg\min_{q\in\Delta(S)} \|q-Z_{i,\cdot}\|_1$ and return $\widehat{Q}$.
\end{enumerate}
For every $u\in[N]$ and $i\in[S]$, let $V_{u,i} := \sum_{t=0}^{T-1} \mathbf 1_{\{L^{(u)}_t=A_i\}} $ and consider the event
\[
\mathcal E
:=
\left\{
V_{u,i}\leq m
\text{ for every }u\in[N]\text{ and }i\in[S]
\right\}.
\]
Note that the event $\mathcal E$ depends only on the data-independent Stage 1 (before step 3) of the reduction and is therefore independent of all the samples $\mathbf Y$. Since the event $\mathcal{E}$ implies that all the $\mathcal{A}$ to $\mathcal{B}$ transitions are at most $m$, the event $\mathcal E$ guarantees that the required sample $Y^{(u)}_{r_{u,i},i}$ always exists. Conditional on $X^{(u)}_t=a_i$,
\[
\mathbb P\left[X^{(u)}_{t+1}=b_j\mid X^{(u)}_t=a_i\right]
=
(1-\eta)Q_{ij}+\frac{\eta}{S}.
\]
Conditional on $X^{(u)}_t=b_j$,
\[
X^{(u)}_{t+1}
=
\begin{cases}
b_j & \text{with probability }1-\eta,\\
a_i & \text{with probability }\eta/S,
\quad i\in[S].
\end{cases}
\]
Thus, conditioned on $\mathcal E$, each simulated trajectory has exactly the law of a length-$T$ trajectory from $P_{\pi_{\min},Q}$ with initial distribution $\mu=\operatorname{Unif}(\mathcal B)$. Since the source users and all simulation randomness are independent,
the $N$ trajectories are independent.

\paragraph{Upper Bound on the Probability of $\overline{\mathcal E}$.}
Fix a user $u\in[N]$. We emphasize that the variables
$\mathbf 1\{L^{(u)}_t=A_i\}$ are not independent across the physical
time index $t$, since every visit to an $A$-state is followed
deterministically by a transition back to the $B$-state. Since the first stage starts in a $B$-state, consider a sequence of independent random variables
\[
C^{(u)}_1,C^{(u)}_2,\ldots
\in\{0,1,\ldots,S\}
\]
where $\mathbb P[C^{(u)}_r=0]=1-\eta$ and $\mathbb P[C^{(u)}_r=j]=\frac{\eta}{S}$ for $j\in[S]$.

At the $r$-th time for which $L_t^{(u)} = B$, the construction uses
$C^{(u)}_r$ to determine $L_{t+1}^{(u)}$: if $C^{(u)}_r=0$, $L_{t+1}^{(u)}$ is in a $B$ state, whereas if $C^{(u)}_r=j$, it moves to $L_{t+1}^{(u)} = A_j$ and returns deterministically to the $B$-state at the following step. For the fixed row $i$, the variables $\mathbf 1_{\{C^{(u)}_r=i\}}$ are therefore independent Bernoulli random variables with parameter $p := \frac{\eta}{S} = \frac{\pi_{\min}}{1-S\pi_{\min}}$.
Let $G^{(u,i)}_\ell$ be the number of $B$-state departure opportunities between the $(\ell-1)$-st and the $\ell$-th visits to
$A_i$. Then $G^{(u,i)}_1,\ldots,G^{(u,i)}_{m+1}
\stackrel{\mathrm{i.i.d.}}{\sim}
\operatorname{Geom}(p)$, and
\[
J_{u,i}
:=
\sum_{\ell=1}^{m+1}G^{(u,i)}_\ell
\]
is the number of transitions from a$B$-state to $A_i$ required to visit
$A_i$ exactly $m+1$ times.

Recall that $V_{u,i} = \sum_{t=0}^{T-1} \mathbf 1\{L^{(u)}_t=A_i\}$ is the number of samples from the $i$-th row distribution required
by the $u$-th simulated trajectory. Let $K_u(T)$ denote the number
of $B$-state departure opportunities occurring during the first
$T$ transitions. Since at most one such opportunity occurs per
physical transition, $K_u(T)\leq T$. Moreover, if $V_{u,i}>m$, then the skeleton visits $A_i$ at least $m+1$ times within the first $T$ transitions. Consequently,
\[
\{V_{u,i}>m\}
\subseteq
\{J_{u,i}\leq K_u(T)\}
\subseteq
\{J_{u,i}\leq T\}.
\]

Since $\pi_{\min}<1/(2S)$, $p = \frac{\pi_{\min}}{1-S\pi_{\min}} \leq 2\pi_{\min}$. With the choice $m=\lceil 8\pi_{\min} T\rceil$, we have $pT \leq 2\pi_{\min} T \leq \frac{m}{4}$ and hence $T \leq \frac{m}{4p} \leq \frac{m+1}{2p}$. Applying Lemma~\ref{lem:geomchernoff} with $m+1$ geometric random
variables and lower-tail parameter $1/2$, we obtain
\begin{align*}
\mathbb P[V_{u,i}>m]
&\leq
\mathbb P[J_{u,i}\leq T] \leq
\mathbb P\left[
J_{u,i}
\leq
\frac{m+1}{2p}
\right] \leq
\exp\left(-c\frac{m+1}{4}\right) \leq
\exp(-c_0m),
\end{align*}
where $c_0>0$ is a universal constant. Taking a union bound over all $N$ users and all $S$ rows gives
\[
\mathbb P[\overline{\mathcal E}]
=
\mathbb P\left[
V_{u,i}>m
\text{ for some }u\in[N],\,i\in[S]
\right]
\leq
NS\exp(-c_0m).
\]

\paragraph{Privacy guarantee.}
Consider two neighboring source databases $\mathbf Y = \left(Y^{(1)},\ldots,Y^{(u^\star)},\ldots,Y^{(N)}\right)$ and $\mathbf Y' = \left(Y^{(1)},\ldots,{Y'}^{(u^\star)},\ldots,Y^{(N)}\right)$, which differ only in the samples of a single source user $u^\star$.

The event $\mathcal E$ and the layer skeletons are independent of the
source data. Conditioned on any fixed collection of skeletons
satisfying $\mathcal E$, changing source user $u^\star$ can change
only the trajectory $X^{(u^\star)}$. All other trajectories remain
unchanged. Thus, the two trajectory databases supplied to
$\mathcal M$ differ in at most one entire trajectory and are
neighboring under user-level adjacency. Consequently, by the user-level privacy guarantee of $\mathcal M$ and
post-processing, for every measurable set $\mathcal S$,
\[
\mathbb P\left[
\mathcal R(\mathbf Y)\in\mathcal S
\mid\mathcal E
\right]
\leq
e^\varepsilon
\mathbb P\left[
\mathcal R(\mathbf Y')\in\mathcal S
\mid\mathcal E
\right]
+\delta.
\]

Since $\mathcal E$ is data-independent and the estimator returns the
fixed matrix $U$ on $\overline{\mathcal E}$,
\begin{align*}
\mathbb P\left[\mathcal R(\mathbf Y)\in\mathcal S\right]
&=
\mathbb P\left[
\mathcal R(\mathbf Y)\in\mathcal S
\mid\mathcal E
\right]\mathbb P[\mathcal E]
+
\mathbf 1_{\{U\in\mathcal S\}}
\mathbb P[\overline{\mathcal E}]\\
&\leq
e^\varepsilon
\mathbb P\left[
\mathcal R(\mathbf Y')\in\mathcal S
\mid\mathcal E
\right]\mathbb P[\mathcal E]
+
\delta
+
e^\varepsilon \mathbf 1_{\{U\in\mathcal S\}}
\mathbb P[\overline{\mathcal E}]\\
&\leq
e^\varepsilon
\mathbb P\left[\mathcal R(\mathbf Y')\in\mathcal S\right]
+\delta.
\end{align*}
Hence, $\mathcal R$ is
$(\varepsilon,\delta)$-user-level differentially private. Note that this argument holds even if $\delta = 0$.

\paragraph{Utility Guarantee.}
Since $\widehat q_i$ is an $\ell_1$ projection and
$q_i\in\Delta(S)$,
\begin{align*}
\|\widehat q_i-q_i\|_1 \leq \|\widehat q_i-Z_{i,\cdot}\|_1 +
\|Z_{i,\cdot}-q_i\|_1 \leq 2\|Z_{i,\cdot}-q_i\|_1.
\end{align*}
Therefore,
\begin{align*}
\|\widehat Q-Q\|_\infty
&\leq
2\|Z-Q\|_\infty =
\frac{2}{1-\eta}
\left\|
\widehat R-\big((1-\eta)Q+\eta U\big)
\right\|_\infty \leq
\frac{2}{1-\eta}
\left\|
\widehat P-P_{\pi_{\min},Q}
\right\|_\infty.
\end{align*}

It follows that, for every $\alpha > 0$,
\begin{align*}
\mathbb P\left[
\|\mathcal R(\mathbf Y)-Q\|_\infty\geq \alpha
\right]
&\leq
\mathbb P\left[
\left\|
\mathcal M(\mathbf X)-P_{\pi_{\min},Q}
\right\|_\infty
\geq
\frac{1-\eta}{2}\alpha
\,\middle|\,
\mathcal E
\right]\mathbb P[\mathcal E] + \mathbb P[\overline{\mathcal E}]\\
&\leq
\mathbb P\left[
\left\|
\mathcal M(\mathbf X)-P_{\pi_{\min},Q}
\right\|_\infty
\geq
\frac{1-\eta}{2}\alpha
\right]
+
NS\exp(-c_0m).
\end{align*}
Here, in the final probability, $\mathbf X$ consists of $N$ independent genuine trajectories from $P_{\pi_{\min},Q}$. If $\pi_{\min}\leq1/(3S)$, then $\eta=\frac{S\pi_{\min}}{1-S\pi_{\min}}\leq\frac{1}{2}$ which implies $1-\eta\geq1/2$ which further implies that
\begin{align*}
\mathbb P\left[
\|\mathcal R(\mathbf Y)-Q\|_\infty\geq \alpha
\right]
&\leq
\mathbb P\left[
\left\|
\mathcal M(\mathbf X)-P_{\pi_{\min},Q}
\right\|_\infty
\geq
\frac{\alpha}{4}
\right]
+
NS\exp(-c_0m).
\end{align*}

If $m\geq C_0\log(NS)$ for a sufficiently large $C_0$, then $NS\exp(-c_0m)\leq\frac{1}{12}$. The accuracy guarantee of $\mathcal{M}$ on ergodic Markov chains proves the claim.

\end{proof}

\begin{theorem}[Complete version of Theorem~\ref{thm:user-level-markov-approx-lower-informal}]
\label{thm:markov-chain-estimation-approx}
Let $0 < \alpha \leq \frac{1}{384\sqrt{2}}$, $0 < \varepsilon < 1$ and $\delta \leq \min\left\{\frac{C_\delta \varepsilon}{S}, \varepsilon^2\right\}$ for some universal constant $C_\delta$. Suppose that there exists an $(\varepsilon, \delta)$-DP learner $\mathcal{M}$ that takes as input $N$ independent length $T$ trajectories $\mathbf{X}$ sampled from a Markov chain with transition matrix $P$  with $\min_{x} \pi(x) = \pi_{\min}$. If
\begin{equation}
\label{eq:high-prob-markov-condition}
    \mathbb P\left[
\left\|
\mathcal M(\mathbf X)-P
\right\|_\infty
\leq
\alpha
\right]
\geq\frac34,
\end{equation}
then for $T \gtrsim \frac{\log(NS)}{\pi_{\min}}$,
\begin{align*}
    N
    \gtrsim \frac{1}{\varepsilon}\min\left\{S^2, \log\left(\frac{1}{\delta}\right)\right\} + \frac{S^{3/2}}{\alpha\eps\sqrt{\pi_{\min}T}\log^{C_\ell}((\alpha\eps)^{-1}edk)} + \frac{S}{\pi_{\min}T\alpha^2},
\end{align*}
users are necessary, where $C_\ell$ is a universal constant.
\end{theorem}

\begin{proof}
Since there exists an $(\varepsilon, \delta)$-DP learner $\mathcal{M}$ that takes $N$ independent length $T$ trajectories $\mathbf{X}$ sampled from a Markov chain with transition matrix $P$ with $\min_{x} \pi(x) = \pi_{\min}$ as input and satisfies~\eqref{eq:high-prob-markov-condition}.
Then, by Theorem~\ref{thm:product-to-markov-reduction}, there exists an algorithm $\mathcal{A}$ that, when given data from $N$ users, each with $m = \left\lceil 8\pi_{\min} T\right\rceil$ samples each from the product distribution corresponding to $\overline{Q}$  satisfies
    \[
        \mathbb P\left[
    \left\|\mathcal A(\mathbf Y)-\overline Q\right\|_\infty
    \leq 4\alpha
    \right]
    \geq \frac23.
    \]
    Since $4\alpha \leq \frac{1}{96\sqrt{2}}$, applying Theorem~\ref{thm:high-prob-approx-final-one-infty-norm} and substituting $m = \Theta\left(\pi_{\min}T\right)$ proves the claim.
\end{proof}

\begin{theorem}
\label{thm:user-level-markov-pure-lower-formal}
Suppose that there exists an $\varepsilon$-DP learner that takes as input $N$ independent length $T$ trajectories $\mathbf{X}$ sampled from a Markov chain with transition matrix $P$ with $\min_{x} \pi(x) = \pi_{\min}$ and pseudo-spectral gap $\gamma_{ps}$. If
\[
\mathbb P\left[
\left\|
\mathcal M(\mathbf X)-P
\right\|_\infty
\leq
\alpha
\right]
\geq\frac34,
\]
then for $T \gtrsim \frac{\log(NS)}{\pi_{\min}}$,
\begin{align*}
    N
    \gtrsim \frac{S^2}{\eps} + \frac{S^{3/2}}{\eps\alpha\sqrt{\pi_{\min}T}} + \frac{S}{\pi_{\min}T\alpha^2}.
\end{align*}
\end{theorem}
\begin{proof}
The proof follows in the same manner as Theorem~\ref{thm:markov-chain-estimation-approx}. For the final step, apply Theorem~\ref{thm:pure-user-level-product-lower-bound} with $\Theta(\pi_{\min}T)$ samples per-user, proving this claim.
\end{proof}

\end{document}